\documentclass[11pt]{article}
\usepackage{amsmath}
\usepackage{amssymb}
\usepackage{cite}
\usepackage{graphicx,bbm,mathrsfs}

\usepackage{geometry}       
\newcommand{\be}{\begin{equation}}  
\newcommand{\ee}{\end{equation}}  
\newcommand{\bea}{\begin{eqnarray}}  
\newcommand{\eea}{\end{eqnarray}}  
\newcommand{\ol}[1]{\overline{#1}}
\newcommand{\hc}{+\,\mathrm{h.c.}}
\newcommand{\vev}[1]{\langle #1 \rangle}

\newcommand{\SU}[1]{\ensuremath{\mathrm{SU}(#1)}}
\newcommand{\U}[1]{\ensuremath{\mathrm{U}(#1)}}
\newcommand{\into}{\ensuremath{\,\rightarrow\,}}

\newcommand{\msq}[1]{\ensuremath{{m^2_{\text{\textit{\tiny\!#1}}}}}}
\newcommand{\hyuk}[1]{\ensuremath{h^{\text{\textit{\tiny#1}}}}}
\newcommand{\yuk}[1]{\ensuremath{y^{\text{\textit{\tiny#1}}}}}
\newcommand{\yukt}[1]{\ensuremath{{\tilde y}^{\text{\textit{\tiny#1}}}}}
\newcommand{\cbm}[1]{\ensuremath{c^{\text{\textit{\tiny#1}}}}}
\newcommand{\fpr}[1]{\ensuremath{f^{\text{\textit{\tiny#1}}}}}
\newcommand{\kap}[1]{\ensuremath{\kappa^{\text{\textit{\tiny#1}}}}}
\newcommand{\atr}[1]{\ensuremath{a^{\text{\textit{\tiny#1}}}}}
\newcommand{\lam}[1]{\ensuremath{\lambda^{\text{\textit{\tiny#1}}}}}
\newcommand{\Ywf}[1]{\ensuremath{{\bf Y}^{\text{\textit{\tiny#1}}}}}

\newcommand{\Phisp}[1]{\ensuremath{{\bf\Phi}^{\text{\textit{\tiny#1}}}}}
\newcommand{\Phitsp}[1]{\ensuremath{{\widetilde{\bf\Phi}}^{\text{\textit{\tiny#1}}}}}
\newcommand{\Phihsp}[1]{\ensuremath{{\widehat{\bf\Phi}}^{\text{\textit{\tiny#1}}}}}
\newcommand{\Lamsp}[1]{\ensuremath{{\bf\Lambda}^{\text{\textit{\tiny\!#1}}}}}
\newcommand{\Lamtsp}[1]{\ensuremath{{\widetilde{\bf\Lambda}}^{\text{\textit{\tiny\!#1}}}}}
\newcommand{\Lamhsp}[1]{\ensuremath{{\widehat{\bf\Lambda}}^{\text{\textit{\tiny\!#1}}}}}
\newcommand{\Delsp}[1]{\ensuremath{{\bf\Delta}^{\text{\textit{\tiny\!#1}}}}}
\newcommand{\Deltsp}[1]{\ensuremath{{\widetilde{\bf\Delta}}^{\text{\textit{\tiny\!#1}}}}}
\newcommand{\Delhsp}[1]{\ensuremath{{\widehat{\bf\Delta}}^{\text{\textit{\tiny\!#1}}}}}
\newcommand{\Ui}{\ensuremath{{{\text{\textit{\tiny U}}}}}}
\newcommand{\Di}{\ensuremath{{{\text{\textit{\tiny D}}}}}}

\newcommand\lsim{\mathrel{\rlap{\lower4pt\hbox{\hskip1pt$\sim$}}
    \raise1pt\hbox{$<$}}}
\newcommand\gsim{\mathrel{\rlap{\lower4pt\hbox{\hskip1pt$\sim$}}
    \raise1pt\hbox{$>$}}}

\newcommand{\Ell}{\mathscr{L}}

\newcommand{\wt}{\widetilde}

\providecommand{\tabularnewline}{\\}

\newcommand{\captionfonts}{\small}

\makeatletter  
\long\def\@makecaption#1#2{%
  \vskip\abovecaptionskip
  \sbox\@tempboxa{{\captionfonts #1: #2}}%
  \ifdim \wd\@tempboxa >\hsize
    {\captionfonts #1: #2\par}
  \else
    \hbox to\hsize{\hfil\box\@tempboxa\hfil}%
  \fi
  \vskip\belowcaptionskip}
\makeatother   

\begin{document}

\vspace*{1.2cm}

\begin{center}

\thispagestyle{empty}

{\Large\bf  The supersymmetric flavour problem in 5D GUTs \\[2mm] and its consequences for LHC phenomenology}\\[10mm]

{\large F.~Br\"ummer$^{\,a}$, S.~Fichet$^{\,b}$, S.~Kraml$^{\,b}$}\\[5mm]

{\it
$^{a}$~Deutsches Elektronen-Synchrotron DESY,\\ 
Notkestra\ss e 85, D-22607 Hamburg, Germany\\[3mm]
$^{b}$~Laboratoire de Physique Subatomique et de Cosmologie, UJF Grenoble 1, 
CNRS/IN2P3, 53 Avenue des Martyrs, F-38026 Grenoble, France
}

\vspace*{12mm}

\begin{abstract}
\noindent We study supersymmetric models with a GUT-sized extra dimension, where both the Higgs fields and the SUSY breaking hidden sector are localized on a 4D brane. Exponential wave function profiles of the matter fields give rise to hierarchical structures in the Yukawa couplings and soft terms. Such structures can naturally explain hierarchical fermion masses and mixings, while at the same time alleviating the supersymmetric flavour problem.
We discuss two sources of supersymmetry breaking, radion mediation and brane fields, and perform a detailed numerical analysis, thoroughly taking into account the proliferation of unknown ${\cal O}(1)$ coefficients that occurs in this class of models.  It turns out that additional assumptions on supersymmetry breaking are necessary to evade the stringent experimental bounds on lepton flavour violation. The favourable regions of parameter space are then examined with regards to their LHC phenomenology. They generically feature heavy gluinos and squarks beyond current bounds. Lepton flavour violation in SUSY cascade decays can give interesting signatures. 
\end{abstract}

\end{center}

\clearpage
%

\section{Introduction}

In the Standard Model, the three generations of quarks and leptons follow a peculiar pattern of hierarchical masses and mixings. Localizing the Standard Model matter fields in the bulk of a compact extra dimension, for instance on a slice of AdS$_5$ \cite{Randall:1999ee}, naturally leads to such flavour hierarchies \cite{Grossman:1999ra,Gherghetta:2000qt}. 

For a strongly warped extra dimension, warping could explain the discrepancy between the Planck scale and the electroweak scale \cite{Randall:1999ee}. In that case the lowest massive Kaluza-Klein modes should have masses around a TeV. In less strongly warped (or even unwarped) models, the KK scale can be only a few orders of magnitude below the Planck scale, and the electroweak hierarchy problem can be solved by TeV-scale supersymmetry. Wave function localization still accounts for the Yukawa hierarchies, and the warped internal space still allows for an interpretation of the 5D model as a holographic dual of some strongly coupled CFT \cite{ArkaniHamed:2000ds}. If the KK scale is not too far from the 4D GUT scale, such a model may even be compatible with standard (logarithmic) MSSM gauge coupling unification, with the GUT group broken at the compactification scale by boundary conditions. 

As an additional benefit of such models, localizing matter fields in 5D can alleviate the SUSY flavour problem (see, for example, \cite{Choi:2003fk,Nomura:2006pn,Nomura:2008pt,Dudas:2010yh}). This is what happens if the Higgs fields are brane fields, and supersymmetry breaking is localized on the same brane (and possibly in the gravitational background). In that case the trilinear soft terms follow a hierarchy structure similar to that of the the Yukawa couplings, and will thus be approximately diagonal in the fermion mass eigenstate basis. Off-diagonal scalar soft masses may also be suppressed. A related mechanism was originally advocated in a 4D setup, with the visible sector fields acquiring large anomalous dimensions from their couplings to a strongly coupled CFT \cite{Nelson:2000sn}. These two pictures can be argued to be related by AdS/CFT duality \cite{Choi:2003fk}. Whether or not FCNCs in such a setup are sufficiently suppressed to evade experimental bounds is however heavily model-dependent. 

This question is the subject of the present paper. We will study, using a concrete example model as our benchmark, to what extent wave-function localization in 5D is enough to suppress flavour-violating processes, and to what extent some additional mechanism or some residual tuning is still needed. We will discuss mass spectra and collider signatures in the favoured regions of parameter space. In the present work we focus on FCNCs and neglect CP violation.

Previous studies of this subject have revealed that wave-function suppression alone is in general not enough to evade the stringent experimental bounds, and that some additional mechanism or residual tuning of the parameters is needed. The authors of \cite{Choi:2003fk} suggested that 5D mass terms, which determine the localization properties of the bulk zero modes, might be required to be quantized in discrete units by the underlying fundamental theory. With suitably chosen discrete 5D masses, and some moderate residual tuning, they argued that all flavour constraints could be satisfied. In \cite{Nomura:2008pt} a $\U 1$ symmetry was used to forbid some of the couplings between the SUSY breaking sector and the visible sector, thus enforcing flavour-diagonal $a$-terms. Flavour violation in the scalar soft masses was then argued to typically be sufficiently suppressed by wave-function localization alone. A recent study of flavour violation in a 5D warped model with gaugino mediation was conducted in \cite{Okada:2011ed}.

Much of our analysis is sufficiently broad to be representative for any 5D SUSY GUT with the Higgs and SUSY breaking sectors localized on the same brane. In particular the results on the low-energy spectrum should generalize, at least qualitatively, to any such model (for instance, those constructed e.g.~in \cite{Nomura:2008pt}). However, when we do need to work with a concrete model for definiteness, we choose the ``holographic GUT'' model of Nomura, Poland and Tweedie (NPT) \cite{Nomura:2006pn}. In the NPT model, there is a warped extra dimension, and the bulk gauge group is $\SU6$. It is broken by boundary conditions to $\SU5\times\U1$ on the UV brane, and by the VEV of an adjoint brane field $\Sigma$ to $\SU4\times\SU2\times\U1$ on the IR brane. This gives essentially the Standard Model gauge group in the 4D effective field theory. Matter fields are localized in the bulk, and boundary conditions are chosen such that their zero modes furnish precisely the matter content of the MSSM. The MSSM Higgs fields are pseudo-Goldstone bosons arising from $\Sigma$ \cite{Inoue:1985cw,Contino:2003ve}, and the symmetry breaking structure results in a GUT-scale degenerate Higgs mass matrix \cite{Brummer:2010gh}, thus automatically solving the $\mu$ problem. Supersymmetry is broken both by some hidden sector chiral superfields on the IR brane, and by the 5D gravitational background (i.e.~by non-vanishing $F$-terms of the radion and chiral compensator superfields in 4D language). Models of this type have been shown to be able to give realistic low-energy mass spectra when assuming flavour-blind supersymmetry breaking \cite{Brummer:2010gh}. Therefore it is interesting to now relax the assumption of flavour-blindness and study the impact on flavour observables.

An important aspect of our work is that we carry out an extensive numerical analysis, giving quantitative results for flavour constraints and collider phenomenology. In that respect we go beyond the previous literature, in particular \cite{Choi:2003fk,Nomura:2006pn,Nomura:2008pt,Dudas:2010yh}, which contain only qualitative discussions. 

To avoid confusion it may be worth noting that our study is conceptually unrelated to work on flavour violation in Randall-Sundrum models.\footnote{See, for instance, \cite{Casagrande:2008hr} and references therein.} Such models are typically non-supersymmetric, since the hierarchy problem is solved by the strongly warped fifth dimension. However, with the Standard Model fermions in the bulk, KK mode exchange represents a potentially dangerous source for FCNCs. In the models which we are concerned with, by contrast, the KK modes are far too heavy to contribute to flavour violation. Flavour-violating processes can instead be mediated by the exchange of MSSM superpartners. 

This work is organized as follows: We start by reviewing first the supersymmetric flavour problem in Section~\ref{sfp}, and then the generation of hierarchical fermion masses and mixings from wave-function localization in Section~\ref{hfroml}. We present our parameterization of supersymmetry breaking in warped 5D models in Section~\ref{susyb}. Estimates for the magnitudes of soft parameters from naive dimensional analysis are presented in~\ref{branefields}, for the case that SUSY breaking mediation is dominated by brane fields. Soft terms for the case that radion mediation dominates are given in~\ref{rmsb}. In Section~\ref{method} we explain details about the numerical analysis which we perform to scan over a large number of models, with the results presented in Section~\ref{results}. Conclusions are given in Section~\ref{conclusions}. Appendix~\ref{modsb} contains a brief review of an example model in which SUSY is broken and the radius of the extra dimension is stabilized, both of which we assume to be the case in the main text without providing an actual mechanism. Appendix~\ref{appendix_epsilon_expansions} contains some analytic expressions for the diagonalization of hierarchical Yukawa matrices, which were used in the numerical analysis.

\section{The SUSY flavour problem}\label{sfp}

We start with a brief recapitulation of the flavour problem in supersymmetry, see e.g.~\cite{Feng:2007ke}, to set up our notation and terminology. The Lagrangian for the flavour sector of the R-parity symmetric MSSM reads
\be\begin{split}
{\cal L}=&\int d^4\theta\,\left({Q_i}^\dag Q_i+{U_i}^\dag U_i+{D_i}^\dag D_i+{E_i}^\dag E_i+{L_i}^\dag L_i\right)\\
&+\int d^2\theta\,\left(\yuk{U}_{ij}\,H_u Q_iU_j+ \yuk{D}_{ij}\,H_d Q_iD_j+\yuk{E}_{ij}\,H_d L_i E_j\right)\hc\\
&+ \msq{Q}_{ij}\,q^\dag_i q_j+\msq{U}_{ij}\,u^\dag_i u_j+\msq{D}_{ij}\,d^\dag_id_j+\msq{L}_{ij} l_i^\dag l_j+\msq{E}_{ij} e^\dag_i e_j\\
&+\left(\atr{U}_{ij}\,h_u q_i u_j+\atr{D}_{ij}\,h_d q_id_j+\atr{E}_{ij}\,h_d l_i e_j\hc\right)\,.
\end{split}
\ee
Here $Q$, $U$ and $D$ are the quark superfields and $E$ and $L$ are the lepton superfields (we are omitting right-handed neutrinos) with $q$, $u$, $d$, $l$ and $e$ their scalar components. $H_u$ and $H_d$ are the Higgs superfields and $h_u$ and $h_d$ their scalar components. The couplings $\yuk{U}$, $\yuk{D}$, $\yuk{E}$, $\msq{Q}$, $\msq{U}$, $\msq{D}$, $\msq{L}$, $\msq{E}$, $\atr{U}$, $\atr{D}$ and $\atr{E}$ are complex $3\times 3$ matrices in flavour space; the soft mass matrices $\msq{X}$  with $X=U,D,Q,L,E$ are restricted to be hermitian. 

Flavour rotations $X\into{\cal U}_X X$, with $X=$ any matter superfield and ${\cal U}_X$ a unitary $3\times 3$ matrix, leave the kinetic terms invariant but change the couplings. By appropriately rotating $E\into{\cal U}_E E$ and $L\into{\cal U}_L L$, the Yukawa matrix $\yuk{E}$ can be diagonalized. 

Likewise, it is possible to choose either $\yuk{D}$ or $\yuk{U}$ diagonal. However, either of these two choices fixes ${\cal U}_Q$ up to a phase, so $\yuk{D}$ and $\yuk{U}$ cannot be chosen diagonal simultaneously: There is CKM mixing in the quark sector of the Standard Model. It is of course possible to go to a field basis where $\SU{2}_L$ is non-linearly realized, by splitting the weak doublet according to $Q=(U_L,\,D_L)^T$ (in a gauge where the Higgs expectation values will eventually be $\vev{h_u}=(0,\;v^u)^T$ and $\vev{h_d}=(v_d,\;0)^T$). One can then perform independent flavour rotations on $U_L$ and $D_L$  and thus diagonalize both up-type and down-type mass matrices simultaneously; this is the ``super-CKM basis'', in which the quarks are mass eigenstates but no longer weak interaction eigenstates.

In the limit of vanishing soft terms, flavour-changing processes are suppressed in the MSSM as they are in the Standard Model. In particular, there are no FCNCs at the tree level. Processes such as neutral meson mixing
$K\leftrightarrow\ol K$ and $D\leftrightarrow\ol D$ are present at one loop, but by the GIM mechanism the corresponding box diagrams are suppressed by $(m_c^2-m_u^2)/m_W^4$ or $(m_s^2-m_d^2)/m_W^4$ (where $m_{u,d,c,s}$ are the quark masses, which of course coincide with the squark masses in the supersymmetric limit). 

By contrast, in general the soft terms $\msq{X}$ and $\atr{X}$ will be non-diagonal in the CKM basis. Furthermore, in general the squark masses for the first two generations have no reason to be small or near-degenerate, so the GIM suppression is lost. Generic soft mass matrices and trilinear terms will therefore give unacceptably large contributions to strongly constrained processes such as flavour-violating lepton or meson decays, or neutral meson mixing.

This problem should be addressed by imposing some specific structure on the soft masses and trilinear soft terms.\footnote{Of course the sfermions might as well be very heavy, some tens of TeV at least, such that they effectively decouple. This scenario, however, is disfavoured by naturalness arguments, since for SUSY to provide a natural solution to the electroweak hierarchy problem, the superpartner masses should not be too far above the electroweak scale.} For instance, if the $a$-terms $\atr{U}$, $\atr{D}$, $\atr{E}$ are proportional to the respective Yukawa matrices, then they will evidently be diagonal in the CKM basis. Furthermore, if $\msq{X}\sim\mathbbm{1}$ for all $X$, then the soft mass matrices are unaffected by flavour rotations and remain diagonal in the CKM basis, with degenerate entries. Exact or approximate patterns like these can be the consequence of some flavour-blind mechanism of supersymmetry breaking mediation, or they can result from horizontal symmetries, or from wave-function localization in an extra dimension. In this work we are investigating the latter mechanism.

\section{Flavour hierarchies from localization}\label{hfroml}

We will now review how localization of matter fields in an extra dimension can naturally generate large hierarchies in fermion masses and mixings, and how it may at the same time ameliorate the supersymmetric flavour problem. 

To be precise, we will consider a 5D model, compactified on an interval such that the KK mass scale is close to the 4D GUT scale $M_{\rm GUT}\approx 2\cdot 10^{16}$ GeV. The zero modes contain the MSSM fields, with chiral matter obtained from 5D bulk hypermultiplets. The 5D SUSY Lagrangian allows for hypermultiplet mass terms, which will not decouple the zero modes, but rather distort their wave function profiles in the fifth dimension. The resulting zero mode wave functions are exponentially localized towards one of the branes. 

The simplest example is given by a flat extra dimension obtained as the $\mathbb{Z}_2$ orbifold of a circle of radius $2\pi R$. A 5D hypermultiplet can be decomposed into two 4D chiral superfields $(X,\,X^{c})$, whose action is \cite{ArkaniHamed:2001tb}

\be
S=\int d^4x\int_0^{\pi R} dy\Biggl[\int d^4\theta\,\left(X^\dag X+X^c {X^c}^\dag\right)+\int d^2\theta\,\left(X^c\partial_y X+M\,XX^c\right)\hc\Biggr]
\ee
If $X^c$ is odd and $X$ is even under the orbifold projection, $X$ will have a zero mode whose profile is given by
\be\label{zmflat}
x_0(y)\sim e^{-M\,y}.
\ee

Similarly, in a warped extra dimension with metric $ds^2=e^{-ky}dx^2+dy^2$, the action for a hypermultiplet reads \cite{Marti:2001iw}
\be\begin{split}
S=\int d^4x\int_0^{\pi R} dy\Biggl[&\int d^4\theta\,e^{-2ky}\left(X^\dag X+X^c {X^c}^\dag\right)\\
&+\int d^2\theta\, e^{-3ky}\left(\frac{1}{2}X^c\partial_yX-\frac{1}{2}X\partial_y X^c+M\,XX^c\right)\hc\Biggr].
\end{split}
\ee 
Defining $c=M/k$, as is common convention, the $X$ scalar zero mode has a profile \cite{Gherghetta:2000qt}
\be\label{zmwarped}
x_0(y)= \frac{1}{\sqrt{\pi R}}e^{(\frac{3}{2}-c)ky}.
\ee 
For later convenience, we also define the profile factor $f$ to be $x_0(y)$ evaluated on the $y=\pi R$ brane and with one power of the warp factor absorbed,
\be\label{profile}
f=e^{-\pi kR}\,x_0(\pi R)=\frac{1}{\sqrt{\pi R}} e^{(\frac{1}{2}-c)\pi kR}\,.
\ee
After integrating over $y$ one obtains for the zero mode kinetic action in four dimensions
\be
S_4=\int d^4x\int d^4\theta\,{\bf Y}\,X_0^\dag X_0\,,
\ee
where
\be\label{wfnorm}
{\bf Y}=\frac{e^{(\frac{1}{2}-c)\,2\pi kR}-1}{\left(\frac{1}{2}-c\right)\,2\pi kR}\,.
\ee
For $c>\frac{1}{2}$ the zero mode is localized towards the $y=\pi R$ (IR) brane, whereas for $c<\frac{1}{2}$ it is localized towards the $y=0$ (UV) brane. For $c=\frac{1}{2}$ it has a flat profile. In the following we always assume $c\gtrsim -\frac{1}{2}$, because for $c<-\frac{1}{2}$ some massive modes with twisted boundary conditions could become exponentially light. With $c\gtrsim -\frac{1}{2}$, all KK excitations decouple around the scale $ke^{-\pi kR}$, which we choose around $M_{\rm GUT}$. 

One may also introduce purely 4D chiral superfields which are entirely localized on one of the branes. For an IR-brane Higgs superfield $H$ the action can contain, besides the kinetic terms, also Yukawa couplings between $H$ and the even 4D chiral parts $X_i$ of 5D hypermultiplets $(X_i,X_i^c)$:
\be\label{generalyuk}
S_{\rm brane}=\int d^4x\int dy\,\left[\int d^4\theta e^{-2ky}H^\dag H+\int d^2\theta\;e^{-3ky} h_{ij}\, HX_i X_j\hc\right]\,\delta(y-\pi R)
\ee
The effective 4D Yukawa coupling for canonically normalized zero modes reads
\be\label{normyuk}
y_{ij}=h_{ij}\frac{f_i\,f_j}{\left({\bf Y}_i {\bf Y}_j\right)^{1/2}}\qquad\text{(no summation)}\,,
\ee
where the ${\bf Y}_{i}$ are defined analogously as in Eq.~\eqref{wfnorm}, and the $f_{i}$ as in Eq.~\eqref{profile}. Since the $f_{i}$ depend exponentially on the $c_{i}$, hierarchical Yukawa matrices are obtained from ${\cal O}(1)$ $c$-parameters and anarchical $h_{ij}$ coefficients.

As a concrete example, let us review the flavour sector of the holographic GUT model of \cite{Nomura:2006pn}. The gauge symmetry in the bulk is $\SU 6$, broken to $\SU 5\times\U 1$ by boundary conditions on the UV brane. The bulk zero modes are contained in three 4D chiral superfields ${\cal F}_i$ (each containing a $\ol{\bf 5}$ of $\SU 5$), three chiral superfields ${\cal T}_i$ (each containing a $\bf 10$), and three chiral superfields ${\cal N}_i$ (each containing a singlet). These zero modes are identified with the MSSM matter fields. On the IR brane the boundary conditions preserve the $\SU 6$ bulk gauge symmetry. However, there is an IR brane superfield $\Sigma$ in the adjoint of $\SU 6$ whose expectation value breaks $\SU 6$ spontaneously to $\SU 4\times\SU 2\times\U 1$. The low-energy gauge group is given by the intersection of $\SU 5\times\U 1$ and $\SU 4\times\SU 2\times\U 1$ in $\SU 6$, which is the Standard Model gauge group (apart from an extra $\U 1$ which is Higgsed on the UV brane). Two weak doublet components of $\Sigma$ remain massless, and are identified with the MSSM Higgs doublets. The other $\Sigma$ components either acquire supersymmetric masses directly from the scalar potential, or are eaten by the Higgs mechanism. The field content is listed in Table \ref{fieldcontent}.

\begin{table}
\begin{center}
\begin{tabular}{c|c|c|c}
field & $\SU 6$ rep. & massless mode & localization\\ \hline
${\cal F}_i$ & $\ol{\bf 70}$ & $\ol{\bf 5}$ & bulk \\
${\cal F}^c_i$ & ${\bf 70}$ & -- & bulk \\ \hline
${\cal T}_i$ & ${\bf 20}$ & ${\bf 10}$ & bulk \\
${\cal T}^c_i$ & $\ol{\bf 20}$ & -- & bulk \\ \hline
${\cal N}_i$ & ${\bf 56}$ & ${\bf 1}$ & bulk \\
${\cal N}^c_i$ & $\ol{\bf 56}$ & -- & bulk \\ \hline
$\Sigma$ & ${\bf 35}$ & $\begin{array}{c} {\bf 2}_{1/2}\oplus{\bf 2}_{-1/2} \\ \text{of } \SU 2\times\U 1 \end{array}$ & IR brane 
\end{tabular}
\caption{Field content of the NPT model, listing the $\SU 6$ representations and the massless modes' $\SU 5$ representations (if applicable).}\label{fieldcontent}
\end{center}
\end{table}

A possible choice for the scales in this model is to fix the cutoff scale of the 5D theory to be around the 4D reduced Planck scale, $M_*= 2\cdot 10^{18}$ GeV say, and the KK scale $k\,e^{-\pi kR}$ to be slightly below the GUT scale, $k\,e^{-\pi kR}= 10^{16}$ GeV. The AdS curvature $k$ is chosen to lie in between, $k=2\cdot 10^{17}$ GeV, and $kR=1$. This gives a somewhat large size of the extra dimension in units of the cutoff, around $\pi R M_*=30$.

The Yukawa couplings arise from
\be\label{hgutyuk}
{\cal L}\supset \delta(y-\pi R)\int d^2\theta\,\left(h^{\cal T}_{ij}\,\Sigma{\cal T}_i{\cal T}_j+h^{\cal F}_{ij}\,\Sigma{\cal T}_i{\cal F}_j+h^{\cal N}_{ij}\,\Sigma{\cal F}_i{\cal N}_j\right)\hc
\ee
The IR brane couplings $h_{ij}$ can be estimated using naive dimensional analysis; their typical magnitude up to ${\cal O}(1)$ uncertainty is $6\pi^2/M_*$. 

For illustration consider the following choice of $c$-parameters: Setting
\be\label{simplecs}
c_{{\cal T}_1}\approx\frac{5}{2},\quad c_{{\cal T}_2}\approx c_{{\cal F}_1}\approx c_{{\cal F}_2}\approx c_{{\cal F}_3}\approx\frac{3}{2},\quad c_{{\cal T}_3}\approx c_{{\cal N}_1}\approx c_{{\cal N}_2}\approx c_{{\cal N}_3}\approx\frac{1}{2}
\ee
we obtain in the quark sector
\be
\yuk{U}=\left(\begin{array}{ccc}
\lam{U}_{11}\epsilon^4 & \lam{U}_{12}\epsilon^3 & \lam{U}_{13}\epsilon^2\\
\lam{U}_{12}\epsilon^3 & \lam{U}_{22}\epsilon^2 & \lam{U}_{23}\epsilon\\
\lam{U}_{13}\epsilon^2 & \lam{U}_{23}\epsilon & \lam{U}_{33}\\
\end{array}\right),\qquad 
\yuk{D}=\left(\begin{array}{ccc}
\lam{D}_{11}\epsilon^3 & \lam{D}_{12}\epsilon^3 & \lam{D}_{13}\epsilon^3\\
\lam{D}_{21}\epsilon^2 & \lam{D}_{22}\epsilon^2 & \lam{D}_{23}\epsilon^2\\
\lam{D}_{31}\epsilon & \lam{D}_{32}\epsilon & \lam{D}_{33}\epsilon\\
\end{array}\right)\,,
\ee
where $\epsilon\approx e^{-\pi kR}\approx\frac{1}{20}$, and where the $\lam{U,D}_{ij}$ are of the order $\lam{U,D}_{ij}={\cal O}\left(6\pi/M_* R\right)\approx{\cal O}(1)$. Note that $\yuk{U}$ is symmetric, because it arises from the symmetric $\Sigma{\cal T}_i{\cal T}_j$ coupling in Eq.~\eqref{hgutyuk}. In the lepton sector one has
\be
\yuk{E}=\left(\begin{array}{ccc}
\lam{E}_{11}\epsilon^2 & \lam{E}_{12}\epsilon & \lam{E}_{13}\\
\lam{E}_{21}\epsilon^2 & \lam{E}_{22}\epsilon & \lam{E}_{23}\\
\lam{E}_{31}\epsilon^2 & \lam{E}_{32}\epsilon & \lam{E}_{33}\\
\end{array}\right),\qquad 
\yuk{N}=\left(\begin{array}{ccc}
\lam{N}_{11}& \lam{N}_{12} & \lam{N}_{13}\\
\lam{N}_{21} & \lam{N}_{22} & \lam{N}_{23}\\
\lam{N}_{31} & \lam{N}_{32}& \lam{N}_{33}\\
\end{array}\right)\,.
\ee
Hierarchical Yukawa matrices of this type are well-known to roughly give the observed masses and mixings. Unwanted $\SU 5$ relations can be avoided by taking into account contributions from higher-dimensional operators, in particular higher powers of $\Sigma$.

The assignment Eq.~\eqref{simplecs} can be refined to even better reproduce the known fermion masses and CKM angles. 
We will  describe in detail how we fit the $c$-parameters and the $\lambda_{ij}$ in Section \ref{method}.

\section{Supersymmetry breaking}\label{susyb}

SUSY breaking can be parameterized by $F$- and $D$-type spurions in the K\"ahler potential, which we generally denote by $\widetilde{\bf \Phi}$ and $\widetilde{\bf \Delta}$, and by $F$-type spurions in the superpotential denoted by $\wt{\bf \Lambda}$. Omitting the neutrinos from now on, the 4D Lagrangian can be brought into the form
\be
\begin{split}\label{spurionL}
{\cal L}=&\int d^4\theta\Biggl[\left({\bf Y}_{H_u}+\left(\wt{\bf \Phi}_{H_u}\theta^2\hc\right)+\wt{\bf \Delta}_{H_u}\theta^4\right)H_u^\dag H_u+\left(H_u\leftrightarrow H_d\right)\\
&\qquad+\left(\Ywf{U}_{ij}+\left(\Phitsp{U}_{ij}\theta^2\hc\right)+\Deltsp{U}_{ij}\theta^4\right)U^\dag_i U_j+\left(U\leftrightarrow \{D,\,Q,\,L,\,E\}\right)\Biggr]\\
&+\int d^2\theta\Biggl[\left(\yukt{U}_{ij}+\Lamtsp{U}_{ij}\theta^2\right)H_u U_i Q_j+\left( \yukt{D}_{ij}+\Lamtsp{D}_{ij}\theta^2\right)H_d D_i Q_j\\
&\qquad\qquad+\left(\yukt{E}_{ij}+\Lamtsp{E}_{ij}\theta^2\right)H_d E_i L_j+\tilde\mu H_u H_d\Biggr]\\
&\qquad\hc
\end{split}
\ee
Here the ${\bf Y}$, $\wt{\bf \Delta}$, $\wt{\bf \Phi}$ and $\wt{\bf \Lambda}$ are c-number functions of the compactification radius and of the expectation values of hidden sector fields. In particular, in the absence of brane kinetic terms (which are generically subdominant at large volume), the wave-function coefficient ${\bf Y}$ for the bulk fields is as in Eq.~\eqref{wfnorm}:
\be\label{wfcoeffs}
\Ywf{X}_{ij}=\delta_{ij}\frac{e^{(\frac{1}{2}-\cbm{X}_i)2\pi kR}-1}{(\frac{1}{2}-\cbm{X}_i)2\pi kR}\,,
\ee
where $X=U,\,D,\,Q,\,L,\,E$.

Again for $X=$ any matter field, we define rescaled quantities by
\be\begin{split}\label{rescaled}
&\Phisp{X}_{ij}=\frac{\Phitsp{X}_{ij}}{\left(\Ywf{X}_{ii}\Ywf{X}_{jj}\right)^{1/2}},\quad{\bf\Phi}_{H_{u,d}}=\frac{\wt{\bf\Phi}_{H_{u,d}}}{{\bf Y}_{H_{u,d}}},\quad\Delsp{X}_{ij}=
\frac{\Deltsp{X}_{ij}}{\left(\Ywf{X}_{ii} \Ywf{X}_{jj}\right)^{1/2}},\\
&\Lamsp{U,D}_{ij}=\frac{\Lamtsp{U,D}_{ij}}{\left(\Ywf{U,D}_{ii} \Ywf{Q}_{jj}{\bf Y}_{H_{u,d}}\right)^{1/2}},\quad\Lamsp{E}_{ij}=\frac{\Lamtsp{E}_{ij}}{\left(\Ywf{E}_{ii} \Ywf{L}_{jj}{\bf Y}_{H_d}\right)^{1/2}}\,.
\end{split}\ee
The Yukawa matrices for canonically normalized fields are then
\be\label{canyuk}
\yuk{U}_{ij}=\frac{\yukt{U}_{ij}}{\left(\Ywf{U}_{ii}\Ywf{Q}_{jj}{\bf Y}_{H_u}\right)^{1/2}},\quad \yuk{D}_{ij}=\frac{\yukt{D}_{ij}}{\left(\Ywf{D}_{ii}\Ywf{Q}_{jj}{\bf Y}_{H_d}\right)^{1/2}},\quad \yuk{E}_{ij}=\frac{\yukt{E}_{ij}}{\left(\Ywf{E}_{ii}\Ywf{L}_{jj}{\bf Y}_{H_d}\right)^{1/2}}\,.
\ee
The scalar soft masses for matter fields are 
\be
\msq{X}=\left(\Phisp{X}\right)^\dag\Phisp{X}-\Delsp{X}\,,
\ee
and the trilinear terms are given by
\be\begin{split}\label{tril}
&\atr{U}=(\Phisp{Q})^T \yuk{U}+\yuk{U}(\Phisp{U})^T+{\bf \Phi}_{H_u}\,\yuk{U}-\Lamsp{U}\,,\\
&\atr{D}=(\Phisp{Q})^T \yuk{D}+\yuk{D}(\Phisp{D})^T+{\bf \Phi}_{H_d}\,\yuk{D}-\Lamsp{D}\,,\\
&\atr{E}=(\Phisp{L})^T \yuk{E}+\yuk{E}(\Phisp{E})^T+{\bf \Phi}_{H_d}\,\yuk{E}-\Lamsp{E}\,.
\end{split}
\ee

There are two natural possibilities for the origin of supersymmetry breaking in our 5D setup: Supersymmetry breaking by the radion multiplet \cite{Chacko:2000fn}, or supersymmetry breaking by additional brane fields. In radion-mediated SUSY breaking, the radion superfield (whose lowest component sets the radius of the extra dimension) acquires an $F$-term expectation value. This scenario is quite predictive since all the couplings are essentially determined from geometry. By contrast, if there are additional SUSY-breaking fields $Z_I$ on the branes, their couplings to the visible sector are additional free parameters. The general case will be a mixture of these two; see the Appendix for a concrete model in which both brane field SUSY breaking and radion mediation contribute to the soft terms.

In order to alleviate the flavour problem, we make the crucial assumptions that the MSSM Higgs fields are brane fields (as is the case e.g.~in the NPT model) and that the $Z_I$ are localized on the same brane. 
With the profile functions of Eq.~\eqref{zmwarped}, and with the dependence on the radion multiplet restored \cite{Marti:2001iw}, the 4D Lagrangian is
\be\begin{split}\label{4DLradion}
{\cal L}&=\int d^4\theta\;\varphi\ol\varphi\sum_i \frac{e^{(\frac{1}{2}-\cbm{U}_i)k\pi(T+\ol T)}-1}{(\frac{1}{2}-\cbm{U}_i)2\pi kR}\;U_i^\dag U_i\\
&\qquad+\left(U\leftrightarrow \{D,\,Q,\,L,\,E\}\right)\\
&+\int d^4\theta\;\varphi\ol\varphi\, e^{-k\pi(T+\ol T)}\left(1+\left(\widehat{\bf\Phi}_{H_u}(Z_I)\theta^2\hc\right)+\widehat{\bf\Delta}_{H_u}(Z_I)\theta^4\right)H_u^\dag H_u\\
&\qquad +\left(H_u\leftrightarrow H_d\right)\\
&+\int d^4\theta\;\varphi\ol\varphi\sum_{ij}\frac{e^{\frac{1}{2}(1-\cbm{U}_i-\cbm{U}_j)k\pi(T+\ol T)}}{\pi R}\left(\Phihsp{U}_{ij}(Z_I)\theta^2\hc+\Delhsp{U}_{ij}(Z_I)\theta^4\right)U_i^\dag U_j\\
&\qquad+\left(U\leftrightarrow \{D,\,Q,\,L,\,E\}\right)\\ 
&+\int d^2\theta\;\varphi^3\,W_{\rm brane}\hc
\end{split}
\ee
where
\be\begin{split}
W_{\rm brane}=\sum_{ij}\Bigl[&\left(\hyuk{U}_{ij}+\Lamhsp{U}_{ij}(Z_I)\theta^2\right)\frac{e^{-(\cbm{U}_i+\cbm{Q}_j)\pi kT}}{\pi R}\; H_u U_i Q_j\\
&+\left(\hyuk{D}_{ij}+\Lamhsp{D}_{ij}(Z_I)\theta^2\right)\frac{e^{-(\cbm{D}_i+\cbm{Q}_j)\pi kT}}{\pi R}\;H_d D_i Q_j\\
&+\left(\hyuk{E}_{ij}+\Lamsp{E}_{ij}(Z_I)\theta^2\right)\frac{e^{-(\cbm{E}_i+\cbm{L}_j)\pi kT}}{\pi R}\;H_d E_i L_j\Bigr]+e^{-3\pi kT}\hat\mu H_u H_d\,.
\end{split}
\ee
Here $T=R+iB_5+F^T\theta^2+\text{(fermions)}$ is the radion multiplet, with $B_5$ the fifth component of the graviphoton. $\varphi=1+F^\varphi\theta^2$ is the chiral compensator, an auxiliary chiral superfield whose $F$-term component $F^\varphi$ is the scalar auxiliary of the 4D gravitational multiplet. SUSY breaking in the combined gravitational sector is parameterized by $F^T\neq 0$ and $F^\varphi\neq 0$. Note the non-standard dimensions of these fields: In our conventions $T$ and $\varphi$ have mass dimensions $-1$ and $0$ respectively, so $F^T$ and $F^\varphi$ have mass dimension $0$ and $1$. The spurions $\widehat{\bf\Phi}$, $\widehat{\bf\Delta}$ and $\widehat{\bf\Lambda}$ capture the effects of SUSY breaking by the IR brane fields $Z_I$ (which we also take as a background). We have assumed negligible brane-kinetic terms for the bulk fields.

The matter wave function normalization coefficients are as in Eq.~\eqref{wfcoeffs}, while the Higgs wave function normalization reads
\be\label{higgswfcoeffs}
{\bf Y}_{H_{u,d}}=e^{-2\pi kR}\,.
\ee
For the spurionic coefficients one obtains, using profile factors $\fpr{X}_i$ as defined in Eq.~\eqref{profile},
\be\begin{split}\label{spurcoeffs1}
\wt{\bf\Phi}_{H_{u,d}}&=\left(F^\varphi-\pi k F^T+\widehat{\bf\Phi}_{H_{u,d}}\right)e^{-2\pi kR}\,,\\
\Phitsp{X}_{ij}&=\delta_{ij} F^\varphi\frac{e^{(\frac{1}{2}-\cbm{X}_i)2\pi kR}-1}{(\frac{1}{2}-\cbm{X}_i)2\pi kR}+\delta_{ij}F^T\frac{e^{(\frac{1}{2}-\cbm{X}_i)2\pi kR}}{2R}+\Phihsp{X}_{ij}\,\fpr{X}_i\,\fpr{X}_j\,,\\
\Deltsp{X}_{ij}&=\delta_{ij}\,\frac{e^{(\frac{1}{2}-\cbm{X}_i)2\pi kR}}{(\frac{1}{2}-\cbm{X}_i)2\pi kR}\,\left|F^\varphi+k\pi\left(\frac{1}{2}-\cbm{X}_i\right)F^T\right|^2-\delta_{ij}|F^\varphi|^2\frac{1}{(\frac{1}{2}-\cbm{X}_i)2\pi kR}\\
&\qquad +\left(\left(\frac{1}{2}(1-\cbm{X}_i-\cbm{X}_j)k\pi\,\ol{F}^{\ol T}+\ol{F}^{\bar\varphi}\right)\Phihsp{X}_{ij}\hc+\Delhsp{X}_{ij}\right)\,\fpr{X}_i\,\fpr{X}_j
\end{split}\ee
and
\be
\begin{split}\label{spurcoeffs2}
\Lamtsp{U}_{ij}&=\left(\left(3F^\varphi-(\cbm{U}_i+\cbm{Q}_j)k\pi F^T\right)\hyuk{U}_{ij}+\Lamhsp{U}_{ij}\right)\,e^{-k\pi R}\,\fpr{U}_i\,\fpr{Q}_j\,,\\
\Lamtsp{D}_{ij}&=\left(\left(3F^\varphi-(\cbm{D}_i+\cbm{Q}_j)k\pi F^T\right)\hyuk{D}_{ij}+\Lamhsp{D}_{ij}\right)\,e^{-k\pi R}\,\fpr{D}_i\,\fpr{Q}_j\,,\\
\Lamtsp{E}_{ij}&=\left(\left(3F^\varphi-(\cbm{E}_i+\cbm{L}_j)k\pi F^T\right)\hyuk{E}_{ij}+\Lamhsp{E}_{ij}\right)\,e^{-k\pi R}\,\fpr{E}_i\,\fpr{L}_j\,.
\end{split}
\ee
Here we have not listed the spurions $\wt{\bf\Delta}_{H_{u,d}}$, which do not contribute to sfermion soft masses.
The non-canonical Yukawa couplings $\tilde y_{ij}$ can be read off to be
\be\begin{split}
\yukt{U}_{ij}=\hyuk{U}_{ij}\,e^{-\pi kR}\,\fpr{U}_i\,\fpr{Q}_j\,,\\
\yukt{D}_{ij}=\hyuk{D}_{ij}\,e^{-\pi kR}\,\fpr{D}_i\,\fpr{Q}_j\,,\\
\yukt{E}_{ij}=\hyuk{E}_{ij}\,e^{-\pi kR}\,\fpr{E}_i\,\fpr{L}_j\,.\\
\end{split}
\ee
Using Eq.~\eqref{canyuk} and the wave function normalizations from Eqs.~\eqref{wfcoeffs} and \eqref{higgswfcoeffs}, this gives canonical Yukawa couplings as in Eq.~\eqref{normyuk} (note that the Higgs wave function normalization cancels the explicit warp factor).

In order to discuss MSSM spectra, we also need gaugino masses. The 4D gauge field Lagrangian is
\be\label{gauge}
{\cal L}\supset\frac{1}{4}\sum_a\int d^2\theta\,\left(\frac{1}{g_{\rm UV}^2}+\frac{\pi T}{g_5^2}+\widehat{\bf\Omega}^a(Z_I)\theta^2\right) W^{a\alpha} W^a_\alpha\hc+\ldots
\ee
We have omitted terms irrelevant for gaugino masses, and neglected possible effects from bulk Chern-Simons terms \cite{Hebecker:2008rk,Brummer:2009ug}. In this expression $a=1,2,3$ labels the Standard Model gauge factors, $g_5$ is the bulk gauge coupling, and the $1/g_{\rm UV}^2$ term originates from a gauge kinetic term on the UV brane. Note that this term is universal with respect to the Standard Model gauge fields, at least in models where the UV brane preserves $\SU 5$. IR brane terms are irrelevant to our discussion and will be omitted. The gaugino masses are then given by 
\be\label{gauginos}
M_a=\frac{1}{2}g_4^2\left(\widehat{\bf\Omega}^a+\frac{\pi F^T}{g_5^2}\right)\,,
\ee
with $1/g_4^2=1/g_{\rm UV}^2+\pi R/g_5^2$. In our numerical analysis we take the gaugino masses to be equal, 
\be
M_1=M_2=M_3\equiv M_{1/2}\,.
\ee
Universal gaugino masses are induced by the leading contributions in the NPT model, since in this model the GUT group is broken only by expectation values on the IR brane. Higher-dimensional operators involving powers of the GUT Higgs $\Sigma$ can lead to gaugino mass splittings, but we assume that they are sufficiently suppressed. In a more general setup where the gauge symmetry is broken by boundary conditions on the SUSY breaking brane, the universality condition could be relaxed.

It is evident that, in general, the SUSY breaking soft terms will not be flavour-preserving. If SUSY breaking is dominated by radion mediation, the scalar soft masses will be diagonal in the basis we have been using, but their eigenvalues will be non-degenerate; rotating to the CKM basis will therefore induce off-diagonal terms. The trilinear couplings, likewise, are not proportional to the Yukawa couplings. If SUSY breaking soft terms are induced predominantly by brane fields, there are even fewer constraints, since the brane spurions $\widehat{\bf\Phi}$, $\widehat{\bf\Delta}$, $\widehat{\bf\Lambda}$ are generally anarchic.

However, a key assumption for this class of models was that the hierarchies in the fermion masses and mixings are originating mainly from wave-function localization. The same localization effects also leave their imprints the soft terms, and will induce similar hierarchies in the mass matrices and trilinears; such scenarios have been dubbed ``flavourful supersymmetry''  \cite{Nomura:2007ap}.
It is then reasonable to argue that the flavour problem should at least be alleviated, if not solved, in wave-function localization models. For instance, the basis changes used to switch to the CKM basis will approximately also diagonalize the trilinear terms, up to higher-order terms which are exponentially small. An extreme limiting case of this scenario would be to keep only the sfermion soft terms associated with the third generation, or even only those associated with the stop (and with additional flavour-blind contributions to the soft terms induced by RG running). The flavour constraints in this limiting case (which we previously studied in \cite{Brummer:2010gh}) are evidently far less severe.

One of the aims of this work is to test the assertion that, moving away from this limiting case, wave-function localization still gives sizeable FCNC suppression, and thus substantially reduces the tuning required to get a realistic model. From the known Yukawa hierarchies we fix the 5D mass parameters $\cbm{X}_i$. Using anarchical textures for the unknown couplings of the 5D theory, we then calculate the GUT-scale soft parameters, evolve them to the electroweak scale using their renormalization group equations, and calculate the resulting masses and mixings. These are finally compared to the existing bounds on flavour observables.

\subsection{SUSY breaking dominated by brane fields}\label{branefields}

In the case that the dominant source for the soft terms are brane fields $Z_I$, there is a large number of unknown coupling parameters. If, in accordance with the holographic interpretation, the theory is taken to be strongly coupled on the IR brane and in the bulk, then the magnitude of these parameters can be estimated by naive dimensional analysis. We use loop factors $\ell_5=24\pi^3$ for 5D superfields (of dimension $3/2$) and $\ell_4=16\pi^2$ for 4D superfields (of dimension $1$) \cite{Chacko:1999hg}. Up to ${\cal O}(1)$ uncertainty, the coefficient for a superpotential term composed of $n$ bulk zero modes and $m$ brane fields is
\be
\frac{M_*^3}{16\pi^2}\left(\frac{\sqrt{\ell_5}}{M_*^{3/2}}\right)^n\left(\frac{\sqrt{\ell_4}}{M_*}\right)^m\,,
\ee
while for a K\"ahler potential term it is
\be
\frac{M_*^2}{16\pi^2}\left(\frac{\sqrt{\ell_5}}{M_*^{3/2}}\right)^n\left(\frac{\sqrt{\ell_4}}{M_*}\right)^m\,.
\ee
Here $M_*$ is again the cutoff scale, taken to be close to the reduced Planck scale in 4D. The 5D Yukawa couplings are thus estimated to be
\be
h_{ij}\approx\frac{6\pi^2}{M_*}\,.
\ee
To estimate the IR brane spurion coefficients, we define a Goldstino superfield $Z$ with expectation value $\vev Z=F^Z\theta^2$ to be the combination of SUSY breaking brane fields which couples to the visible sector. $Z$ will in general be composite, but we can treat it as a single, elementary, canonically normalized field for our purposes, absorbing any compositeness factors of $4\pi$ or warp factors in the definition of $F^Z$. We obtain
\be\begin{split}\label{hatted}
\widehat{\bf\Phi}_{H_{u,d}}&\approx 4\pi\frac{F^Z}{M_*},\\
\Phihsp{X}_{ij}&\approx\frac{6\pi^2}{M_*}\frac{F^Z}{M_*},\\
\Delhsp{X}_{ij}&\approx\frac{24\pi^3}{M_*}\left|\frac{F^Z}{M_*}\right|^2,\\
\Lamhsp{U,D,E}_{ij}&\approx\frac{24\pi^3}{M_*}\frac{F^Z}{M_*}.\\
\end{split}\ee

It is now convenient to define matrices
\be\label{kappa}
\kap{X}_{ij}=\frac{6\pi^2}{M_*}\frac{\fpr{X}_i\,\fpr{X}_j}{\left(\Ywf{X}_i\,\Ywf{X}_j\right)^{1/2}}\,.
\ee
The $\kap{X}_{ij}$ are dimensionless and hierarchical, with their largest entries at most of order of the top Yukawa coupling (as their structure is determined by the same profile factors as the Yukawa matrices). With Eqns.~\eqref{kappa}, \eqref{hatted}, \eqref{spurcoeffs1}, \eqref{spurcoeffs2}, and \eqref{rescaled} one obtains, up to ${\cal O}(1)$ uncertainty,
\be\begin{split}\label{scmassbrane}
\Phisp{X}_{ij}\approx\kap{X}_{ij}\frac{F^Z}{M_*}\,,\qquad \Delsp{X}_{ij}\approx 4\pi\kap{X}_{ij}\,\left|\frac{F^Z}{M_*}\right|^2\,.
\end{split}\ee
Hence the soft masses $\msq{X}={\Phisp{X}}^\dag\Phisp{X}-\Delsp{X}$ are dominated by $\Delsp{X}$.

For the trilinear soft terms of Eq.~\eqref{tril} we find
\be
\begin{split}\label{trilbrane}
\atr{U}_{ij}\approx(\kap{Q}\yuk{U})_{ij}\frac{F^Z}{M_*}+(\yuk{U}\kap{U})_{ij}\frac{F^Z}{M_*}+4\pi\,\yuk{U}_{ij}\frac{F^Z}{M_*}-\Lamsp{U}_{ij}\,,\\
\end{split}
\ee
where
\be
\Lamsp{U}_{ij}\approx 4\pi \yuk{U}_{ij}\frac{F^Z}{M_*}\,.
\ee
Clearly $\atr{U}_{ij}$ is dominated by the last two terms in Eq.~\eqref{trilbrane}. Provided that there are no accidental cancellations taking place between them, it is of the order
\be\label{trilbrane2}
\atr{U}_{ij}\approx 4\pi \yuk{U}_{ij}\frac{F^Z}{M_*}\,.
\ee
Analogous statements hold for the other $a$-terms. Comparing Eqs.~\eqref{scmassbrane} and \eqref{trilbrane2}, the largest $a$-terms will be around a factor of $\approx\sqrt{4\pi}$ larger than the largest scalar soft masses.

The NDA estimate for $\widehat{\bf\Omega}^a$ in Eq.~\eqref{gauge} is \cite{Buchmuller:2005rt}
\be
\widehat{\bf\Omega}^a\approx\frac{6\pi^2}{\sqrt{C}\,g_5^2}\frac{F^Z}{M_*^2}\,,
\ee
with $C$ a group-theoretical factor. For the NPT model with bulk gauge group $\SU 6$ we set $C=C_2\left(\SU 6\right)=6$. The canonically normalized gaugino masses of Eq.~\eqref{gauginos} are thus
\be\label{ndagauginos}
M_a\approx \frac{3\pi^2}{\sqrt{6}}\frac{g_4^2}{g_5^2}\;\frac{F^Z}{M_*^2}\,.
\ee
The four-dimensional gauge coupling gets contributions from both the 5D bulk gauge coupling and a UV brane-localized kinetic term (cf.~Eq.~\eqref{gauge}):
\be\label{gaugecouplings}
\frac{1}{g_4^2}=\frac{1}{g_{\rm UV}^2}+\frac{\pi R}{g_5^2}\,.
\ee
The bulk gauge coupling can also be estimated from naive dimensional analysis; this gives
\be\label{bulkgauge}
\frac{1}{g_5^2}\approx\frac{C}{24\pi^3} M_*\,.
\ee
The UV brane, in the holographic picture, can be weakly coupled and thus $g_{\rm UV}$ can be smaller than its NDA value. In fact, unless $R$ is rather large, $g_{\rm UV}$ must be small in order to obtain the proper unified gauge coupling $g_4^2\approx 0.5$ from Eq.~\eqref{gaugecouplings}. 

Eqns.~\eqref{bulkgauge} and \eqref{ndagauginos} lead to suppressed gaugino masses:
\be
M_a\approx \frac{\sqrt{6}}{8\pi} g_4^2\frac{F^Z}{M_*}\approx 0.05\frac{F^Z}{M_*}\,.
\ee
Evidently, to have gaugino masses which are comparable with the other soft terms, IR brane terms cannot be the only source of SUSY breaking. Alternatively one could consider models which are weakly coupled also in the bulk and where, consequently, there is no large NDA suppression as in Eq.~\eqref{bulkgauge}.

Finally, we impose that the Higgs mass matrix be degenerate at the GUT scale:
\be\label{dhmm}
m_{H_u}^2+|\mu|^2=m_{H_d}^2+|\mu|^2=|B_\mu|\,.
\ee 
This is the case in a large class of models \cite{Brummer:2010gh}, in particular in the NPT model which we choose as our benchmark. The relation Eq.~\eqref{dhmm} has the advantage of constraining our free parameters somewhat more; it could however be relaxed in a more general setting. The NDA estimate for the Higgs mass parameters is
\be\label{ndahiggs}
m_{H_{u,d}}^2+|\mu|^2=|B_\mu|\approx 16\pi^2\left|\frac{F^Z}{M_*}\right|^2\,.
\ee

\subsection{SUSY breaking dominated by radion mediation}\label{rmsb}

It is instructive to also consider the opposite extreme case, where brane contributions to soft masses are negligible and the dominant source is radion mediation. The soft terms for bulk scalars in this case are (see, for example, \cite{Marti:2001iw, Choi:2003fk})
\be\label{rmsbsfermions}
\msq{X}_{ij}=\delta_{ij}\left|\frac{F^T}{2R}\right|^2\,\left(\frac{(\frac{1}{2}-\cbm{X}_i)\pi kR}{\sinh\left((\frac{1}{2}-\cbm{X}_i)\pi kR\right)}\right)^2\,,
\ee
and 
\be\begin{split}\label{rmsbtrils}
\atr{U}_{ij}&=\frac{F^T}{2R}\left(\frac{(\frac{1}{2}-\cbm{U}_i)2\pi kR}{e^{(\frac{1}{2}-\cbm{U}_i)2\pi kR}-1}+\frac{(\frac{1}{2}-\cbm{Q}_j)2\pi kR}{e^{(\frac{1}{2}-\cbm{Q}_j)2\pi kR}-1}\right)\yuk{U}_{ij}\,,\\
\atr{D}_{ij}&=\frac{F^T}{2R}\left(\frac{(\frac{1}{2}-\cbm{D}_i)2\pi kR}{e^{(\frac{1}{2}-\cbm{D}_i)2\pi kR}-1}+\frac{(\frac{1}{2}-\cbm{Q}_j)2\pi kR}{e^{(\frac{1}{2}-\cbm{Q}_j)2\pi kR}-1}\right)\yuk{D}_{ij}\,,\\
\atr{E}_{ij}&=\frac{F^T}{2R}\left(\frac{(\frac{1}{2}-\cbm{E}_i)2\pi kR}{e^{(\frac{1}{2}-\cbm{E}_i)2\pi kR}-1}+\frac{(\frac{1}{2}-\cbm{L}_j)2\pi kR}{e^{(\frac{1}{2}-\cbm{L}_j)2\pi kR}-1}\right)\yuk{E}_{ij}\,.
\end{split}
\ee

It is well known that soft masses for brane-localized scalars cannot be induced by $F^T$ or $F^\varphi$ at the tree level. Together with the Higgs mass degeneracy condition which holds in the NPT model, as discussed earlier, the Higgs mass parameters are then \cite{Nomura:2006pn, Brummer:2010gh}
\be\label{rmsbhiggs}
m_{H_u}^2=m_{H_d}^2=0,\qquad \mu=F^T k\pi-F^\varphi,\qquad |B_\mu|=|\mu|^2\,.
\ee
As explained in \cite{Brummer:2010gh}, it is difficult to obtain realistic spectra with this condition.

Finally, the gaugino mass is
\be
M_{1/2}=\frac{\pi}{2}\frac{ g_4^2}{g_5^2}F^T\,.
\ee
As discussed in the previous Section, the gauge fields could be predominantly UV brane fields (if the theory is strongly coupled in the bulk, $R$ is only moderately large, and therefore the dominant contribution to the 4D gauge coupling in Eq.~\eqref{gaugecouplings} comes from the UV brane term $1/g_{\rm UV}^2$). In that case the radion-mediated gaugino mass will be suppressed, and since the brane contribution to $M_{1/2}$ is also small, no realistic phenomenology can be obtained. We are thus led to focus on the other case where $R$ is large enough to overcome the NDA suppression of Eq.~\eqref{bulkgauge}, $RM_*\sim{\cal O}(100)$. Then the second term on the RHS of Eq.~\eqref{gaugecouplings} contributes sizeably to $1/g_4^2$; in the extreme case where the UV brane term can be neglected, we obtain
\be\label{rmsbgauginos}
M_{1/2}=\frac{F^T}{2R}\,,
\ee
comparable with the largest other radion-mediated soft masses.

\section{Parameterization and constraints}\label{method}

As we showed in the previous section, if SUSY breaking is dominated by brane sources alone, then the gaugino masses will be relatively suppressed (to the extent that naive dimensional analysis is valid). In order to pass the constraints on chargino and gluino searches, the remaining soft terms would then have to be in the multi-TeV range. This scenario is clearly disfavoured on from the naturalness point of view, and would probably be impossible to probe at the LHC. Furthermore, large $a$-terms and soft masses as predicted NDA tend to lead to tachyonic sfermions in the low-energy spectrum. On these grounds we will dismiss the possibility that soft terms are induced by brane sources alone, and instead focus on the case where radion mediation gives a significant contribution. 

However, if the soft terms are exclusively generated by radion mediation and brane sources can be entirely neglected, we do not find realistic electroweak symmetry breaking. In fact previous analyses \cite{Choi:2003kq,Brummer:2009ug} have shown that it is difficult to reconcile minimal radion-mediated scenarios with a GUT-scale degenerate Higgs mass matrix. While more refined scenarios \cite{Hebecker:2008rk,Brummer:2009ug} can give realistic TeV-scale physics, in our case the Higgs sector is subject to the even stronger condition Eq.~\eqref{rmsbhiggs}, which turns out to be too restrictive. 

In short, radion mediation is necessary to provide sizeable gaugino masses, while brane sources are necessary to avoid vanishing Higgs soft masses. We will therefore study the general case where both sources of supersymmetry breaking are present. Their relative importance will evidently depend on the relative size of $F^T/2R$ and $F^Z/M_*$. It will also depend on $\tan\beta$, since we will determine the $c$-parameters from the Yukawa couplings which are fixed by $\tan\beta$ and the known fermion masses, and the soft masses depend on the $c_i$.

The phenomenologically most problematic soft terms are the trilinear $a$-terms originating from brane-localized SUSY breaking. Note that these are enhanced over the other soft terms according to the NDA estimate of Section~\ref{branefields}. Large $a$-terms lead not only to large flavour violation but also to tachyonic third-generation sfermions; we find that to avoid these, the brane contribution should be subdominant with respect to the radion-mediated contribution,
\be 
\frac{F^Z}{M_*}\lesssim\,0.2\,\frac{F^T}{2R}\,.
\ee
Even with this condition satisfied, anarchic $a$-terms still tend to induce unacceptably large flavour violation---unless, again, the overall scale of SUSY breaking is unnaturally large. We therefore choose to set the brane trilinear term of Eq.~\eqref{trilbrane} to zero, which is justified if the SUSY breaking fields on the brane are charged under some symmetry (for a comprehensive discussion see e.g.~\cite{Nomura:2008pt, Nomura:2008gg}).

To now investigate supersymmetric flavour violation, we decompose the brane-induced soft mass matrices into a hierarchical part which depends on the $c$-parameters and an ${\cal O}(1)$ part:
\be
\msq{X}_{ij,{\rm brane}}=4\pi\,\,\left|\frac{F^Z}{M_*}\right|^2\kap{X}_{ij}\,\lambda^{m_X^2}_{ij}\qquad(X=U,D,Q,E,L)\,,\label{sfermions}
\ee
where $\lambda_{ij}^{m_X^2}$ are dimensionless hermitian matrices with ${\cal O}(1)$ entries. Recall that
$\kap{X}_{ij}$ was defined as
\be
\kap{X}_{ij}=\frac{6\pi^2}{M_*}\frac{\fpr{X}_i\,\fpr{X}_j}{\left(\Ywf{X}_i\,\Ywf{X}_j\right)^{1/2}}\,,
\ee
with $\fpr{X}_i$ and $\Ywf{X}_i$ as in Eqns.~\eqref{profile} and \eqref{wfnorm}.
Similarly, we write the Yukawa matrices $y_{ij}$ as 
\be\label{yukawas}
\begin{split}
\yuk{U}_{ij} &= \frac{6\pi^2}{M_*}\frac{\fpr{U}_i\,\fpr{Q}_j}{\left(\Ywf{U}_i\,\Ywf{Q}_j\right)^{1/2}} \;\lam{U}_{ij}\,,\\ \
\yuk{D}_{ij} &= \frac{6\pi^2}{M_*}\frac{\fpr{D}_i\,\fpr{Q}_j}{\left(\Ywf{D}_i\,\Ywf{Q}_j\right)^{1/2}}\;\lam{D}_{ij}\,,\\ 
\yuk{E}_{ij} &= \frac{6\pi^2}{M_*}\frac{\fpr{E}_i\,\fpr{L}_j}{\left(\Ywf{E}_i\,\Ywf{L}_j\right)^{1/2}}\;\lam{E}_{ij}\,.
\end{split}
\ee
Here $\lam{U,D,E}_{ij}$ are dimensionless ${\cal O}(1)$ matrices. Such a parameterization can be applied to any model which predicts the Yukawa matrices to be hierarchical, with power-suppressed entries, up to a priori unknown anarchical ${\cal O}(1)$ coefficients. Examples include, besides our wave function localization scheme, also Froggatt-Nielsen type models.

Even though the structure of the Yukawa matrices is dominated by the hierarchical part, it is important to take also the anarchical $\lambda$-coefficients properly into account. An adequate framework for this is Bayesian statistics. We will treat the $\lambda$-parameters as random variables with associated probability density functions (PDFs) $f(\lambda)$. The choice of PDF reflects our theoretical bias (for instance, that all matrix entries should be ${\cal O}(1)$), so this PDF constitutes a ``prior'' in the usual Bayesian vocabulary. The predictions of the model, such as masses or low-energy observables, will then also be PDFs. This approach will enable us to compute Bayesian credibility intervals, the equivalent of confidence intervals in the frequentist approach.

We restrict ourselves to real couplings since the CP problem is not the subject of our study. It should however be kept in mind that, depending on the model, CP-violating observables may provide constraints which are just as stringent as those coming from FCNCs. We also neglect the neutrino sector, because it would induce additional model dependence. We focus on lepton flavour violation in the charged lepton sector, where the constraints are most stringent \cite{Amsler:2008zzb}. 
The wave-function suppression factors $\fpr{X}_i$, and consequently the hierarchy structure, are determined by the six 5D bulk mass parameters $c_{{\cal T}_i}$ and $c_{{\cal F}_i}$ as described in Section \ref{hfroml}. In addition to these, the following unknown ${\cal O}(1)$ flavour coefficients enter the analysis: Five symmetric $3\times 3$ matrices $\lambda^{m_X^2}$ for the squark and slepton masses in Eq.~\eqref{sfermions}, a symmetric Yukawa coefficient matrix $\lam{U}$,
and two unconstrained $3\times 3$ Yukawa coefficient matrices $\lam{D,E}$ in Eq.~\eqref{yukawas}; or a total of 54 additional parameters.

In the Standard Model subsector, we have the six $c$-parameters and the 24 Yukawa coefficients $\lam{U,D,E}_{ij}$.  On the other hand, the experimental observables are the nine Standard Model fermion masses and the three CKM angles. 
Simply setting all $|\lam{U,D,E}_{ij}|=1$ and adjusting the six $c_{{\cal T}_1}$, $c_{{\cal T}_2}$, $c_{{\cal T}_3}$, $c_{{\cal F}_1}$, $c_{{\cal F}_2}$, $c_{{\cal F}_3}$ would not reproduce the Standard Model data with reasonable accuracy, so some deviation of the $\lambda^{u,d,e}$ parameters from unity is clearly needed. However, allowing all 24 $\lam{U,D,E}_{ij}$ to vary and attempting a full Bayesian analysis would require us to take into account the entire SUSY model, because of SUSY threshold corrections to the Yukawa couplings. This would be computationally very involved, and the results rather unwieldy. 
Instead we will take a simplified approach, which still allows us to extract the essential information.

To quantify matrix anarchy, we allow the $|\lambda_{ij}|$ to vary independently within the range 
\be\label{Ell}
|\lambda_{ij}|\in [1/\Ell, \Ell]\,,
\ee
where $\Ell\geq1$ is a constant which is universal for all $|\lambda_{ij}|$. 
With logarithmic weighting, the prior PDF is
\be\label{prior}
f(\log|\lambda_{ij}|)=U(-\log \Ell,\log \Ell)\,,
\ee
$U(a,b)$ being the uniform distribution on the interval $[a,b]$.  We also allow the signs of the $\lambda_{ij}$ to be independently $\pm 1$ (subject to certain restrictions; see below). As stated above, to satisfy the experimental constraints at some reasonable level of precision (e.g., $1\sigma$--$3\sigma$), a minimum $\Ell>1$ denoted by $\Ell_{\rm min}$ is necessary.

We also need to take experimental uncertainties into account. Each MSSM observable has an associated experimental PDF $f_{\rm ex}$, characterizing the uncertainty with which it is measured. On the other hand, any MSSM observable can be expressed in terms of our model parameters $\lambda_{ij}$, which for any given $\Ell$ defines its theoretical PDF $f_{\Ell}$. The combination of these two PDFs gives the total PDF for any given observable.\footnote{In the present situation, the PDFs are combined simply by convolution. For example, the top Yukawa coupling is given by $y_t=|\lam{U}_{33}|e^{-2\pi k R (c_{\mathcal{T}_3}-1/2 )}$ to leading order (cf.~Appendix \ref{appendix_epsilon_expansions}), hence $-\pi k R (c_{\mathcal{T}_3}-1/2 )=(\log y_t-\log|\lam{U}_{33}|)/2$. The PDF of $\log y_t-\log|\lam{U}_{33}|$ is the convolution of the $\log y_t$ and $\log|\lam{U}_{33}|$ PDFs, in other words, of $f_{\rm ex}(\log y_t)$ and $f_{\Ell}=U(-\log\Ell,\log\Ell)$. The $\lambda_{ij}$ can be regarded as nuisance parameters, the parameter of interest being $c_{\mathcal{T}_3}$. For the other couplings, the $f_{\Ell}$ are more complicated because the dependence on the $\lambda_{ij}$ is more involved (see Appendix \ref{appendix_epsilon_expansions}).}

For each independent experimental constraint (or equivalently, each independent observable), there is now a a characteristic value of $\Ell$, denoted by $\Ell^*$, above which the total PDF is dominated by $f_\Ell$. This happens when  $f_{\rm ex}$ and $f_\Ell$ have roughly the same width. We take the $f_{\rm ex}$ to be normal distributions of variance $\sigma^2_{\rm ex}$, with GUT-scale propagated errors from \cite{Ross:2007az}; explicitly, the mean values and relative uncertainties are given in Table \ref{fex}. The $\Ell^*$ associated to each constraint is estimated to be $\log\Ell^*\sim\sigma_{\rm ex}$. While this is only a rough estimate, it will be sufficient for our purposes.

\begin{table}
\begin{center}
\begin{tabular}{|c|c|c|c|}
\hline 
GUT-scale parameter & $\tan\beta=1.3$ & $\tan\beta=10$ & $\tan\beta=50$\tabularnewline
\hline
\hline 
$y_{t}$ & $4\,(75\%)$ & $0.48\,(4.2\%)$ & $0.51\,(5.9\%)$\tabularnewline
\hline 
$y_{b}$ & $0.0113\,(89\%)$ & $0.051\,(3.9\%)$ & $0.37\,(5.4\%)$\tabularnewline
\hline 
$y_{\tau}$ & $0.0114\,(2.6\%)$ & $0.07\,(4.3\%)$ & $0.51\,(7.8\%)$\tabularnewline
\hline 
$y_{u}/y_{c}$ & \multicolumn{3}{c|}{$0.0027\,(22\%)$}\tabularnewline
\hline 
$y_{d}/y_{s}$ & \multicolumn{3}{c|}{$0.051\,(14\%)$}\tabularnewline
\hline 
$y_{e}/y_{\mu}$ & \multicolumn{3}{c|}{$0.0048\,(4.2\%)$}\tabularnewline
\hline 
$y_{c}/y_{t}$ & $0.0009\,(111\%)$ & $0.0025\,(8\%)$ & $0.0023(8.7\%)$\tabularnewline
\hline 
$y_{s}/y_{b}$ & $0.014\,(29\%)$ & $0.019\,(11\%)$ & $0.016\,(12\%)$\tabularnewline
\hline 
$y_{\mu}/y_{\tau}$ & $0.059\,(3.4\%)$ & $0.059\,(3.4\%)$ & $0.05\,(4\%)$\tabularnewline
\hline 
$A$ & $0.56\,(61\%)$ & $0.77\,(2.6\%)$ & $0.72\,(2.8\%)$\tabularnewline
\hline 
$\lambda$ & \multicolumn{3}{c|}{$0.227\,(0.4\%)$}\tabularnewline
\hline 
$|\rho+i\eta|$ & \multicolumn{3}{c|}{$0.397\,(16\%)$}\tabularnewline
\hline
\end{tabular}
\caption{Mean values and propagated relative uncertainties 
of the experimental PDFs $f_{\rm ex}$
associated to the GUT-scale Yukawa matrix eigenvalues and CKM matrix
Wolfenstein parameters. These propagated values are extracted from \cite{Ross:2007az}.}\label{fex}
\end{center}
\end{table}

\begin{figure}
\centering \includegraphics[trim = 0cm 3cm 0cm 5cm, clip,width=12cm]{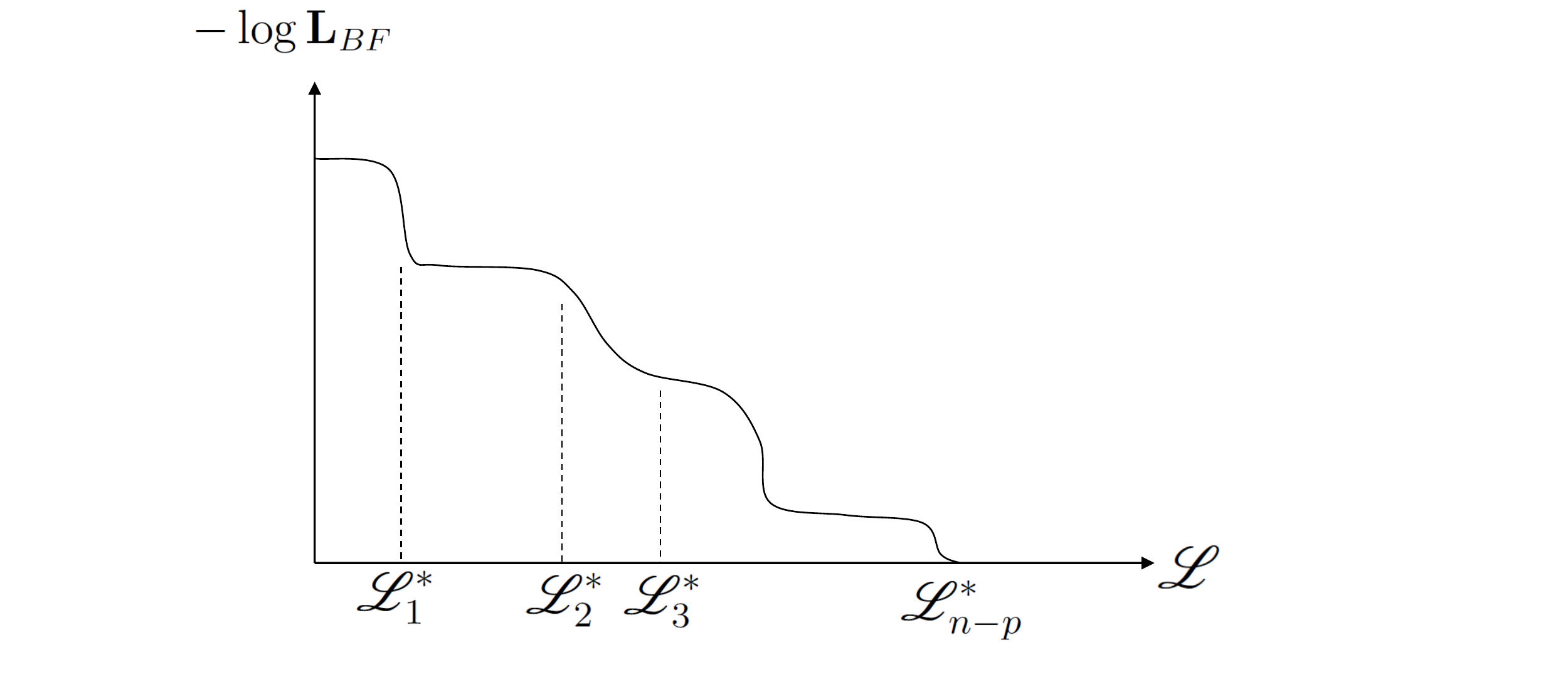}
\caption{Logarithm of the likelihood of the best fit point, $-\log{\bf L}_{\rm BF}$ as a function of $\Ell$, for a fit involving $n$ constraints and $p$ parameters. The constraints weaken and eventually decouple as $\Ell$ increases. The fit becomes perfect, i.e.~$-\log{\bf L}_{\rm BF}=0$, once only $p$ of the $n$ constraints remain.  \label{fig:sketch_L}}
\end{figure}

The constraints with the largest $\Ell^*$ correspond to those that are the ``hardest to fulfill'' (i.e.~to observables that are the hardest to fit). 
For a given constraint, once $\Ell\geq\Ell^*$, its width (dominated by $f_\Ell$) increases with $\Ell$, which makes the constraint easier to fulfill, until it effectively decouples from the fit. In a fit involving $n$ constraints and $p$ parameters, whose best-fit point has likelihood ${\bf L}_{\rm BF}$, the quantity $-\log{\bf L}_{\rm BF}$ will decrease as $\Ell$ increases. The fit becomes perfect, with $-\log{\bf L}_{\rm BF}=0$, once only $p$ of the $n$ constraints remain.
This is illustrated in Figure \ref{fig:sketch_L}.

These observations can be used to determine $\Ell_{\rm min}$ as follows. 
We first perform a fit involving the six $c$-parameters, and the seven constraints with largest $\Ell^*$. This gives a best-fit point with likelihood ${\bf L}_{\rm BF} <1$.
Using the PDF of the seventh constraint, we can then deduce the value of $\Ell$ necessary to increase ${\bf L}_{\rm BF}$ until the required level of precision is reached. This value of $\Ell$ gives $\Ell_{\rm min}$, provided that the constraints which were previously not taken into account can also be satisfied with this $\Ell$. If one of these constraints is not satisfied, 
we repeat the procedure, including this additional constraint in the fit.

The estimation of $\Ell_{\rm min}$ depends on the precision required to fit the Standard Model values (masses and CKM mixing angles).  It also depends on the running  of the Yukawa couplings, which in turn depends on $\tan\beta$ and on threshold corrections. Finally, it depends on the efficiency of finding the best fit point. We find
\be
\Ell_{\rm min}=1.2\,\div\, 1.5\,.
\ee
The value $\Ell_{\rm min}=1.2$ is obtained for a  fit of $3\sigma$ precision, while $\Ell_{\rm min}=1.5$ corresponds to a more conservative  $1\sigma$ fit. It is reasonable to also impose an upper bound on $\Ell$, our starting point being that the $\lambda$s should all be ${\cal O}(1)$, but this bound is of course much less rigorously defined. 
In the analysis of Section \ref{results}, $\Ell$ will be allowed to vary within the range 
\be
1.2\,\leq\,\Ell\,\leq\, 3\,.
\ee
Flavour matrix anarchy being an essential ingredient in our framework, $\Ell$ can in a sense be regarded as a model parameter which measures the allowed deviation from the superimposed hierarchical structure.

Finally, to determine the $c$-parameters we use the constraints with largest $\Ell^*$, which turn out to be the quark Yukawa couplings. The $c_{{\cal T}_i}$ are then deduced from the expressions given in Appendix~\ref{appendix_epsilon_expansions}, subsection~\ref{sect:fermion_masses}. The $c_{{\cal F}_i}$, on the other hand, are not hierarchical, $c_{{\cal F}_1}\simeq c_{{\cal F}_2}\simeq c_{{\cal F}_3}$, so they cannot be determined from the analytic expressions for the down-type Yukawa couplings. We therefore set $c_{{\cal F}_1}= c_{{\cal F}_2}= c_{{\cal F}_3}\equiv c_{{\cal F}}$, with $c_{{\cal F}}$ determined by the bottom Yukawa coupling. We have checked that this choice does not sensitively influence the mass spectrum, the mixings, or the rates of flavour violating processes.

We close this section with some remarks on the signs of the $\lambda_{ij}$. In the limit $\Ell\into 1$, such that all $\lambda_{ij}$ are $\pm 1$, some sign combinations give rise to accidental cancellations when rotating to the mass eigenstate basis, by which one or two of the fermion masses vanish exactly. For instance, if all $\lambda_{ij}=+1$ then there is only one mass eigenstate with nonzero mass. 
When ultimately taking $\Ell>1$, the previously vanishing masses will acquire widespread, $\Ell$-dependent PDFs. For greater predictivity we thus restrict our analysis to sign combinations which are non-singular in the limit $\Ell\into 1$.

Furthermore, we expect the $\yuk{D}$ and $\yuk{E}$ Yukawa matrices to obey GUT relations at the level of signs. This is because all Yukawa couplings originate on the IR brane, where $\SU 6$ is broken only spontaneously. Higher-dimensional operators involving some powers of $\vev\Sigma$ can lead to violation of the GUT relations, but the leading contributions which determine the signs are given by the $\SU{6}$-symmetric second term in Eq.~\eqref{hgutyuk}, which implies ${\rm sign}\,\left(\lam{D}_{ij}\right)={\rm sign}\,\left(\lam{E}_{ji}\right)$. The same argument holds for soft masses in the $\cal T$ and $\cal F$ sector, that is ${\rm sign}\,\left(\msq{Q}_{ij}\right)={\rm sign}\,\left(\msq{U}_{ij}\right)={\rm sign}\,\left(\msq{E}_{ij}\right)$ and ${\rm sign}\,\left(\msq{L}_{ij}\right)={\rm sign}\,\left(\msq{D}_{ij}\right)$.

\section{Results}\label{results}

We are finally in a position to describe our numerical analysis. The dominant flavour constraint comes from the lepton sector, more precisely from $\textrm{BR}(\mu\rightarrow e \gamma)$, as is the case for many other SUSY GUT models -- see for instance~\cite{Buras:2010pm}. The leading contribution to $\mu\rightarrow e \gamma$ contains $\tan\beta$-enhanced pieces \cite{Hisano:1995nq}, so that in the mass-insertion approximation the branching ratio can roughly be estimated as
\be
{\rm BR}(\mu\into e\gamma)\approx\frac{\alpha^3}{G_F^2}\frac{1}{m_{\rm S}^8}\left[\left(\left|(\msq{L})_{12}\right|^2+\left|(\msq{E})_{12}\right|^2\right)\,\tan^2\beta+\left|\frac{m_{\rm S}\,\atr{E}_{12}}{y_\mu}\right|^2\right]\,,
\ee
where $m_{\rm S}$ is a typical sparticle mass, and $\msq{E}$, $\msq{L}$ and $\atr{E}$ are in the super-CKM basis.\footnote{In this formula the first two terms reflect the contributions of diagrams 1(c) and 2 of Reference \cite{Hisano:1995nq}, while the last term corresponds to a single $LR$ flavour-changing mass insertion.} For our purposes, however, this estimate is not precise enough, and the branching ratio is instead determined by a full one-loop calculation \cite{Arganda:2005ji}. The lepton sector comprises $21$ relevant $\lambda$-parameters, $9$ from the Yukawa matrix and $6$ each from the soft mass matrices $\msq{E}$ and $\msq{L}$. The soft terms in the quark sector  
enter only at higher loop order, so these are all the $\lambda$-parameters which need to be fixed. 

Despite the fact that $\Ell$ should be larger than ${\Ell}_{\rm min}$ in order to fit the Standard Model, we initially set $\Ell=1$, so the only variables are the signs of the $\lambda_{ij}=\pm 1$. There are $3\cdot 2^{13}$ physically inequivalent relevant sign combinations: By field redefinitions one can choose five of the signs in the Yukawa coefficients $\lam{E}_{ij}$ to be positive, and of the remaining $16$ combinations, only $6$ lead to non-vanishing fermion masses for all three generations. Six independent signs in each $\lambda^{m_E^2}_{ij}$ and $\lambda^{m_L^2}_{ij}$ can be chosen independently, hence we have $6\cdot 2^6\cdot 2^6=3\cdot 2^{13}=24576$ combinations. We first scan over these sign combinations while keeping $\Ell=1$, and subsequently allow for $\Ell>1$. 

For any given $\tan\beta$, the $c$-parameters are now determined as described in Section~\ref{method}. 
For any given scale of radion-mediated SUSY breaking $F^T/2R$ and of brane-source mediated SUSY breaking $F^Z/M_*$, the soft terms at the GUT scale are calculated from 
Eqns.~\eqref{ndahiggs}--\eqref{rmsbhiggs}, \eqref{rmsbgauginos}, and \eqref{sfermions}. 
The sparticle mass spectrum, mixing matrices, and low-energy flavour observables are computed using \verb!SPheno3! \cite{Porod:2003um}, appropriately modified to handle, in particular, the DHMM condition Eq.~\eqref{dhmm}.

\begin{figure}
\centering \includegraphics[width=\textwidth]{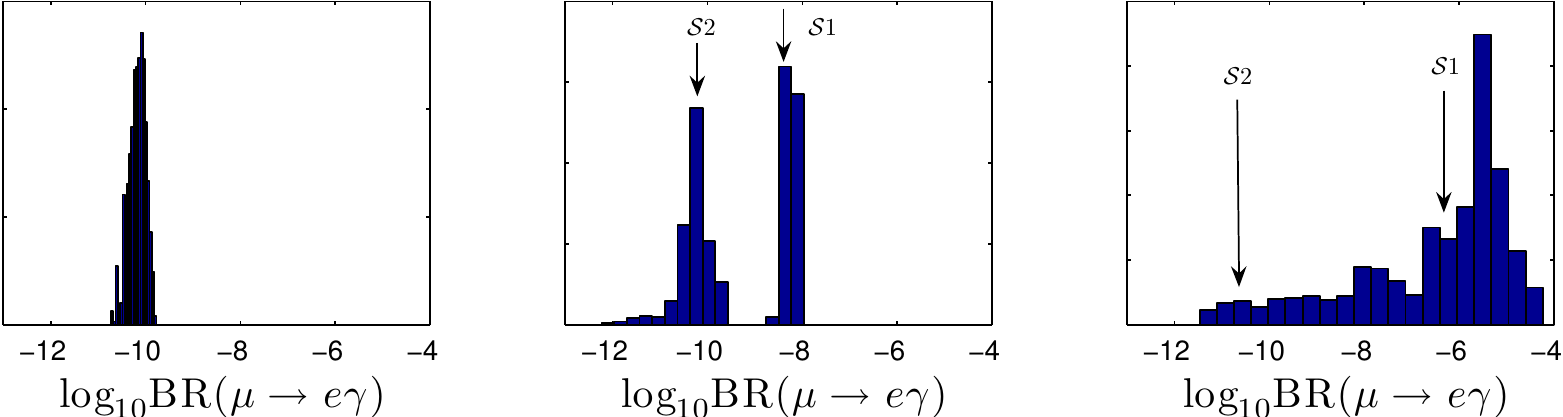}
\caption{Distributions of $\textrm{BR}(\mu\rightarrow e \gamma)$ for all permissible sign combinations in the $\lambda$ matrices at $\Ell=1$. The model parameters are $F^T/2R=F^Z/M^*=1500~\textrm{GeV}$, and $\tan\beta=5,\,10,\,30$ (from left to right). The current experimental bound is $\textrm{BR}(\mu\rightarrow e \gamma)<1.2\cdot 10^{-11}$ \cite{Brooks:1999pu}. The arrows indicate the positions of two benchmark sign combinations S1 and S2 used later in our analysis.  \label{fig:sign_dist}}
\end{figure}

Figure~\ref{fig:sign_dist} shows some sample distributions for  $\textrm{BR}(\mu\rightarrow e \gamma)$ at $\Ell=1$ for various $\tan\beta$. At low $\tan\beta$, the dominant contribution to $\textrm{BR}(\mu\rightarrow e \gamma)$ comes from the trilinear term  $\atr{E}$. By assumption, the trilinears are induced only by radion mediation and so do not depend on the $\lambda$-parameters. This explains why in the left panel of Fig.~\ref{fig:sign_dist} the value of $\textrm{BR}(\mu\rightarrow e \gamma)$ is hardly sensitive to the sign combination. By contrast, at large $\tan\beta$, $\atr{E}$ and the brane soft masses $\msq{E}_{\rm brane}$, $\msq{L}_{\rm brane}$ (defined in Eq.~\eqref{sfermions}) give comparable contributions to $\textrm{BR}(\mu\rightarrow e \gamma)$. Since these depend on the $\lambda^{m^2_E}_{ij}$ and $\lambda^{m^2_L}_{ij}$, the distributions span a much wider range at large $\tan\beta$.

To also allow for $\Ell>1$, and hence for a good Standard Model fit, we choose two benchmark sign combinations in the $\lambda$ matrices, which we denote by S1 and S2 (indicated by arrows in Fig.~\ref{fig:sign_dist}). Explicitly, we have
\be\begin{split}
\lam{E}=\left(\begin{array}{ccc}+&+&+\\+&+&-\\+&-&- \end{array}\right),\qquad
\lambda^{m_E^2}=\left(\begin{array}{ccc}+&+&+\\+&+&+\\+&+&+ \end{array}\right),\qquad
\lambda^{m_L^2}=\left(\begin{array}{ccc}+&+&+\\+&+&+\\+&+&+ \end{array}\right)
\end{split}
\ee
for S1, and 
\be\begin{split}
\lam{E}=\left(\begin{array}{ccc}+&+&+\\+&+&-\\+&-&+ \end{array}\right),\qquad
\lambda^{m_E^2}=\left(\begin{array}{ccc}+&+&+\\+&+&+\\+&+&+ \end{array}\right),\qquad
\lambda^{m_L^2}=\left(\begin{array}{ccc}+&+&+\\+&+&+\\+&+&+ \end{array}\right)\,.
\end{split}
\ee
for S2. As is evident from Fig.~\ref{fig:sign_dist}, with the sign combination S2 the $\mu\rightarrow e \gamma$ decay rate is suppressed even for large $\tan\beta$, whereas the S1 point exhibits increasingly large flavour violation. We then scan several values of $\tan\beta$, $F^T/2R$, and $F^Z/M_*$ for these two sign combinations, and allow the $|\lambda_{ij}|$ to deviate from $1$. The values of the $|\lambda_{ij}|$ are drawn from their prior defined by  Eq.~\eqref{prior}. 

\begin{figure}
\centering \includegraphics[width=4.5cm]{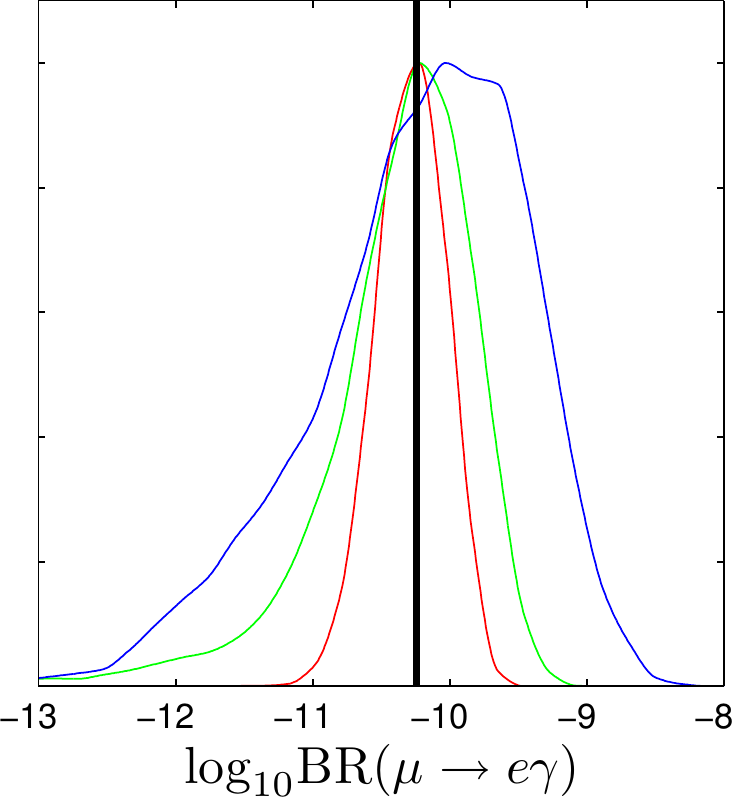}
\includegraphics[width=4.5cm]{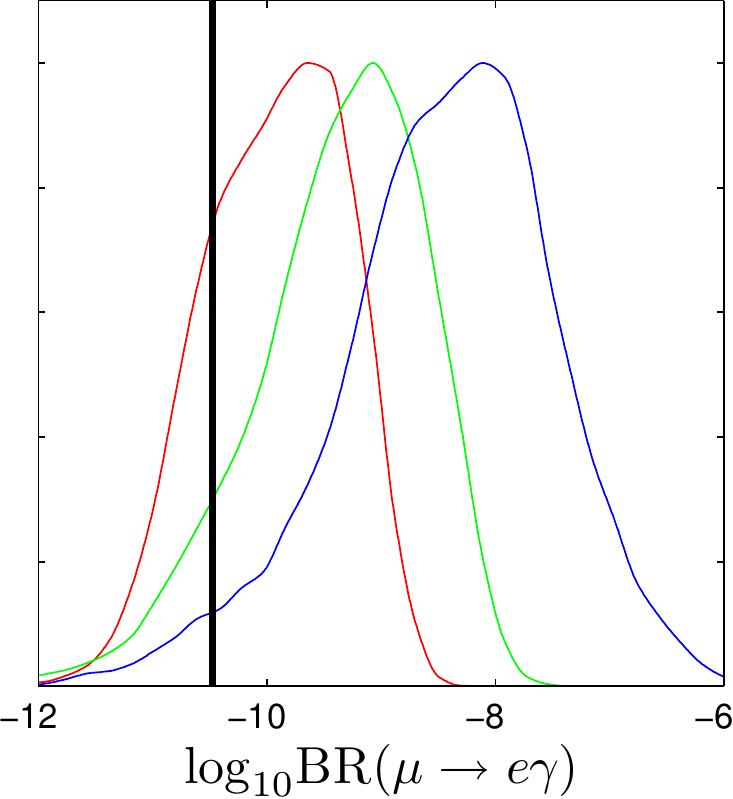}
\includegraphics[width=4.5cm]{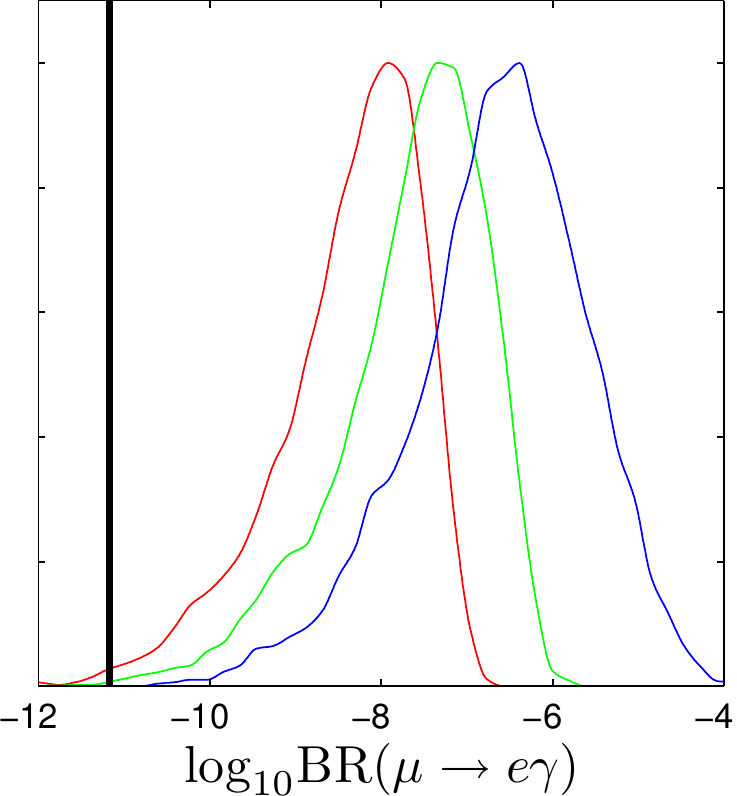}
\caption{Probability density functions of $\textrm{BR}(\mu\rightarrow e \gamma)$ for the sign combination S2, with $F^T/2R=F^Z/M_*=1500~\textrm{GeV}$, and  $\tan\beta=5$, $10$, $30$ from left to right. The red, green, blue lines correspond to $\mathscr{L}=1.2$, $1.5$, $3$ respectively, with the vertical black lines showing the values at $\mathscr{L}=1$.
Radion mediation dominates at $\tan\beta=5$, while the brane source dominates at $\tan\beta=30$. The PDFs are normalized to have the same maximum.
 \label{fig:lambda_dist}}
\end{figure}

Figure \ref{fig:lambda_dist} shows a typical example for the sign combination S2 and $F^T/2R=F^Z/M_*$, such that $\atr{E}$ (i.e.\ radion mediation) dominates at low $\tan\beta$, and brane soft masses dominate at large $\tan\beta$. 
In this case, $\textrm{BR}(\mu\rightarrow e \gamma)$ is suppressed at large $\tan\beta$. One can also see that the width of the PDF increases with $\mathscr{L}$ as expected. Note, moreover, that at large $\tan\beta$, where the brane contribution dominates,  the mean value of the PDF is shifted towards larger values when increasing $\mathscr{L}$. The reason for this $\mathscr{L}$-dependent shift is that certain cancellations occur between the brane contributions for this sign combination in the limit $\Ell\into 1$.
At low $\tan\beta$, where radion mediation dominates, this effect is less important. 

Regardless of the chosen sign combination, when radion mediation dominates, the $\textrm{BR}(\mu\rightarrow e \gamma)$ constraint is weakened by at least an order of magnitude, depending on $\mathscr{L}$.
By contrast, when the brane-induced soft terms dominate, there are large contributions to $\textrm{BR}(\mu\rightarrow e \gamma)$. 

\begin{figure}[t]
\centering \includegraphics[width=6.4cm]{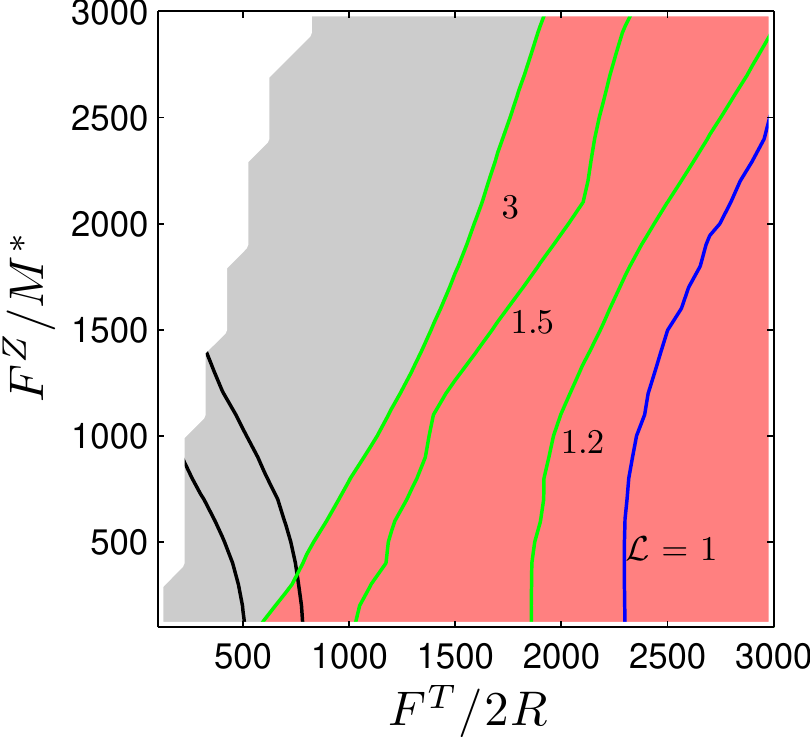}
\includegraphics[width=6.4cm]{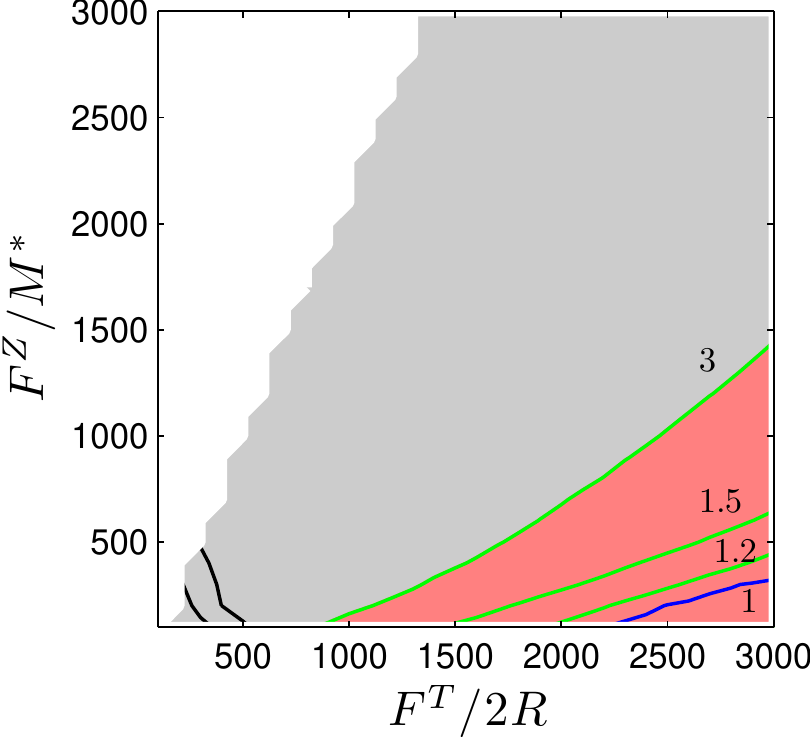}
\caption{ The dominant constraints in the $(F^T/2R,~F^Z/M_*)$ plane for the sign combination S1, $\tan\beta=5$ (left) and $\tan\beta=30$ (right). $F^T/2R$ and $F^Z/M_*$ are in GeV units. The black lines on the bottom left are $m_h=111~\textrm{GeV}$ and $m_h=114~\textrm{GeV}$ isolines. The blue lines show the $\textrm{BR}(\mu\rightarrow e \gamma)$ constraint for  $\mathscr{L}=1$, while green lines show $95\%$ BCIs of the same constraint for $\mathscr{L}=1.2$, $1.5$, $3$.  The red regions, towards large $F^T/2R$,  satisfy $m_h>111~\textrm{GeV}$ and $\textrm{BR}(\mu\rightarrow e \gamma)<1.2\cdot 10^{-11}$ for at least one value of $\mathscr{L}$. In the white regions, a too large $F^Z/M_*$ leads to tachyonic sleptons.  
\label{fig:isolines1}}
\end{figure}

\begin{figure}
\centering 
\includegraphics[width=6.4cm]{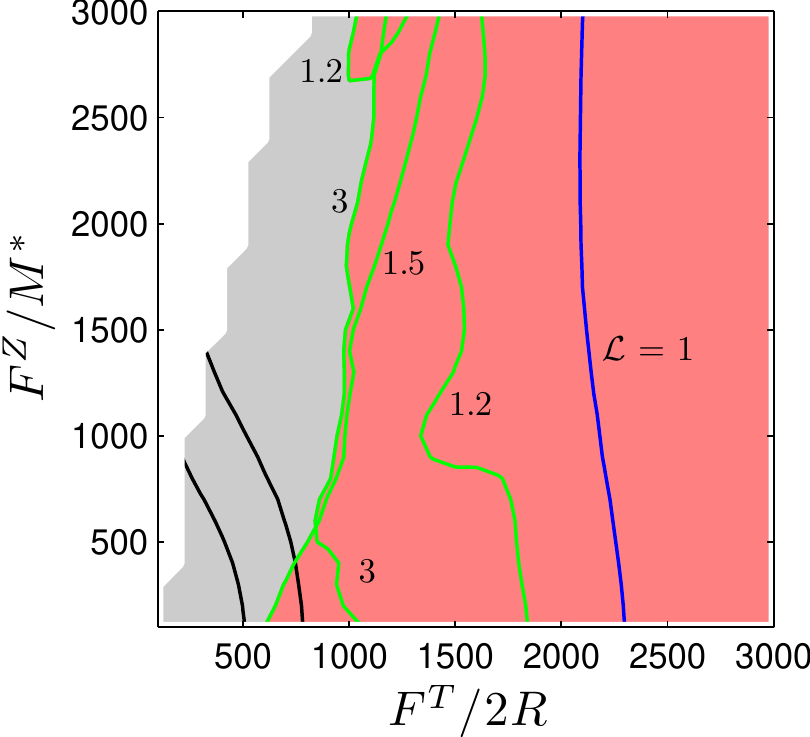}
\includegraphics[width=6.4cm]{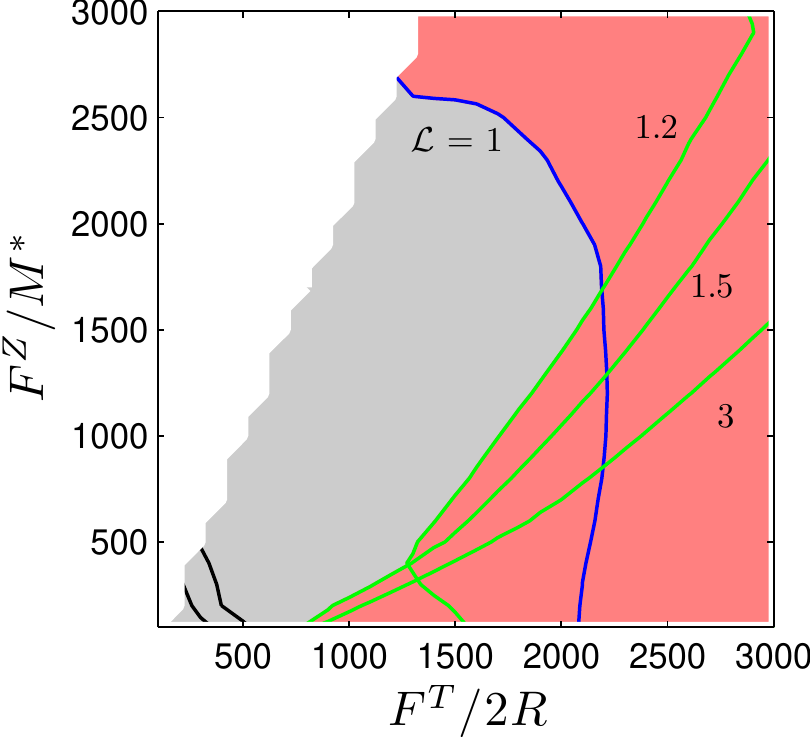}
\caption{Same as Fig.~\ref{fig:isolines1} but for the sign combination S2.
\label{fig:isolines2}}
\end{figure}

In Figs.~\ref{fig:isolines1} and \ref{fig:isolines2} we show the lower bounds on $\textrm{BR}(\mu\rightarrow e \gamma)$, given  by $95\%$ Bayesian credibility intervals (BCIs), for various $\mathscr{L}\geq\mathscr{L}_{\rm min}$. 
Figure~\ref{fig:isolines1} is for combination S1, while  Fig.~\ref{fig:isolines2} is for combination S2. 
For any given value of $\Ell$, the red regions to the right of the corresponding $\Ell$-line pass the experimental constraints.\footnote{We consider $m_h>114$~GeV and $m_h>111$~GeV to account for the $\sim3$~GeV theoretical uncertainty on the light Higgs mass.} 
In the left panel, where radion mediation dominates, the bound on $F^T/2R$ is weakened by a factor 2--3 when increasing $\mathscr{L}$. Since we have $M_{1/2}\sim F^T/2R$, the gluino mass is typically $m_ {\tilde{g}} \sim 2\,(F^T/2R)$ due to RG running. Moreover, squark masses are dominated by the gluino contribution to their RGEs, so we have $m_{\tilde q}\approx m_{\tilde g}$. 
Therefore, the weakening of the $\textrm{BR}(\mu\rightarrow e \gamma)$ bound  opens a part of the parameter space which is relevant for the production of SUSY particles at the LHC.

We conclude from Figs.~\ref{fig:isolines1} and \ref{fig:isolines2} that there are regions of the parameter space 
which pass the flavour constraints, and where SUSY 
particles are within discovery reach~\cite{ATLAS:1999fr,Ball:2007zza} at the LHC. 
ATLAS and CMS searches with about 1~fb$^{-1}$ of data at 7~TeV already 
put lower limits on gluino and squark masses of roughly $m_{\tilde g,\tilde q}\gtrsim 1$~TeV 
for $m_{\tilde q}\simeq m_{\tilde g}$~\cite{ATLAS:EPS,CMS:EPS}.  
It is worthwhile noting that in our case the $\textrm{BR}(\mu\rightarrow e \gamma)$ constraint 
forces the SUSY spectrum to be heavy, generically beyond the current LHC limits. In that sense the most severe constraints on our class of models still originate from flavour precision experiments, rather than from direct superpartner searches.

At this point a comment is in order concerning the effects of subdominant flavour constraints.
The next-to-dominant constraint turns out to be $\textrm{BR}(\mu\rightarrow 3e )$, which is strongly correlated to $\textrm{BR}(\mu\rightarrow e \gamma)$, such that a weakening of the latter weakens also the former. We do not show this or other subdominant constraints here, since little would be gained by taking them into account. 

Production cross sections at the LHC are very similar to those of the mSUGRA case with $m_{\tilde q}\approx m_{\tilde g}$, 
and can be characterized by the gluino--squark mass scale, see e.g.~\cite{Baer:1995nq,Baer:2006rs}. 
At $m_{\tilde g,\tilde q}\approx1$, 2 and 3~TeV, the overall SUSY cross section is of the order of 1~pb, 10~f\/b 
and 1~f\/b, respectively. 
For  $m_{\tilde g,\tilde q}\approx1$~TeV, the cross section is dominated by gluino--gluino, squark--squark and gluino--squark production. For heavier masses, $m_{\tilde g,\tilde q}\approx2$--3~TeV, squark--squark and electroweak ino-ino (mainly $\tilde\chi^\pm_1\tilde\chi^0_2$) production dominate, while 
gluino production becomes negligible. However, interesting LHC signatures arise from the slepton mass patterns and mixings, leading to chargino/neutralino and slepton decays that are specific to the setup studied here.

Let us therefore next discuss details of the spectrum, of mixings, and of LHC phenomenology.
Two cases will be distinguished.  
First, we will treat the case where the effects of $F^Z/M_*$ in the scalar soft masses are negligible
compared to those of $F^T/2R$. In this scenario, which we call radion dominated, the only effect of brane-localized SUSY breaking is to provide nonzero Higgs soft masses. Since our model has a GUT-scale degenerate Higgs mass matrix, Eq.~\eqref{dhmm}, it is convenient to fix the Higgs soft masses by this condition, and to set $F^Z/M_*$ to zero for the purposes of calculating sfermion soft terms. We have checked that a small $F^Z/M_*$, of the order of $\sim 0.02\,(F^T/2R)$, is sufficient to generate the necessary Higgs soft masses but has negligible effect in the sfermion sector. 

Second, we will discuss the situation where the scalar soft masses receive contributions both from  $F^Z/M_*$ and from $F^T/2R$, which we call mixed brane--radion scenario.
$F^Z/M_*$ should be bounded from above, because it can induce tachyonic sleptons through RG running if it is too large, and because it can enhance  $\textrm{BR}(\mu\rightarrow e \gamma)$ depending on the signs of the $\lambda$-parameters. This can also be seen in Figs.~\ref{fig:isolines1} and \ref{fig:isolines2}.

\subsection{Radion dominated scenario }\label{radion_domination}

\begin{figure}
\centering \includegraphics[trim = 0cm 7.5cm 0cm 8cm, clip,width=12cm]{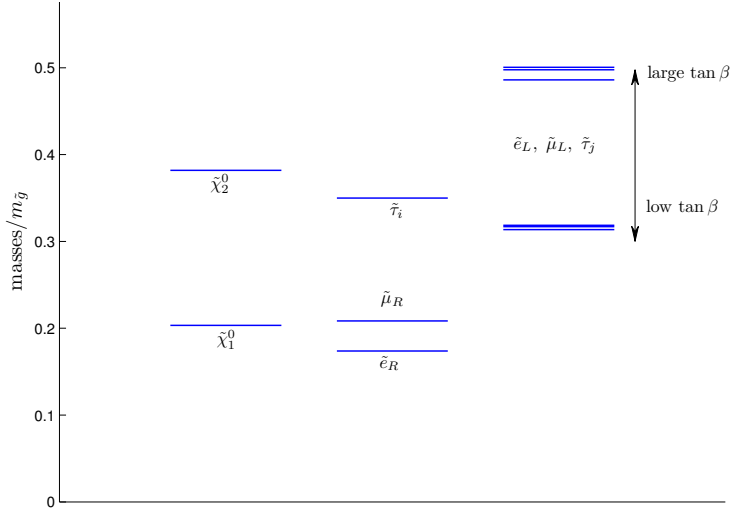}
\caption{Spectrum of SUSY particles, normalized to the gluino mass, when radion mediation dominates. The lightest state is almost completely a right selectron $\tilde{e}_R$. The left-handed slepton masses increase with $\tan\beta$, due to the $y^e$ enhancement (see text).
For large $\tan\beta$ and large $\mathscr{L}$, the RGE effects
of flavour anarchy cause the slepton mass PDFs to grow wider, except the one for $m_{\tilde{e}_R}$.
  \label{fig:RMSB_spectrum}}
\end{figure}

In the case where scalar soft terms are dominated by radion mediation, the effects of flavour matrix anarchy appear only in Yukawa couplings, not in the soft terms.
In other words, the soft terms do not directly depend on the $\lambda$-parameters.
They will be sensitive to flavour matrix anarchy, i.e to $\mathscr{L}$, only through RGE effects.
The mass ordering  depends  mainly on $\tan\beta$, and is charted in Fig.~\ref{fig:RMSB_spectrum}. 
Here we denote the six slepton mass eigenstates $\tilde{l}_{1\ldots6}$ by their dominant components, for example $\tilde{l}_1\sim \tilde{e}_R$, $\tilde{l}_2\sim \tilde{\mu}_R$, and so forth. 
The $\tilde{l}_1\sim \tilde{e}_R$ turns out to be the lightest state of the spectrum, followed by the lightest neutralino $\tilde\chi^0_1$ and the $\tilde{l}_2\sim \tilde{\mu}_R$.
The masses of left-handed sleptons and sneutrinos increase with $\tan\beta$. This is because, as $\tan\beta$ grows, the charged lepton Yukawa couplings $y^e$ increase, so the $c$-parameters shrink and the corresponding soft terms are enhanced. 

At small $\tan\beta$, the PDFs are strongly peaked at the values shown in Fig.~\ref{fig:RMSB_spectrum}. At large $\tan\beta$ and large $\mathscr{L}$ (about $1.5\lesssim\mathscr{L}\lesssim 3$), the RGE effects of flavour anarchy widen the slepton mass PDFs, again due to the $y^e$ enhancement at large $\tan\beta$. This effect is  hierarchical: For the $m_{\tilde{\tau}}$ PDF it is larger than for the $m_{\tilde{\mu}_R}$ PDF, while $m_{\tilde{e}_R}$ shows almost no sensitivity. 
Typically, for $\tan\beta\geq30$ and $\mathscr{L}\geq1.5$, this uncertainty is large enough to flip the mass ordering of $\tilde{\mu}_R$ and $\tilde{\chi}_1^0$, and of $\tilde{\tau}_1$ and $\tilde{\chi}_2^0$.
Other mass orderings are  conserved.

A stable charged slepton as the lightest SUSY particle (LSP) is obviously excluded by cosmology. 
In particular it is not a suitable dark matter candidate. Dark matter could instead be composed of gravitinos $\widetilde G$, which are the LSP in many models of warped supersymmetry (such as the one in Appendix \ref{modsb}), or of axinos $\tilde a$. In that case the $\tilde{e}_R$ decays as $\tilde{e}_R^\pm\to e^\pm+{\rm LSP}$ (LSP = $\widetilde G$ or $\tilde a$).
If this decay occurs at the epoch of Big Bang Nucleosynthesis (BBN)~\cite{Iocco:2008va}, $\tau(\tilde{e}_R)\gsim 1$\,sec, it  can alter the yield of light elements. This poses important constraints in particular on the gravitino LSP case.
In case of an axino LSP, the $\tilde{e}_R$ decay is much faster \cite{Brandenburg:2005he},  
so that 
BBN constraints can be evaded easily.  
A detailed discussion of this issue is beyond the scope of this paper; in the following 
we simply assume that $\tilde{e}_R$ is in fact the next-to-LSP and that its abundance and lifetime are small enough to evade cosmological constraints. 
Note, however, that even in the axino LSP case with $\tau(\tilde{e}_R)\ll 1$\,sec, 
the $\tilde{e}_R$ appears as a heavy stable charged particle  \cite{Fairbairn:2006gg,adams:2011} in collider experiments. 
For definiteness, we will refer to the $\tilde{e}_R$ as the ``lightest massive particle'' (LMP) in the following.

Let us now turn to LHC cascade decays. Gluinos and squarks, if produced, decay as $\tilde g\to q\tilde q_{R,L}^{}$, 
  $\tilde q_R^{} \to q\tilde\chi^0_1$ ($\sim 100\%$), and $\tilde q_L^{}\to q'\tilde\chi^\pm_1$  
  ($\sim 65\%$) or $q\tilde\chi^0_2$ ($\sim 30\%$). The $\tilde{\chi}_{1,2}^0$ and $\tilde{\chi}^\pm_1$  decay further, and the decay chains  end with the $\tilde{e}_R$ LMP and an electron. If $m_{\tilde{\mu}_R}<m_{\tilde{\chi}_1^0}$, the $\tilde{\chi}_1^0$ can decay both as 
 $\tilde{\chi}_1^0\rightarrow\tilde{e}_R^\pm e^\mp$ and $\tilde{\chi}_1^0\rightarrow\tilde{\mu}_R^\pm\mu^\mp$. The relative rate between the two decays is dictated by the ratio of mass splittings:
\be\label{eq:brnt1}
  \frac{\textrm{BR}(\tilde{\chi}_1^0\rightarrow\tilde{e}_R^\pm e^\mp)\,}
         {\textrm{BR}(\tilde{\chi}_1^0\rightarrow\tilde{\mu}_R^\pm\mu^\mp)} \approx 
  \left(\frac{m_{\tilde{\chi}_1^0}^2-m_{\tilde{e}_R}^2}{m_{\tilde{\chi}_1^0}^2-m_{\tilde{\mu}_R}^2}\right)^2\,.
\ee  
The $\tilde{\mu}_R$ decays dominantly through the three-body mode $\tilde{\mu}_R\rightarrow\tilde{e}_R e \mu$ 
via a virtual $\tilde\chi^0_1$. The decay width is typically of ${\cal O}$(keV), so there is no displaced vertex. 
(In principle the $\tilde{\mu}_R$ may also decay through the LFV mode $\tilde{\mu}_R\to \tilde{e}_R+Z$ if kinematically allowed, but this is suppressed by a very small coupling.) 
This contrasts with universal scalar mass scenarios, where the lightest slepton is typically the $\tilde{\tau}_1$. Observing an electron or a muon associated to the LMP instead of a $\tau$ at the end of the decay chains would therefore be a hint for non-universality in scalar lepton masses. 

As mentioned, the $\tilde{e}_R$ LMP is stable inside the detector and behaves like a heavy muon. 
This can be triggered on in the muon chambers of the ATLAS  and CMS experiments. The muon chambers 
also allow excellent track reconstruction and time of flight measurements with an accuracy of around 1 ns, 
which should allow to reconstruct the mass of the LMP with good precision \cite{Fairbairn:2006gg}.
Moreover, the rate of energy loss through ionization $(dE/dx)$ may be used to identify the LMP and 
measure its properties \cite{Khachatryan:2011ts}. Given this striking signature, Drell-Yan production of 
$\tilde{e}_R$, even with a low cross section, may be exploited. See \cite{Fairbairn:2006gg} and references therein for more details.
 
A further, ambitious idea is to use a stopper detector to observe the LMP late decay~\cite{Hamaguchi:2004df}. With a sufficient number of events, flavour violating decays could be observed, and eventually used to gain some information on $\msq{E}$.
Since $\msq{E}$ is hierarchical, it induces a hierarchical mixing between the right-handed sleptons. LFV processes are thus suppressed by powers of 
the typical hierarchy factor $e^{-\pi kR}\equiv\epsilon$. In particular, one has roughly 
\[\textrm{BR}(\tilde{e}_R\rightarrow \mu\text{ LSP})\,/\,\textrm{BR}(\tilde{e}_R\rightarrow e\text{ LSP})\sim \epsilon^2~.\]
 Given that $\epsilon$ should be  $\mathcal{O}(10^{-1})$ to reproduce the SM flavour hierarchy, if the branching ratio can be measured to $1\%$ or better, this would provide a rough test for our scenario.

The features discussed above are generic for our class of models. Other aspects of LHC phenomenology depend on the precise mass ordering, which in turn depend on $\tan\beta$.
In the following discussion we will therefore distinguish the cases of small and large $\tan\beta$. 
For concreteness we will use three representative scenarios, one with low $\tan\beta$, and two different configurations with large $\tan\beta$ (one featurimg $m_{\tilde{\chi}_1^0}<m_{\tilde{\mu}_R}$, the other $m_{\tilde{\chi}_1^0}>m_{\tilde{\mu}_R}$). The spectra for our canonical choice of $F^T/2R=1.5$~TeV are given in 
Table~\ref{Tab:points_RM_1500}. Since these points lie at the edge of the LHC discovery reach (total cross sections $\lesssim 1$~f\/b), we also provide in Table~\ref{Tab:points_RM_1000} an analogous set of points for $F^T/2R=1$~TeV, which is more interesting for LHC studies. 
The complete SLHA files, including mass matrices and branching ratios, can be obtained from~\cite{points_web}.

For these benchmark points, we have also checked several observables from quark flavour violation using \verb!SUSY_FLAVOR!. Due to the large theoretical uncertainty, the constraints from flavour violation in the quark sector are all satisfied, so lepton flavour violation provides the strongest constraint as anticipated.

In these points, the GUT-scale soft mass $\msq{L}$ has been set to a universal value for simplicity, see end of Section~\ref{method}, whereas generically the left-handed lepton soft masses may differ by ${\cal O}(1)$ factors (they will, however, not be hierarchical since $c_{\mathcal{F}_1}\approx c_{\mathcal{F}_2}\approx c_{\mathcal{F}_3}$). A universal GUT-scale $\msq{L}$ leads to small, RG-induced mass splittings between $\tilde{\mu}_L$ and $\tilde{e}_L$ of typically $\mathcal{O}(0.1\%)$, and somewhat larger mass splittings between  $\tilde{e}_L/\tilde{\mu}_L$ and $\tilde{\tau}_1$ of $\mathcal{O}(5\%)$. This is relevant because, at low $\tan\beta$, the  $\tilde{\chi}_2^0$ and $\tilde{\chi}_1^\pm$ are mostly wino and thus decay mostly into the left-handed sleptons $\tilde{e}_L$, $\tilde{\mu}_L$, $\tilde{\tau}_L$. The branching ratios of LFV decays are at most $\sim10^{-3}$ in that case, and therefore irrelevant for LHC phenomenology.

The main features can be summarized as follows:

\begin{itemize}

\item The lightest Higgs mass $m_{h^0}$ is always lifted by stop loop corrections to take a value closely above the LEP bound, as is typical for the MSSM with relatively heavy stops.

\item
The $\tilde{\chi}_2^0$ decays as $\tilde{\chi}_2^0\rightarrow\tilde{\tau}_1^\pm \tau^\mp$ ($\sim 20$--$30\%$) or $\tilde{\chi}_2^0\rightarrow\tilde{\mu}_L^\pm \mu^\mp/ \tilde{e}_L^\pm e^\mp$ ($\sim 10$--$15\%$ each); the rest goes into $\tilde{\nu}_i \nu_i$.
The sleptons subsequently decay into $\tilde{\chi}_1^0+e/\mu/\tau$, followed by the $\tilde{\chi}_1^0$ decay to the LMP, $\tilde{\chi}_1^0\rightarrow \tilde{e}_R^\pm e^\mp$.
The resulting signature is $\tilde{\chi}_2^0\rightarrow l_i^\pm l_i^\mp e \tilde{e}_R$, i.e.\ same flavour opposite sign (SFOS) dileptons, plus an electron, plus the LMP which behaves like a heavy muon. There is no $E_T^{\rm miss}$.

\item
The $\tilde{\chi}_1^\pm$ cascade decays via a charged slepton $\tilde l_L$ or sneutrino $\tilde\nu$ 
into $\tilde{\chi}_1^0 \tau^\pm \nu_\tau$ ($\sim 50$--$60\%)$ or into 
$\tilde{\chi}_1^0 e^\pm\nu_e/\mu^\pm\nu_\mu $($\sim 20$--$25\%$ each). 
The decay chain gives rise to dilepton signatures of $e+(e/\mu/\tau)$ with uncorrelated charges, plus the LMP, plus $E_T^{\rm miss}$  from the $\nu$s. This decay can be combined with the $\tilde{\chi}_2^0$ decay on the other branch.

\item 
The masses of the sparticles appearing in the decay chains may be determined from kinematic distributions. 
The simplest observable is the endpoint of the SFOS dilepton invariant-mass distribution, 
$M_{ll}^{\rm max}=m_{\tilde{\chi}_2^0}( 1- m^2_{\tilde{l}}/m^2_{\tilde{\chi}_2^0} )^{1/2}( 1- m^2_{\tilde{\chi}_1^0}/m^2_{\tilde{l}})^{1/2}$ from the $\tilde{\chi}_2^0$ decay.\footnote{Note however the ambiguity in the channel with electrons, 
$\tilde{\chi}_2^0\rightarrow \tilde{e}_L^\pm e^\mp \to  \tilde{e}_R + 3e$.} 
For example, for point A', the $M_{ll}$ endpoints are $M_{ee}^{\rm max}=308.28~\textrm{GeV}$, $M_{\mu_\mu}^{\rm max}=308.61~\textrm{GeV}$ and $M_{\tau\tau}^{\rm max}=373.28~\textrm{GeV}$.
This may be used to obtain information on the masses of the three left-handed sleptons. 

At the LHC, it will most likely be not possible to reconstruct the GUT  
scale parameters through a bottom-up evolution. However, at the one-loop level,
$m_{\tilde{\mu}_L}-m_{\tilde{e}_L}$ is RG invariant, and the  
running of $m_{\tilde{\tau}_L} - m_{\tilde{e}_L}$ depends on only a single parameter combination $X_\tau=2|y_\tau|^2(m^2_{H_d}+m^2_{\tilde{\tau}_L}+m^2_{\tilde{\tau}_R})+2|a_\tau|^2$ (cf.~\cite{Martin:1997ns}). By combining the measurement of left-handed slepton masses with other information, such as the mass of coloured particles and the limits on (SUSY) LFV processes, one could at least carry out a hypothesis test on the structure of  $\msq{L}$.

\end{itemize}

%
%

At large $\tan\beta$, the situation is quite different. The left-handed sleptons and the sneutrinos are heavier, and the charged lepton Yukawa couplings are enhanced. The latter induces two effects through the RGEs. On the one hand, as explained above, the mass PDFs become much wider and either ordering, $m_{\tilde{\chi}_1^0}> m_{\tilde{\mu}_R}$ or $m_{\tilde{\chi}_1^0}< m_{\tilde{\mu}_R}$, can now occur. 
On the other hand, LFV processes in the $\mu-\tau$ sector may be sufficiently large to be observed.
Since $m_{\tilde{l}_L}>m_{\tilde{\chi}_2^0}$, the sleptons relevant for LHC phenomenology are now the $\tilde{\mu}_R$, $\tilde{e}_R$ and $\tilde{\tau}_1$.
The $\tilde{\chi}_2^0$ and $\tilde{\chi}_1^\pm$ decay predominantly into the $h^0$ and, if kinematically allowed, into the $\tilde{\tau}_1$, because it contains left-handed components induced by left-right mixing. The points B and C (B' and C') in Table~\ref{Tab:points_RM_1500} (\ref{Tab:points_RM_1000}) are representative examples. 

\noindent
We will first discuss the leading decays, and then study LFV processes.  
Those features which are unrelated to LFV can be summarized as follows.

\begin{figure}
\centering
\includegraphics[width=5cm]{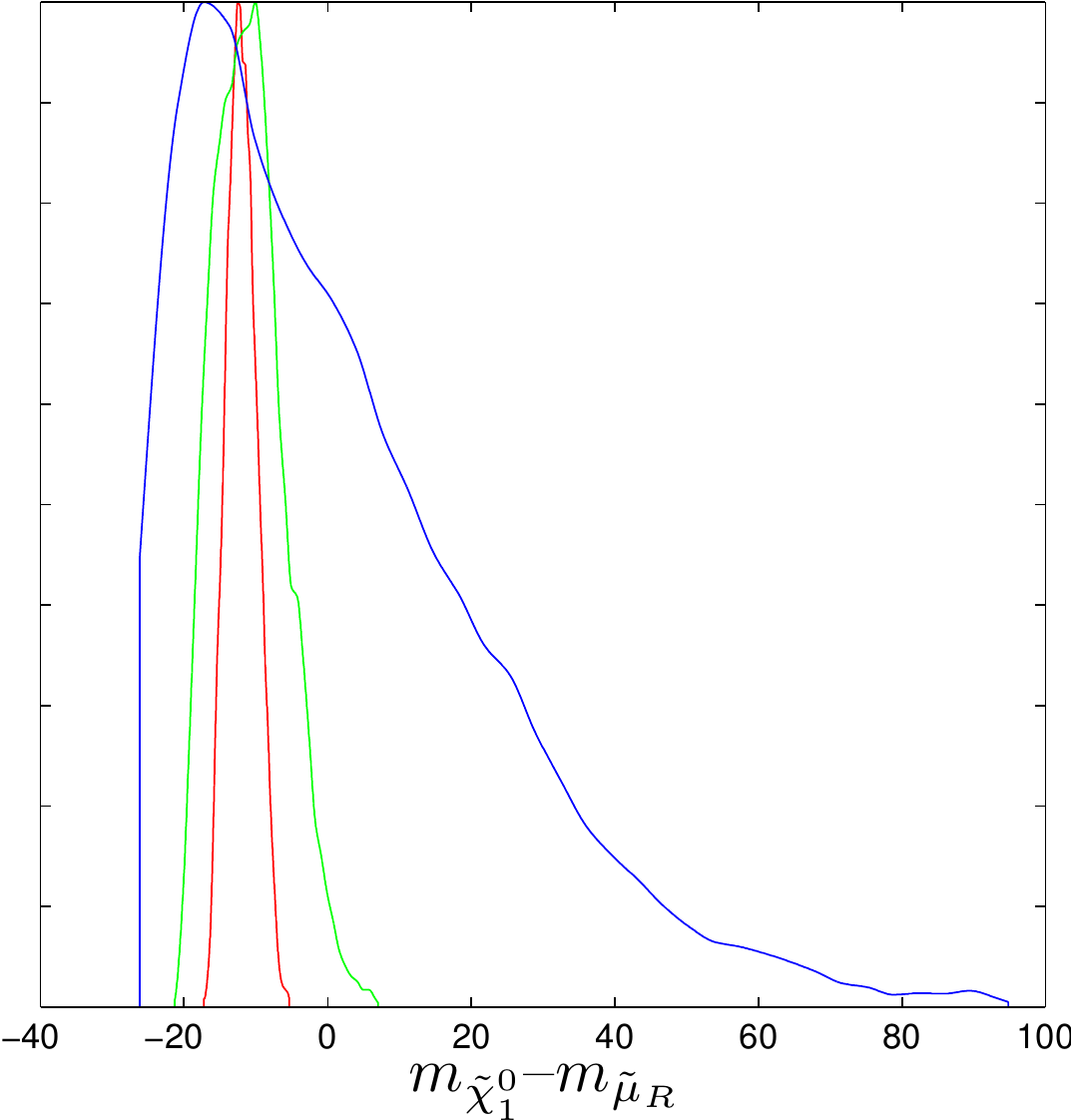}
\caption{PDFs of $\tilde{\chi}_1^0$--$\tilde{\mu}_R$ mass difference, 
for $F^T/2R=1500~\textrm{GeV}$ and $\tan\beta=30$. The red, green and blue lines 
are for $\mathscr{L}=1.2$, 1.5 and 3, respectively.
\label{fig:deltam_pdf}}
\end{figure}

\begin{itemize}

\item Also in these benchmark points, the lightest Higgs mass $m_{h^0}$  is always around $120$ GeV,  closely above the LEP bound.

\item
The $\tilde{\chi}_1^0$ decays into $\tilde{e}_R^\pm e^\mp$ or, if kinematically allowed, also into $\tilde{\mu}_R^\pm \mu^\mp$. If the muonic channel is open, its branching ratio strongly depends on the mass splittings, cf.\ Eq.~\eqref{eq:brnt1}.  
In this case the $\tilde{\mu}_R$ decays predominantly  through a three-body mode $\tilde{\mu}_R \rightarrow \tilde{e}_R e \mu$ as before. The signs of  $e$ and $\mu$ are a priori not correlated, but they are related to the sign of $ \tilde{e}_R$, which may be measured, and the sign of the parent $\tilde{\mu}_R$. 
(If $m_{\tilde{\mu}_R}>m_{\tilde{\chi}_1^0}$, $\tilde{\mu}_R^\pm \to \tilde{\chi}_1^0\mu^\pm$ followed by $\tilde{\chi}_1^0\tilde{e}_R^\pm e^\mp$, giving the same final state.) 
The PDFs of  $m_{\tilde{\chi}_1^0}-m_{\tilde{\mu}_R}$ are shown in Fig.~\ref{fig:deltam_pdf} for different values of $\mathscr{L}$.

\item
The $\tilde{\chi}_2^0$ decays predominantly to a Higgs from $\tilde{\chi}_2^0\to \tilde{\chi}_1^0 h^0$ ($\sim70\%$), if $m_{\tilde{\tau}_1}>m_{\tilde{\chi}_2^0}$. This is the case for points B, B' and C'. If $m_{\tilde{\tau}_1}<m_{\tilde{\chi}_2^0}$ (point C), the leading $\tilde{\chi}_2^0$ decays give a $\tau^+\tau^-$ pair from $\tilde{\chi}_2^0\to \tilde\tau_1^\pm\tau^\mp$, $\tilde\tau_1^\pm\to \tau^\pm\tilde{\chi}_1^0$ 
 ($\sim 85\%$), while $\tilde{\chi}_2^0\to \tilde{\chi}_1^0 h^0$ has $\sim 10\%$ BR. With the $\tilde{\chi}_1^0$ decay, the whole chain produces an OS ditau or a $h^0$, plus the LMP, plus $e$ or $e+\mu^\pm+\mu^\mp$, depending on the $\tilde{\mu}_R/\tilde{\chi}_1^0$ mass ordering and splitting.
The final $e$ or $e+\mu^\pm+\mu^\mp$ should be rather soft compared to the taus.

\item
Analogously, the $\tilde{\chi}_1^\pm$ decays mainly either through $\tilde{\chi}_1^\pm\to \tilde{\chi}_1^0 W^\pm$ ($\sim 65$--$70\%$),
or through $\tilde{\chi}_1^\pm\to \tilde{\tau}_1^\pm\nu_\tau \to \tilde{\chi}_1^0 \tau^\pm \nu_\tau$ ($\sim 85\%$) if kinematically allowed. The latter chain 
gives one hard $\tau$ plus $E_T^{\rm miss}$, plus  the LMP, plus $e$ or $e+\mu^\pm+\mu^\mp$ depending on the $\tilde{\mu}_R/\tilde{\chi}_1^0$ mass ordering and  splitting.
The subleading decay $\tilde{\chi}_1^\pm\rightarrow \tilde{\mu}_R^\pm \nu_\mu$ can have 10--20\% branching ratio and  gives one hard $\mu$ plus $E_T^{\rm miss}$, plus $e$ and  the LMP if  $m_{\tilde{\mu}_R}>m_{\tilde{\chi}_1^0}$, or
one hard $\mu$ plus $E_T^{\rm miss}$, plus the LMP and  $e+\mu$ from the $\tilde{\mu}_R$ three-body decay if  $m_{\tilde{\mu}_R}<m_{\tilde{\chi}_1^0}$.

\item
The invariant-mass distribution of the SFOS ditau can be used to determine the mass of the $\tilde{\tau}_1$.
For measuring the $\tilde{\mu}_R$ mass, one needs to rely on the analysis of chargino decays, or on LFV processes, in which $\tilde{\mu}_R$ appears as an intermediate decay product (see below). 
Knowledge of the masses, or mass splittings, of the $\tilde{\tau}_1$,  $\tilde{\mu}_R$ and $\tilde{e}_R$ 
would now permit to obtain information on $\msq{E}$.
As before, this information combined with other measurements would permit to
carry out a hypothesis test, this time on the structure of $\msq{E}$. In particular, one can check wether the  hierarchical factors $\epsilon$ can be in agreement with values $\mathcal{O}(10^{-1})$ necessary to reproduce the SM flavour hierarchy.

\end{itemize}

%
%

Let us now turn to lepton flavour violation. As mentioned, LFV can be sizeable in the $\mu-\tau$ sector. 
The most interesting decays are those involving $\tilde l_3\simeq \tilde{\tau}_1$ and 
$\tilde l_2\simeq \tilde{\mu}_R$. Figure~\ref{fig:BRs} shows, in the top row, the PDFs for
BR$( \tilde{\tau}_1\rightarrow\tilde{\chi}_1^0\mu)$ and BR$(\tilde{\chi}_2^0 \rightarrow \tilde{\mu}_R \tau)$, 
and for comparison also for BR$(\tilde{\chi}_2^0 \rightarrow \tilde{\mu}_R \mu)$.
The bottom row shows the correlation of these branching ratios with 
$\textrm{BR}(\tau\rightarrow\mu\gamma)$. 
At the chosen scale, $F^T/2R=1500 ~\textrm{GeV}$, the current experimental bound of 
$\textrm{BR}(\tau\rightarrow\mu\gamma)<6.8\times10^{-8}$ is always satisfied.
We observe that $\textrm{BR}(\tau\rightarrow \mu \gamma)$ scales roughly as $ \left(F^T/2R\right)^{(-5)\,\div\, (-4)}$. 
Given the strong correlation, if LFV is observed in slepton or neutralino decays, this leads to a prediction 
for $\tau\rightarrow\mu\gamma$. Our main observations regarding LFV processes at the LHC are:

\begin{figure}
\hspace{-2mm}\begin{tabular}{rrr} 
\includegraphics[width=4.1cm]{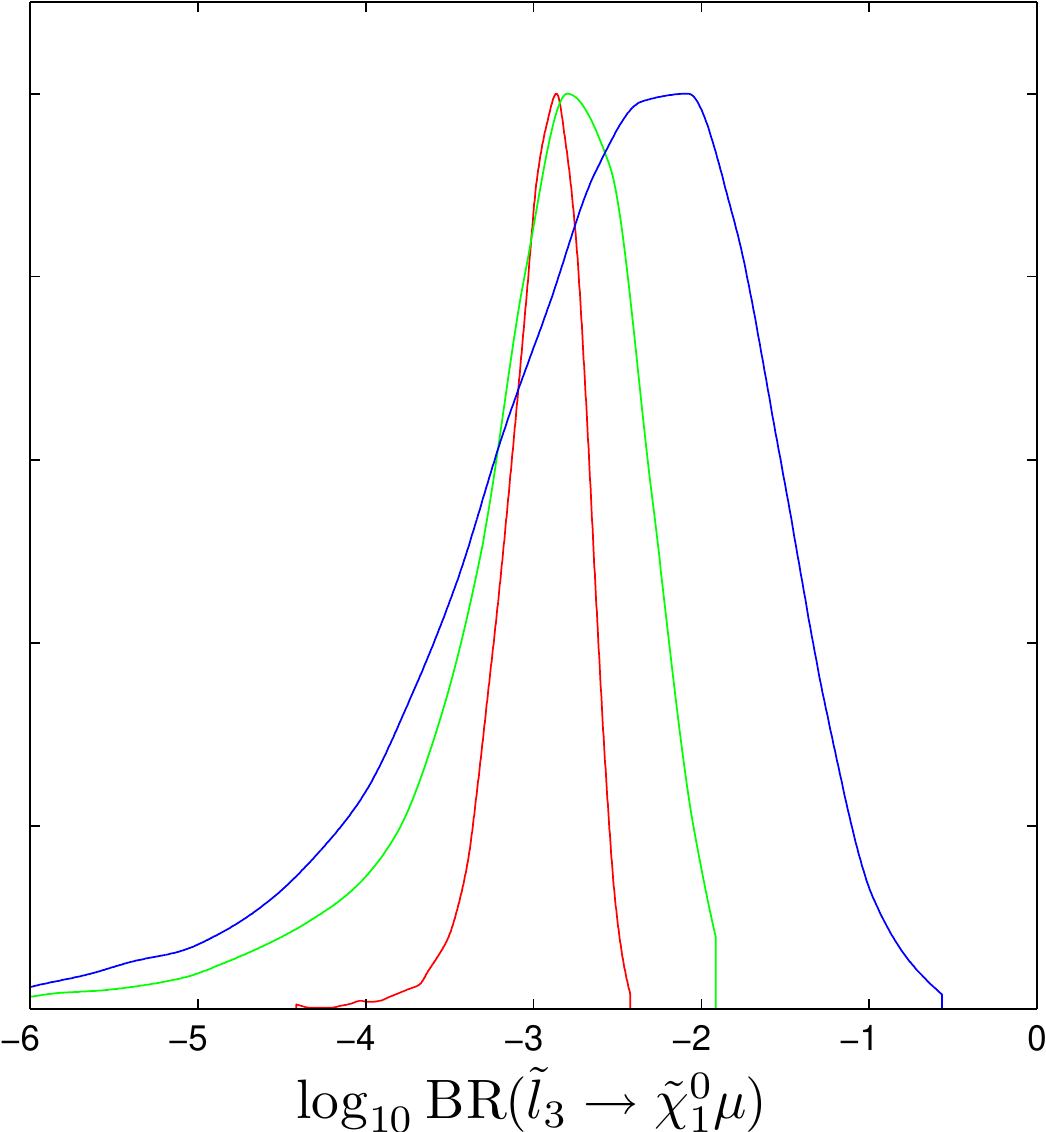}&
\includegraphics[width=4.1cm]{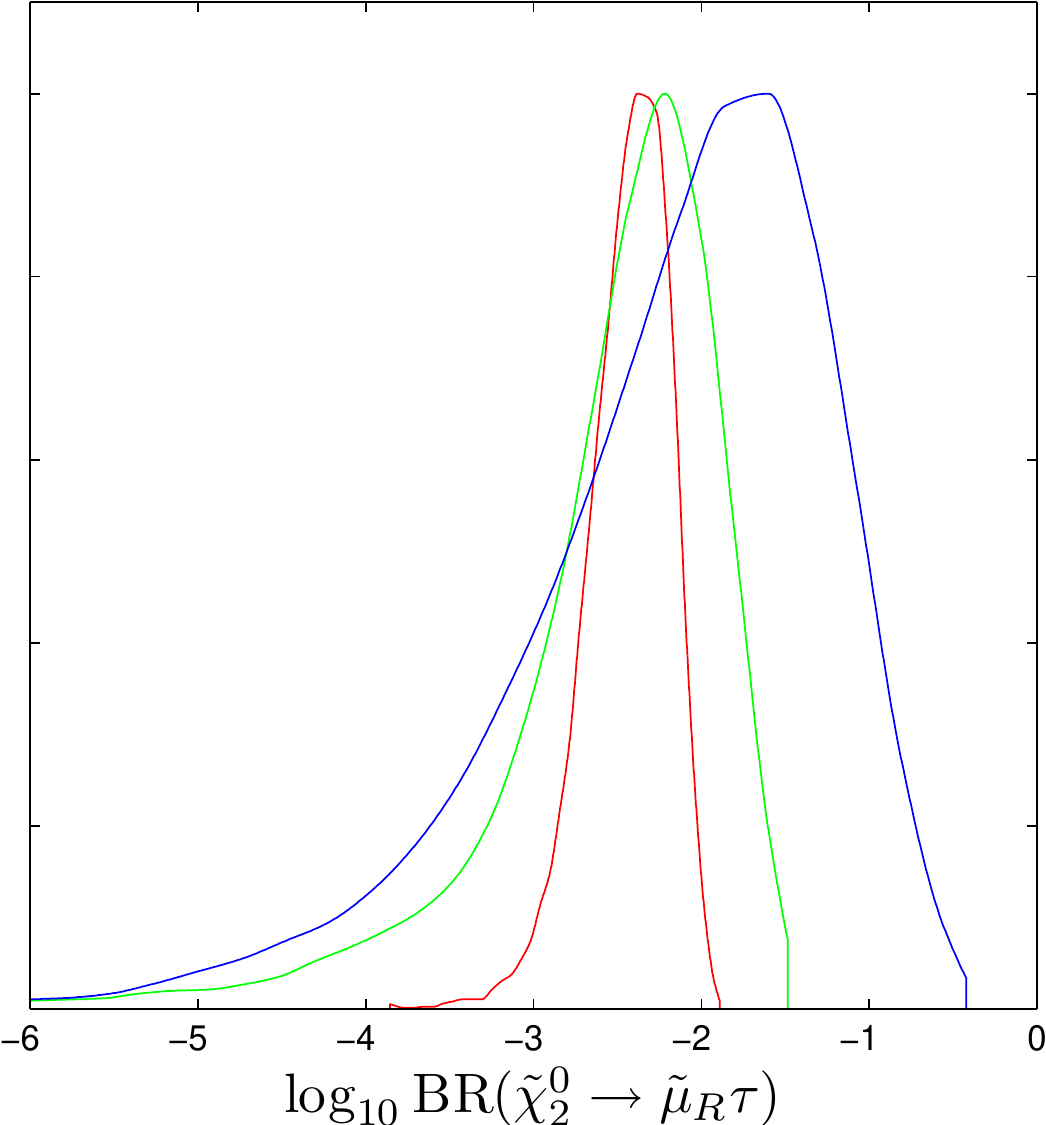}&
\includegraphics[width=4.1cm]{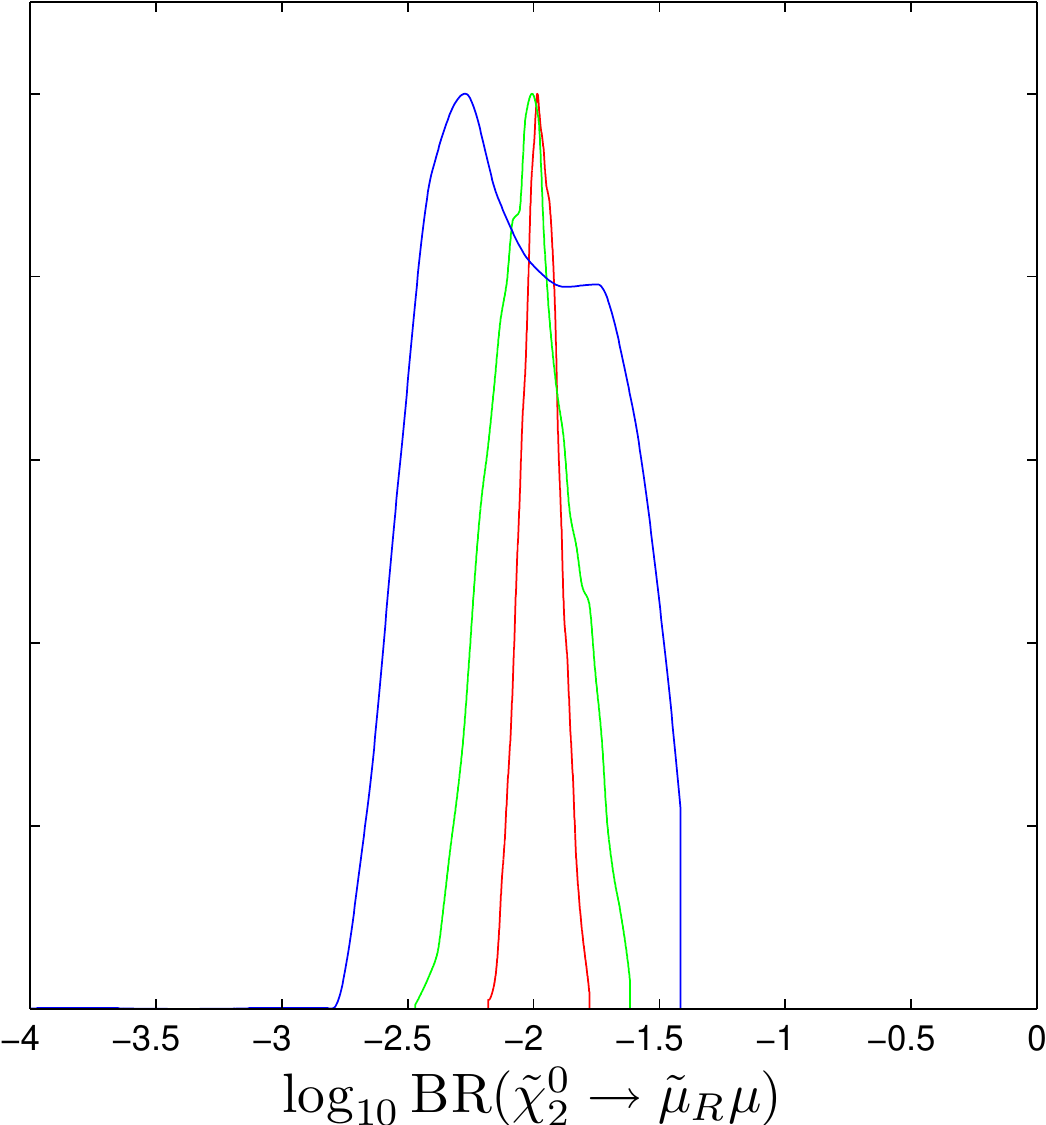}\\[2mm]
\includegraphics[width=4.6cm]{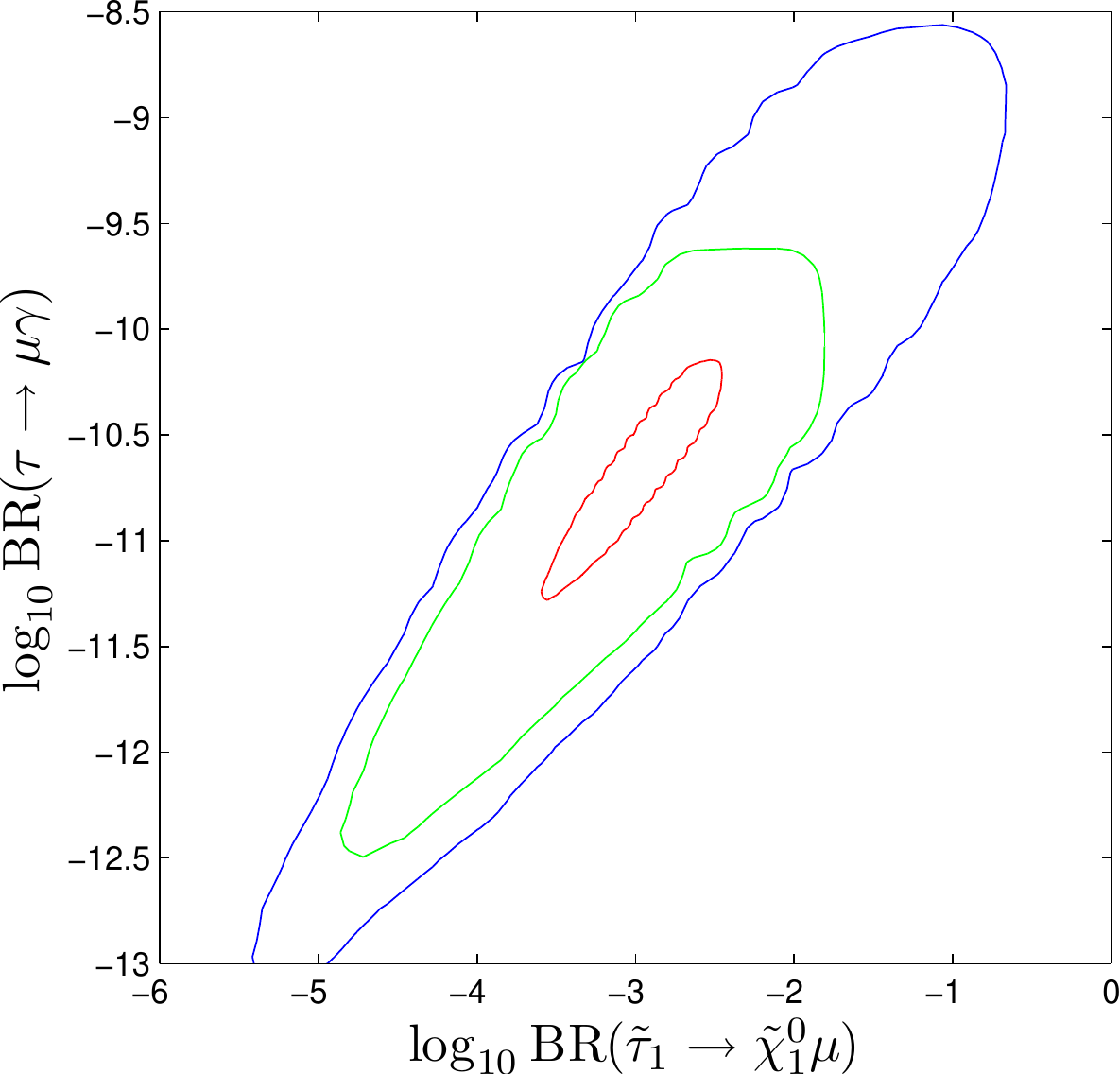}&
\includegraphics[width=4.6cm]{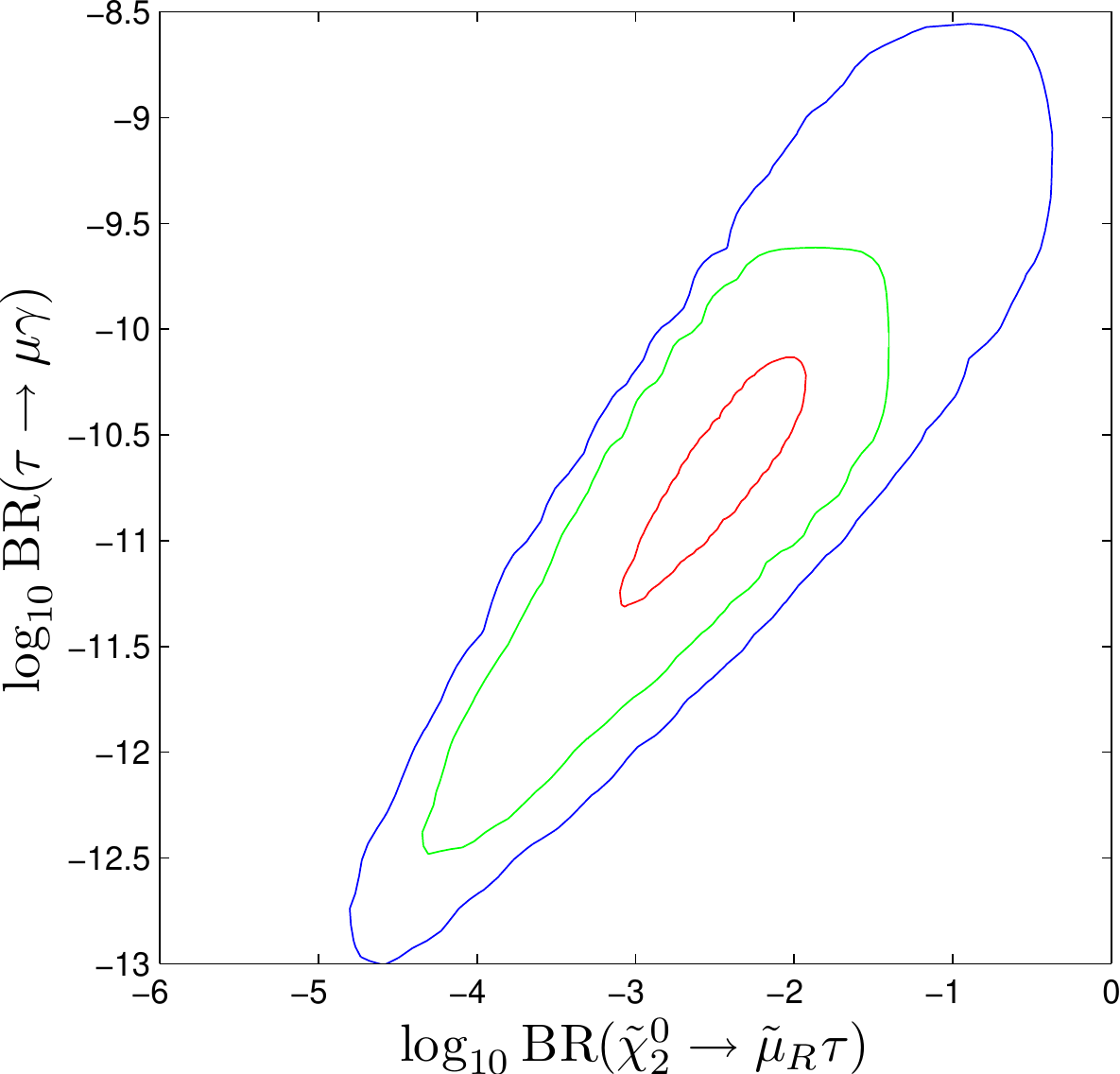}&
\includegraphics[width=4.6cm]{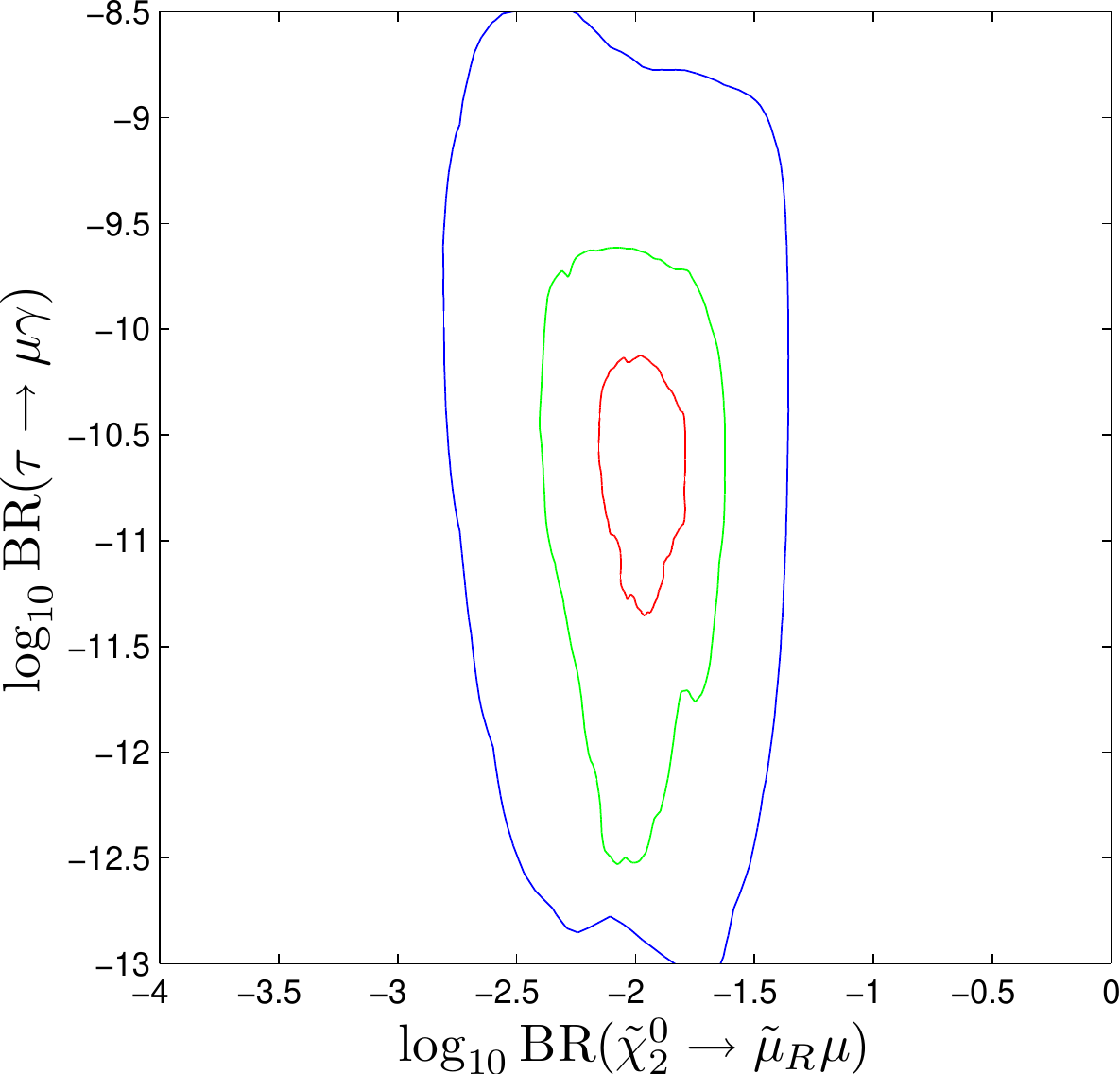}
\end{tabular}
\caption{PDFs of branching ratios relevant for LFV  in the $\mu-\tau$ sector. 
The parameters are $F^T/2R=1500~\textrm{GeV}$ and $\tan\beta=30$. The upper row shows 1D PDFs 
for BR$( \tilde{\tau}_1\rightarrow\tilde{\chi}_1^0\mu)$, BR$(\tilde{\chi}_2^0 \rightarrow \tilde{\mu}_R \tau)$ 
and BR$(\tilde{\chi}_2^0 \rightarrow \tilde{\mu}_R \mu)$. The lower row shows 95\% BC contours of the joint 2D 
PDFs of the these branching ratios with $\textrm{BR}(\tau\rightarrow\mu\gamma)$. In all plots, 
the red, green and blue lines correspond to $\mathscr{L}=1.2$, $1.5$ and $3$, respectively.
\label{fig:BRs}}
\end{figure}

\begin{figure}
\centering
\includegraphics[width=5cm]{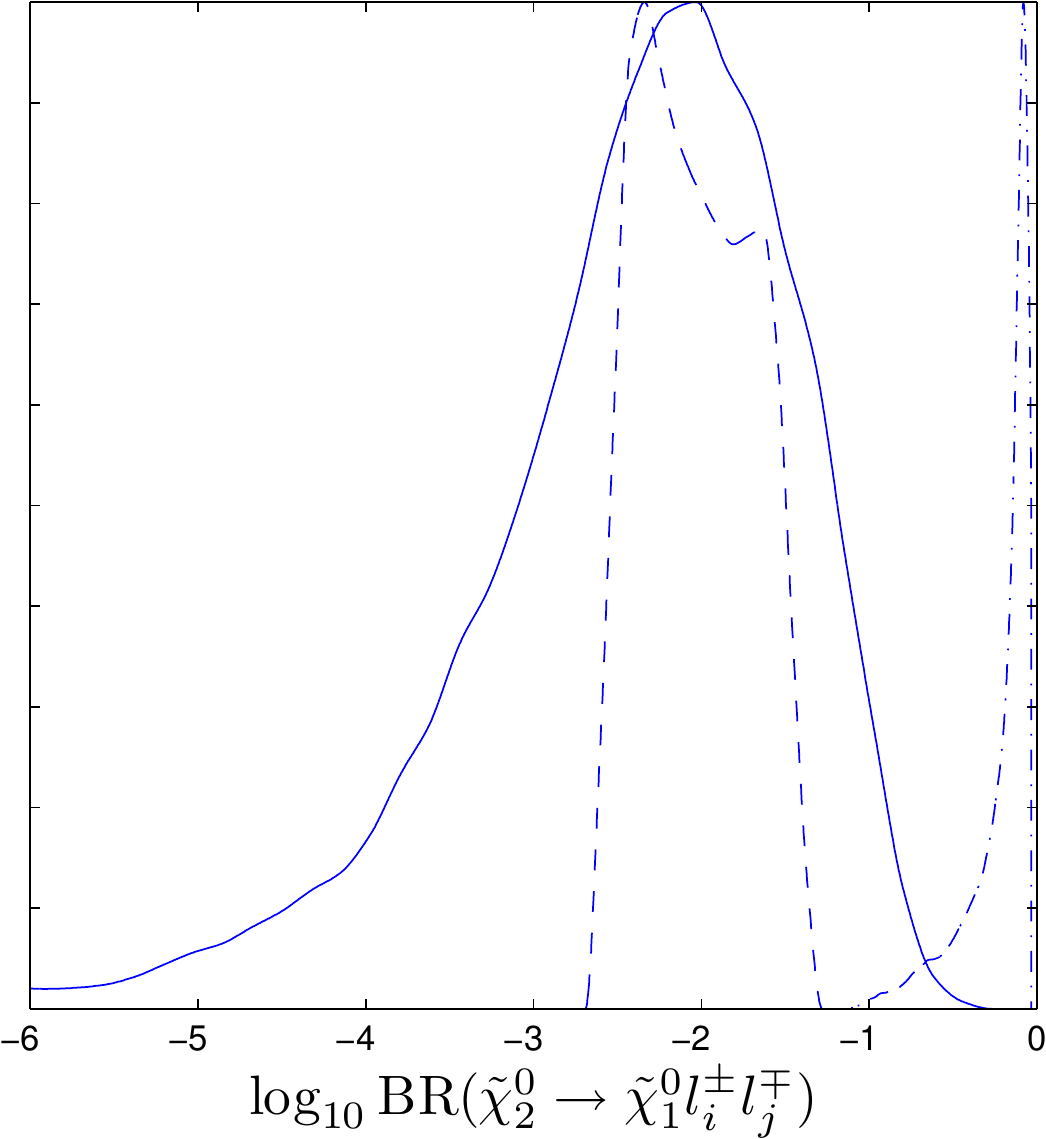}
\caption{PDFs for $\tilde{\chi}_2^0$ cascade decays at $F^T/2R=1500~\textrm{GeV}$, 
$\tan\beta=30$ and $\mathscr{L}=3$; 
dash-dotted line:  $\textrm{BR}(\tilde{\chi}_2^0\rightarrow \tilde{\chi}_1^0 \tau^\pm \tau^\mp)$, 
dashed line: $\textrm{BR}(\tilde{\chi}_2^0\rightarrow \tilde{\chi}_1^0 \mu^\pm \mu^\mp)$, 
solid line: $\textrm{BR}(\tilde{\chi}_2^0\rightarrow \tilde{\chi}_1^0 \tau^\pm \mu^\mp)$.
\label{fig:OS_pdf}}
\end{figure}

\begin{itemize}

\item
While the $\tilde{\tau}_1$ decays mainly to $\tilde{\chi}_1^0\tau$, it can also have a LFV decay $\tilde{\tau}_1\rightarrow\tilde{\chi}_1^0\mu$. The rate of the LFV decay peaks around 1\% but can go up to  $10\%$ or more for $\mathscr{L}=3$, 
see the top-left plot in Fig.~\ref{fig:BRs}.
In the $\tilde{\chi}_2^0$ decay chain, the ditau is then replaced by a $\mu^\pm\tau^\mp$ pair, 
potentially giving rise to an interesting flavour structure in kinematic distributions. 
Kinematic edges with flavour splitting and mixing have very recently been studied in \cite{Galon:2011wh} (see also \cite{Bartl:2005yy}). 
In the $\tilde{\chi}_1^\pm$ decay chain, the single $\tau$ is replaced by a single $\mu$, which must be separated from the non-LFV $\tilde{\chi}_1^\pm\rightarrow \tilde{\mu}_R \nu_\mu$ by kinematics. 
For example, one may exploit subsystem transverse-mass distributions~\cite{Barr:2011xt}
of, e.g., a lepton associated to the `upstream'  jet originating from $\tilde{q}_L$ cascade decays. This may permit to disentangle decay chains involving different slepton mass-eigenstates. 

\item
LFV can also occur directly in the $\tilde{\chi}_2^0$ decays. Indeed the decay $\tilde{\chi}_2^0 \rightarrow \tilde{\mu}_R \tau$ can have a branching ratio of up to ${\cal O}(10\%)$ for $\mathscr{L}=3$, larger than the $\tilde{\chi}_2^0 \rightarrow \tilde{\mu}_R \mu$ rate, cf.\ the middle and top-right plots in Fig.~\ref{fig:BRs}. 
The decay chain then is $\tilde{\chi}_2^0 \to \tilde{\mu}_R \tau \, (\to \tilde{\chi}_1^0\mu\tau) \to \tilde e_Re\mu\tau$ 
with the $\tilde{\chi}_1^0$ being on- or off-shell depending on the $\tilde{\mu}_R/\tilde{\chi}_1^0$ mass ordering.

The PDFs of the branching ratios of $\tilde{\chi}_2^0\rightarrow\tilde{\chi}_1^0\tau^\pm\tau^\mp $,
$\tilde{\chi}_2^0\rightarrow\tilde{\chi}_1^0\tau^\pm\mu^\mp $ and
$\tilde{\chi}_2^0\rightarrow\tilde{\chi}_1^0\mu^\pm\mu^\mp $ 
are shown in Fig.~\ref{fig:OS_pdf}, to illustrate the global rate of LFV expected in $\tilde{\chi}_2^0$ cascade decays.

\end{itemize}

\subsection{Mixed brane-radion scenario}\label{brane_sources}

In the previous subsection we investigated the case $F^Z/M_*\ll F^T/2R$, such that the scalar soft terms were dominated by radion mediation.
We now consider the case where scalar soft terms receive non-negligible contributions from the brane source. 
 The brane source contributions introduce a large uncertainty directly in the GUT-scale scalar soft terms. In addition to the $\lam{U,D,E}_{ij}$, the $\lambda_{ij}^{m^2_X}$ also become relevant. The phenomenology will eventually depend on the signs and magnitudes of all these $\lambda$-parameters. Nevertheless we can still identify some generic features.

Since the case of sizeable $F^Z/M_*$ is much more constrained by $\textrm{BR}(\mu\rightarrow e \gamma)$ at large $\tan\beta$ (see Figs.~\ref{fig:sign_dist}, \ref{fig:isolines1} and \ref{fig:isolines2}), we will focus on the small $\tan\beta$ scenario. 
In this case, the left-right mixing is negligible, because the $a$-terms are not large. It turns out that the right-right mixing is  hierarchical, as in the radion-dominated case. We therefore call the right-handed sleptons  $\tilde{e}_R$, $\tilde{\mu}_R$, and $\tilde{\tau}_R$. On the other hand, the left-left mixing can now be very large, so we denote the left-handed sleptons by $\tilde{l}_{L\,1}$, $\tilde{l}_{L\,2}$, and $\tilde{l}_{L\,3}$.

\begin{figure}
\centering
\includegraphics[width=5cm]{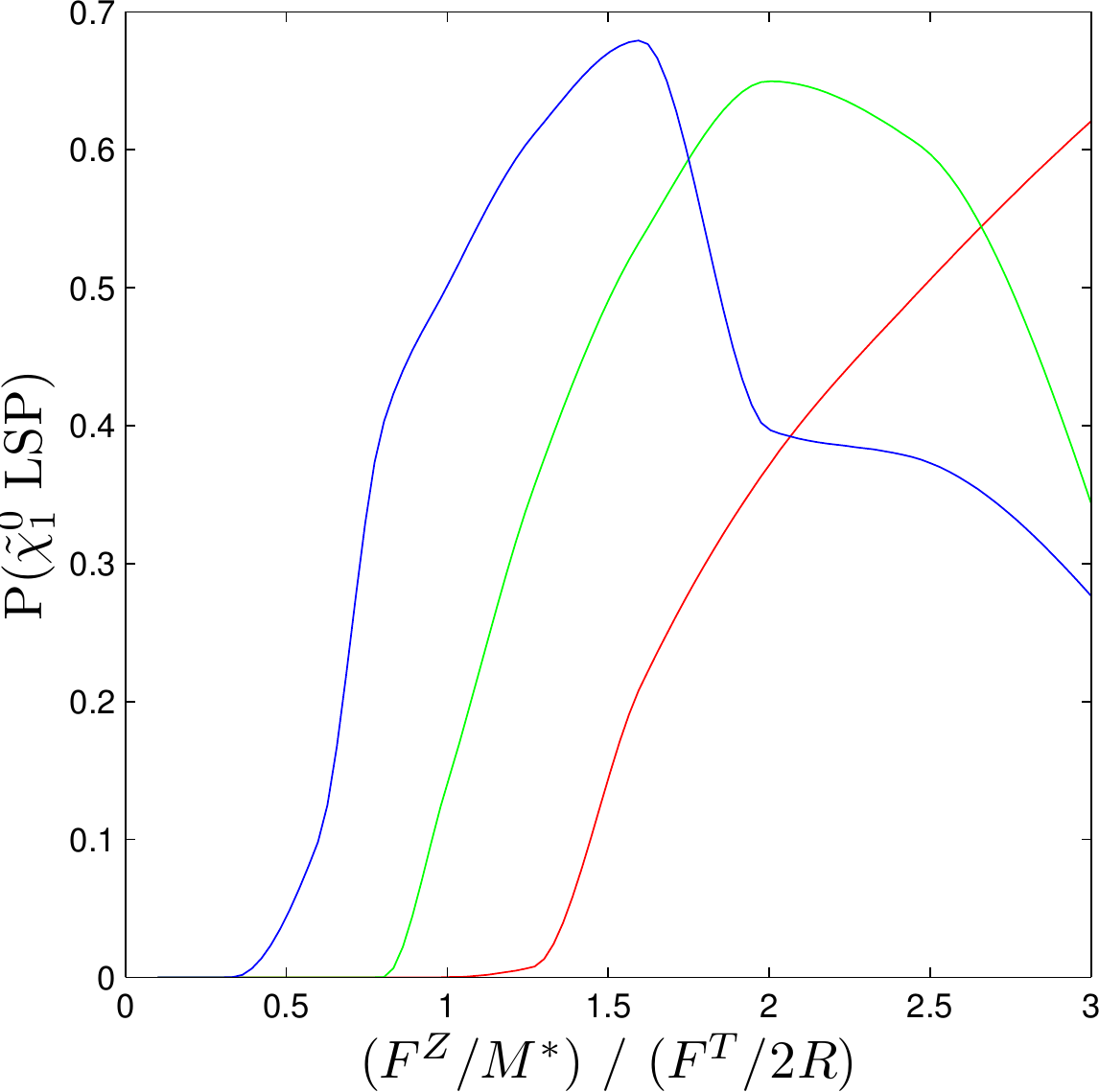}
\caption{ Probability of having a $\tilde{\chi}_1^0$ LSP as a function of $(F^Z/M_*)/ (F^T/2R)$. The red, green, blue lines correspond to $\mathscr{L}=1.2,\,1.5,\,3,$ respectively. The other parameters are $\tan\beta=5$, $F^T/2R=1.5~\textrm{TeV}$.
  \label{fig:chi0_lSP.pdf}}
\end{figure}

The masses of the right-handed sleptons can now span a wide range. All possible mass orderings with respect to the lightest neutralino can appear: $m_{\tilde{e}_R}<m_{\tilde{\chi}_1^0}<m_{\tilde{\mu}_R}$, 
$m_{\tilde{e}_R,\tilde{\mu}_R}<m_{\tilde{\chi}_1^0}$, or $m_{\tilde{\chi}_1^0}<m_{\tilde{e}_R,\tilde{\mu}_R}$.
This last possibility is particularly interesting since, unlike the radion-dominated case,  it features a $\tilde{\chi}_1^0$ LSP which is a viable dark matter candidate.
Matrix anarchy again plays a crucial role in realizing this possibility: With $\mathscr{L}=1$, the RG invariant
\be\label{Sinv}
S\equiv m_{H_u}^2-m_{H_d}^2 + \textrm{Tr}(\msq{Q}-\msq{L}-2\,\msq{U}+\msq{D}+\msq{E})
\ee
 would vanish due to exact $\textrm{SU}(5)$ relations, and the LMP would then be the $\tilde{e}_R$ as in the radion-dominated case. However, with $\mathscr{L}>1$, the $\lambda_{ij}^{m_X^2}$ can be different from one another and induce a non-zero $S$. If $S$ is sufficiently large and negative, the lightest slepton mass can be lifted above the neutralino mass. The probability of finding  a neutralino LSP thus depends  on $\mathscr{L}$, as well as on the ratio $(F^Z/M_*)/ (F^T/2R)$. 
This is illustrated in Figure \ref{fig:chi0_lSP.pdf}, which shows the probability of finding a $\tilde{\chi}_1^0$ LSP as a function of $\mathscr{L}$ and $(F^Z/M_*)/ (F^T/2R)$ for a favourable sign combination of $\lambda_{ij}^{m^2_L}$. This plot is for $\tan\beta=5$ and $F^T/2R=1500~\textrm{GeV}$, but the result is fairly insensitive to the SUSY scale. 

We proceed to discuss the left-handed slepton masses. At small $\tan\beta$, the soft masses $\msq{L}$ are suppressed with respect to the gaugino masses (as explained in subsection \ref{radion_domination}).
The RG running of the masses of the left-handed sleptons is therefore dominated by the gaugino masses and the $S$ parameter. 
For $(F^Z/M_*)/ (F^T/2R)\lesssim1.5$, we obtain the mass ordering 
\be
m_{\tilde{\chi}_1^0}<m_{\tilde{l}_{L}}<m_{\tilde{\chi}_2^0}~.
\label{eq:BS_lepton_ordering}
\ee
This property is particularly interesting for LHC phenomenology.\footnote{
For $(F^Z/M_*)/ (F^T/2R)\gtrsim1.5$ with $\mathscr{L}=3$, the $S$ parameter of Eq.~\eqref{Sinv} can be sufficiently large to make the left-handed sleptons lighter than the right-handed ones. This is why, in Figure \ref{fig:chi0_lSP.pdf}, $P(\tilde{\chi}_1^0~LSP)$ decreases above a value of $(F^Z/M_*)/ (F^T/2R)\gtrsim1.5$. Due to $D$-term splitting between sneutrino and charged slepton masses, a  sneutrino becomes the lightest SUSY particle of the spectrum. Left-handed sneutrino dark matter is strongly constrained by direct detection and cosmology, so one would again have to assume that the actual LSP is the gravitino or an axino. We will not pursue this case any further.}

\begin{figure}
\centering
\includegraphics[width=4.5cm]{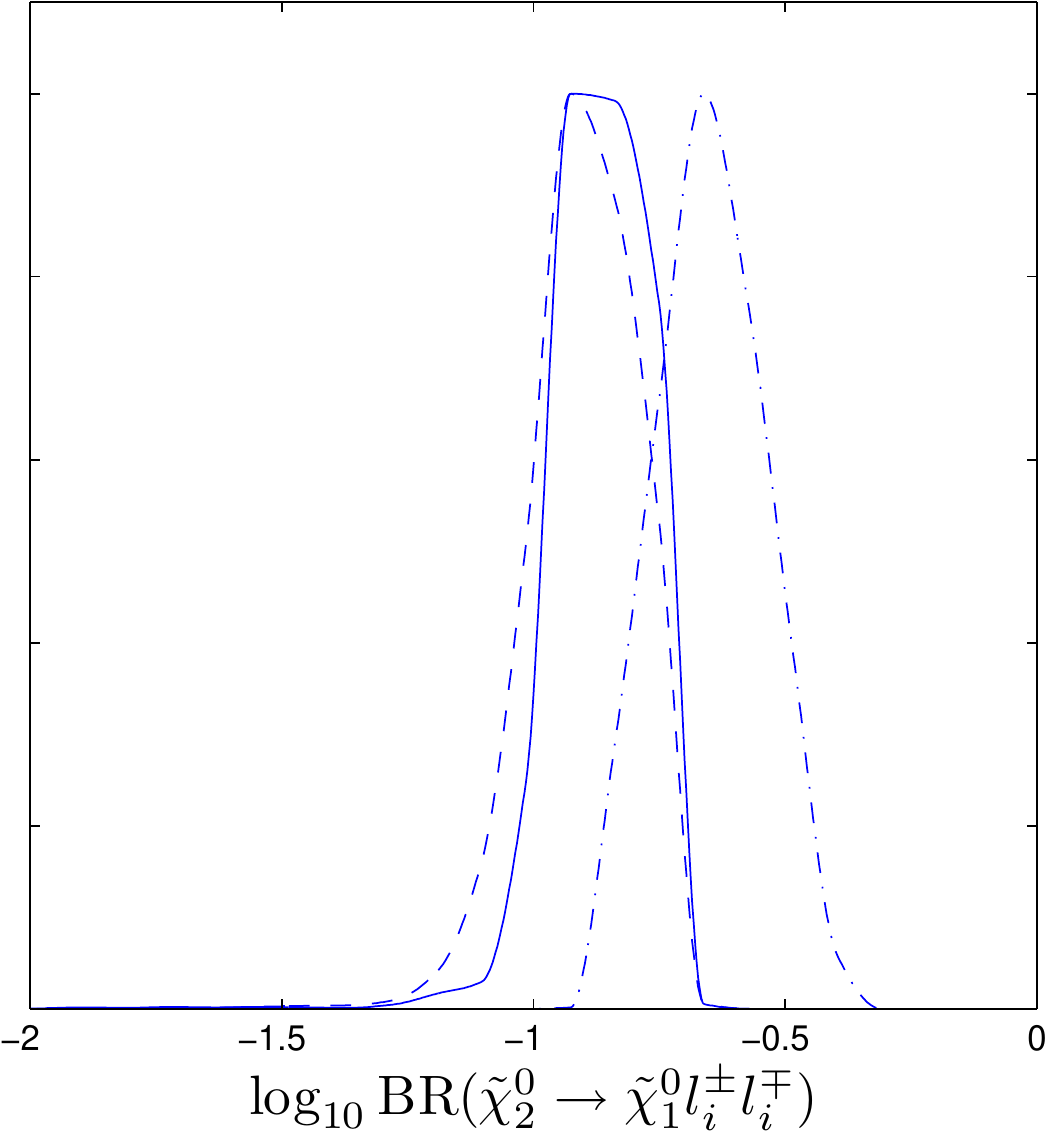}
\includegraphics[width=4.5cm]{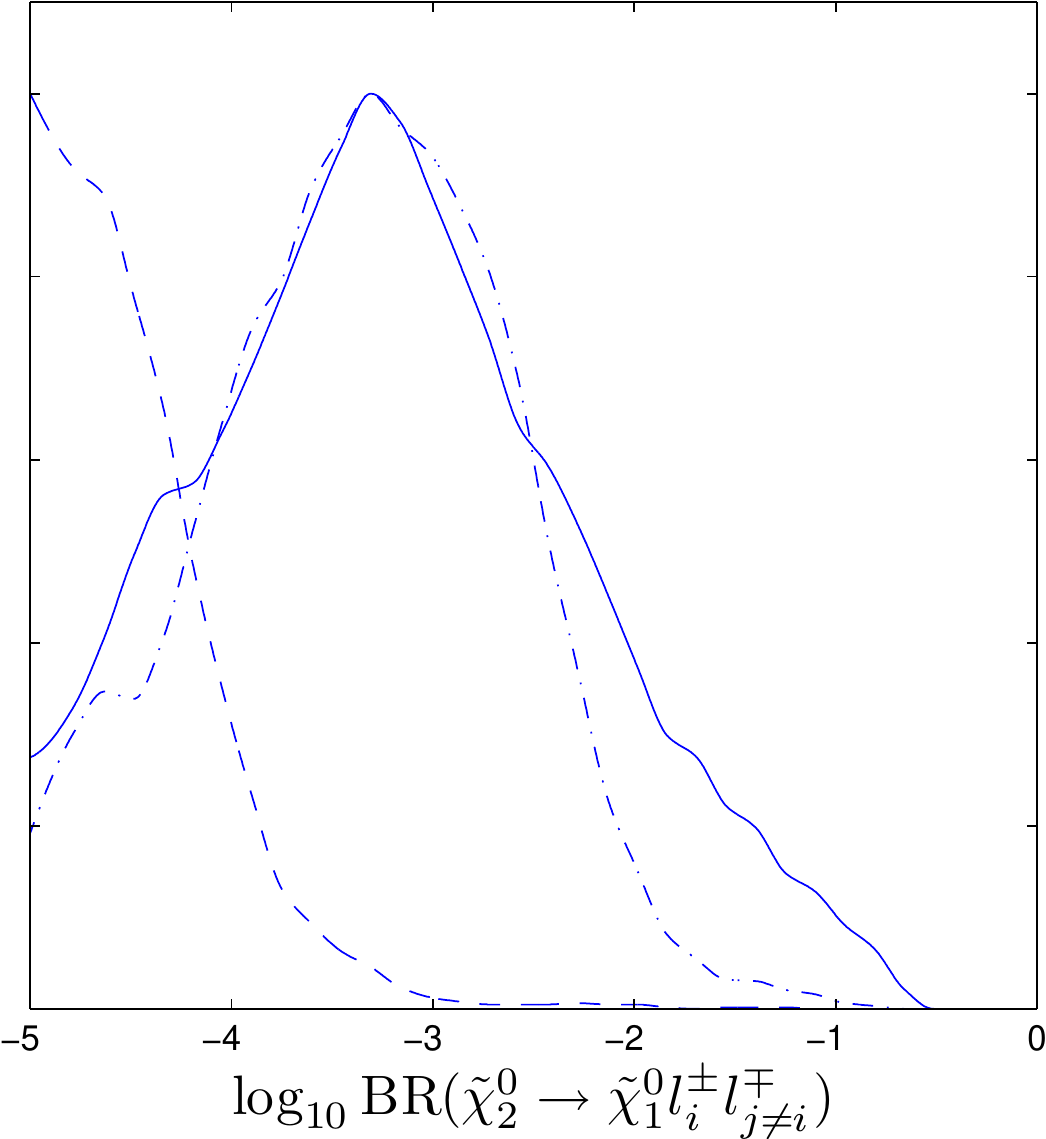}
\includegraphics[width=4.5cm]{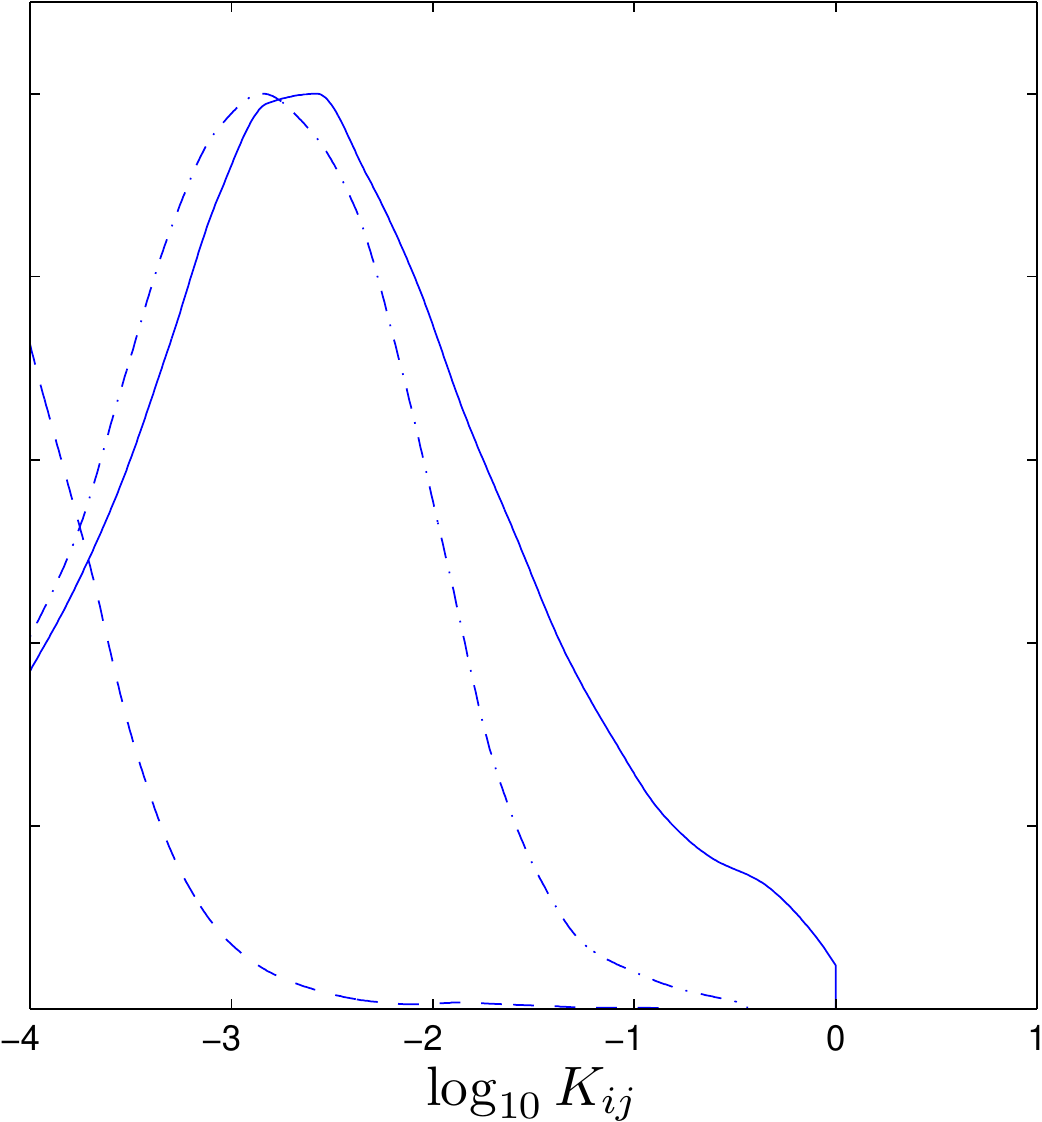}\\[3mm]
\centering
\includegraphics[width=5cm]{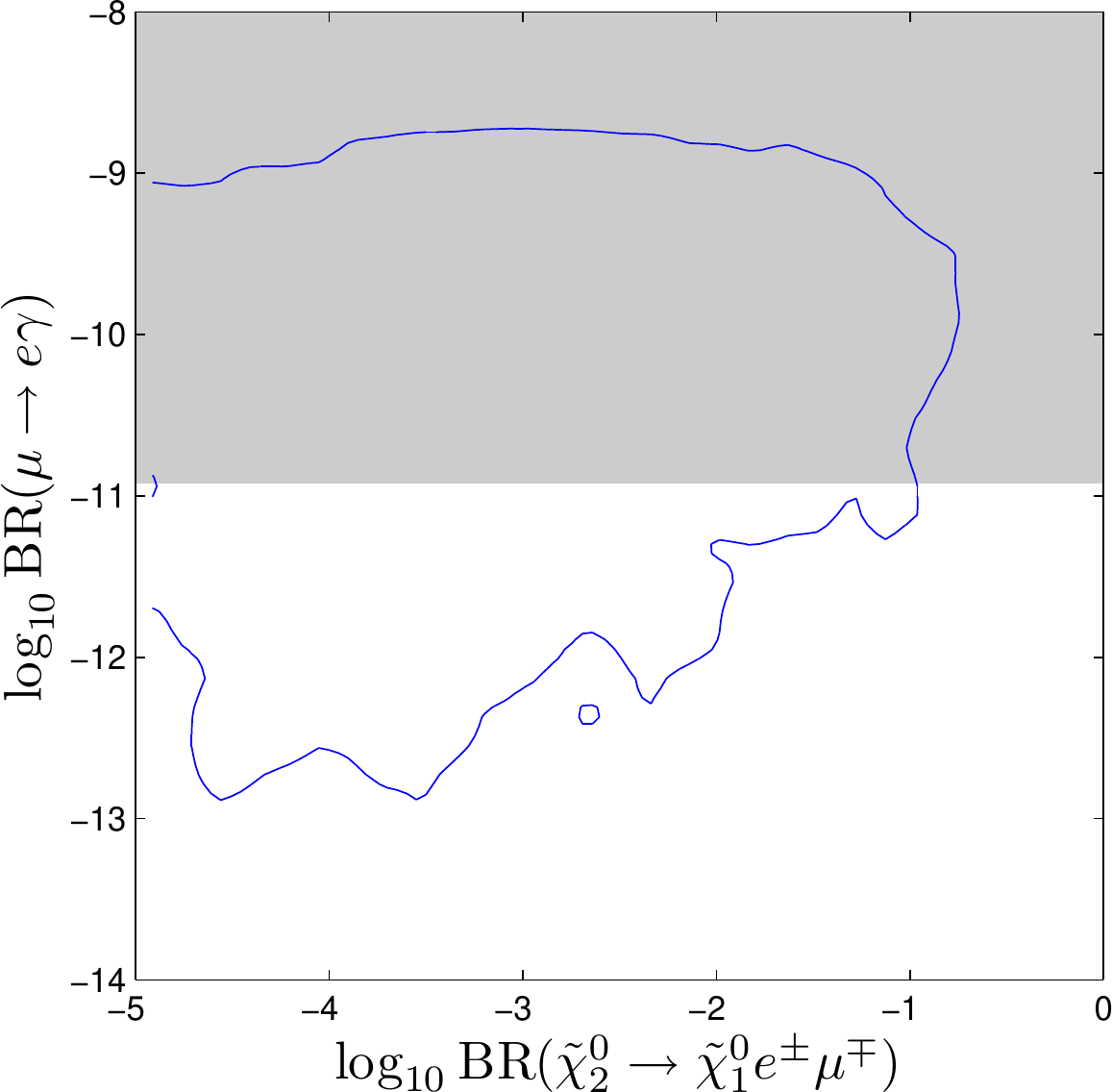}
\includegraphics[width=5cm]{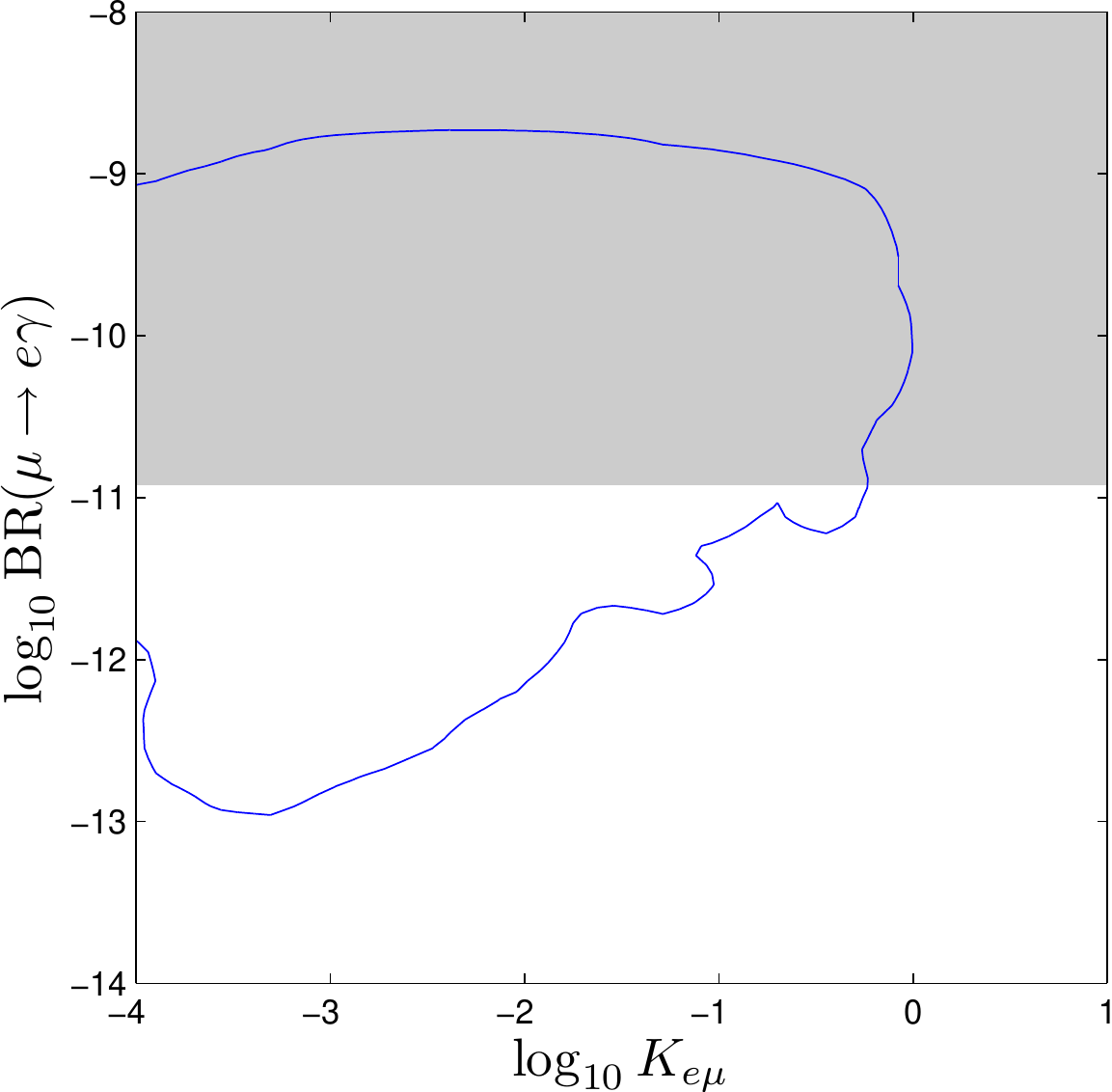}
\caption{
 \textit{Top left:} PDF of $\textrm{BR}( \tilde{\chi}_2^0 \rightarrow l_i l_i \tilde{\chi}_1^0 )$.  Plain, dashed and dash-dotted lines correspond to $l_il_i=ee$, $\mu\mu$, $\tau\tau$ respectively.
\textit{Top center:} PDF of $\textrm{BR}( \tilde{\chi}_2^0 \rightarrow l_i l_j \tilde{\chi}_1^0 )$. Plain, dashed and dash-dotted lines correspond to $l_il_j=e\mu$, $e\tau$, $\mu\tau$ respectively.
\textit{Top right:} Plain, dashed and dash-dotted lines correspond to $K_{e\mu}$, $K_{e\tau}$, $K_{\mu\tau}$ respectively.
\textit{Bottom:} $95$\% BC region of the joint PDF of $\textrm{BR}( \mu\rightarrow e\gamma)$ with $\textrm{BR}( \tilde{\chi}_2^0 \rightarrow e \mu \tilde{\chi}_1^0 )$ \textit{(left)} and $K_{e\mu}$ \textit{(right)}.\newline
All of these distributions are for a fixed sign combination at $F^T/2R=1500~\textrm{GeV}$, $F^Z/M_*=50~\textrm{GeV}$ and $\tan\beta=5$. 
The chosen sign combination represents the typical behaviour of SUSY BRs, while it is somewhat favourable regarding the $\textrm{BR}( \mu\rightarrow e\gamma)$ constraint.
  \label{fig:chi20_decays_BS50}}
\end{figure}

\begin{figure}
\centering
\includegraphics[width=4.5cm]{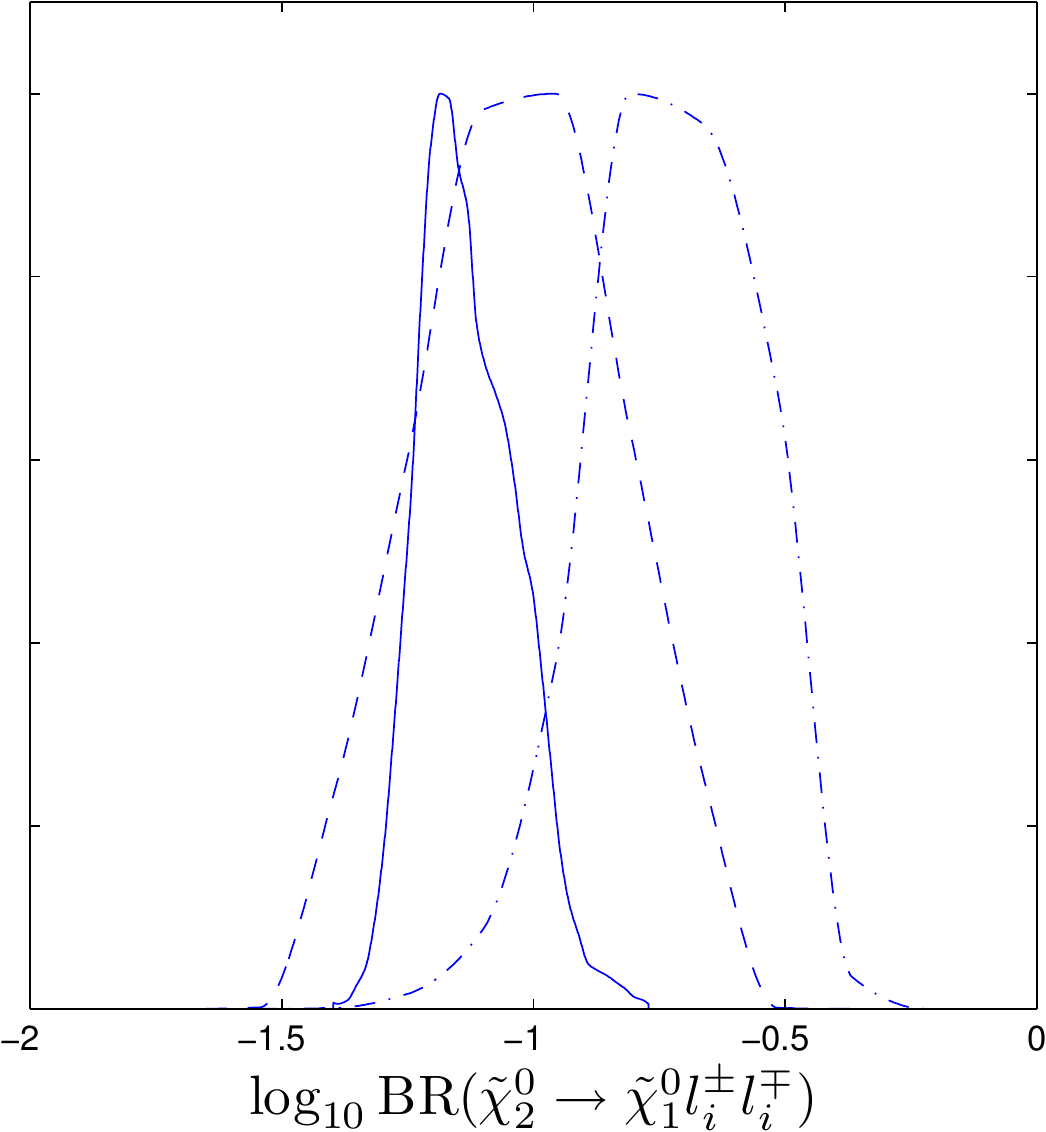}
\includegraphics[width=4.5cm]{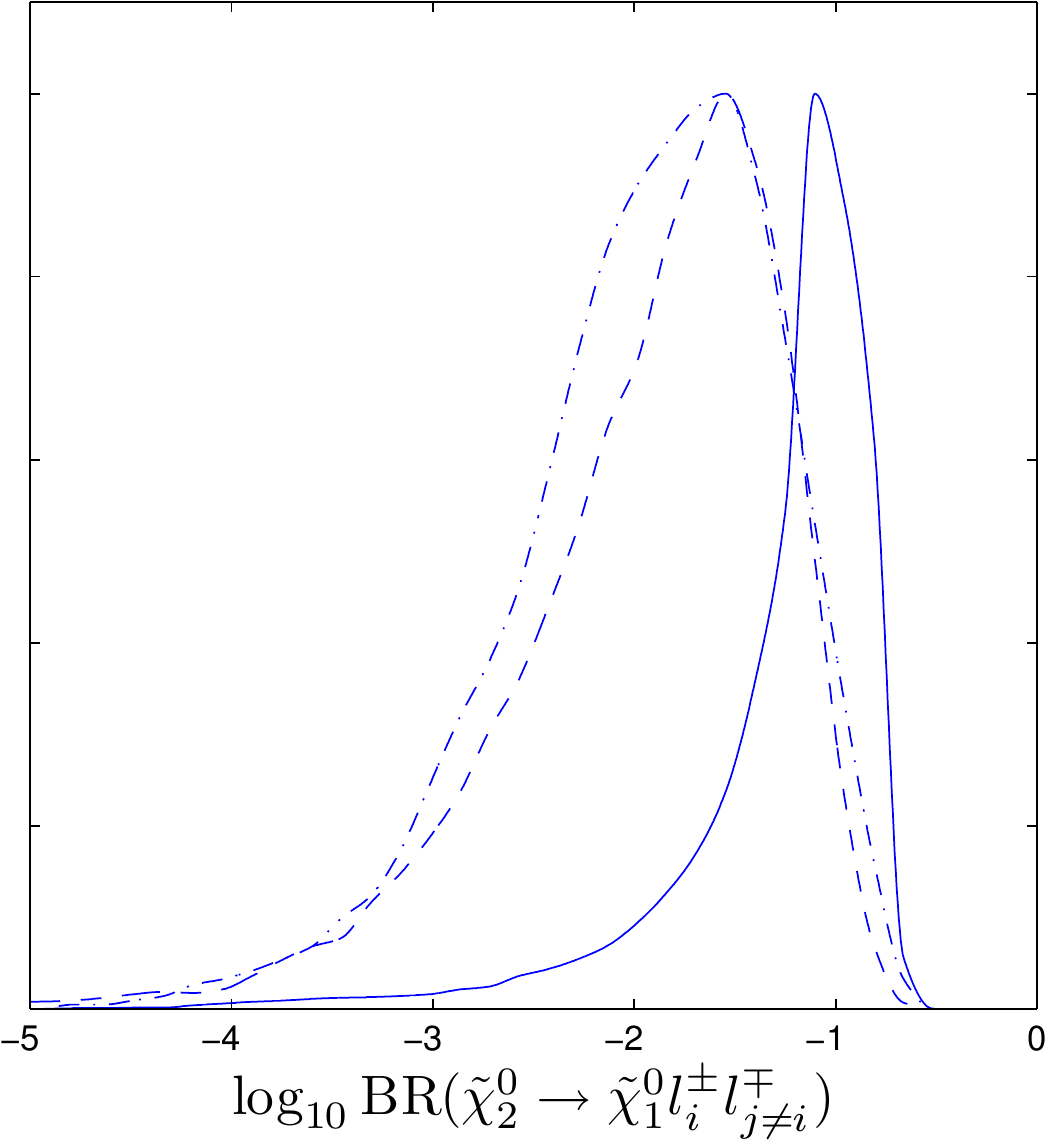}
\includegraphics[width=4.5cm]{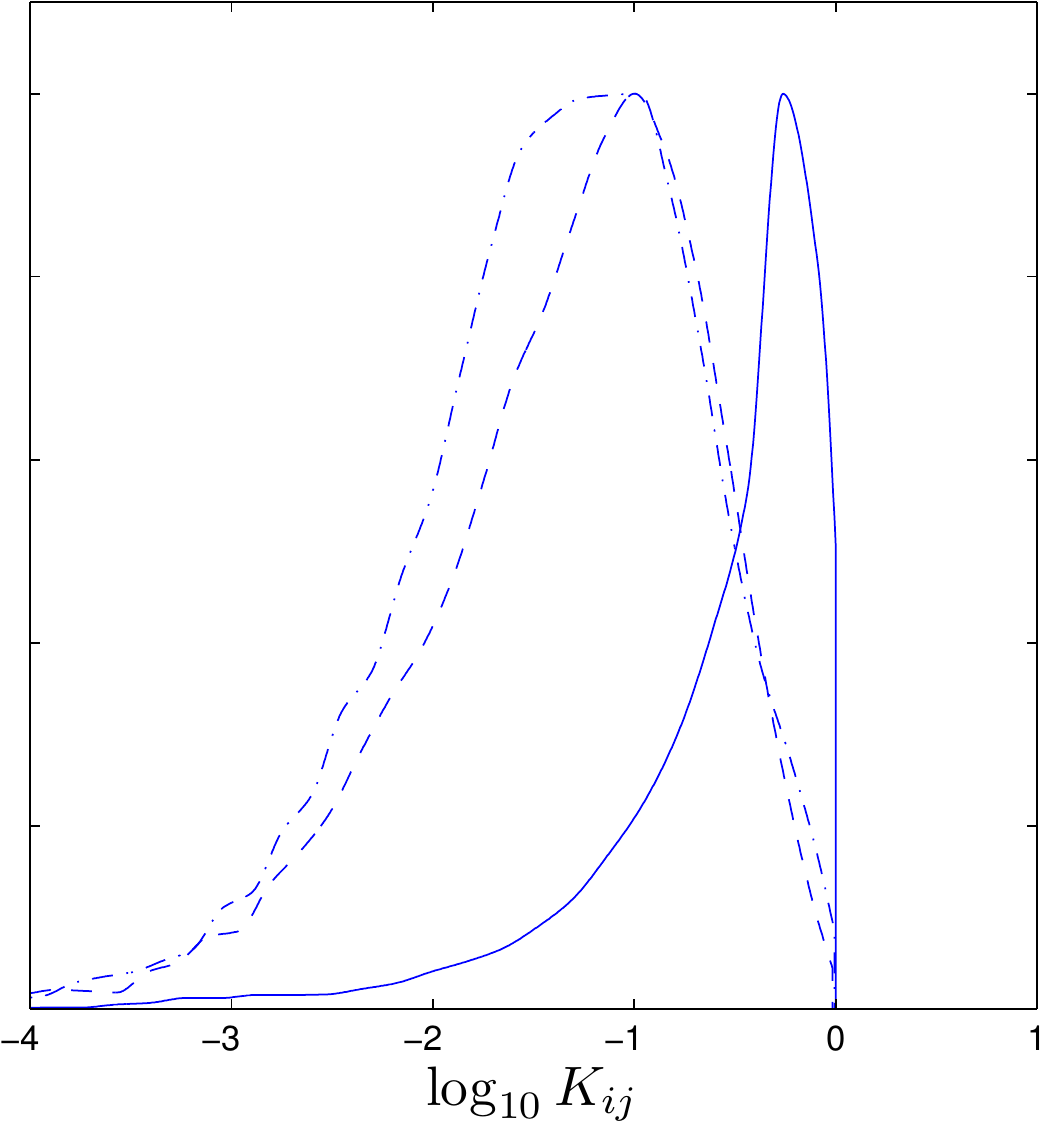}\\[3mm]
\centering
\includegraphics[width=5cm]{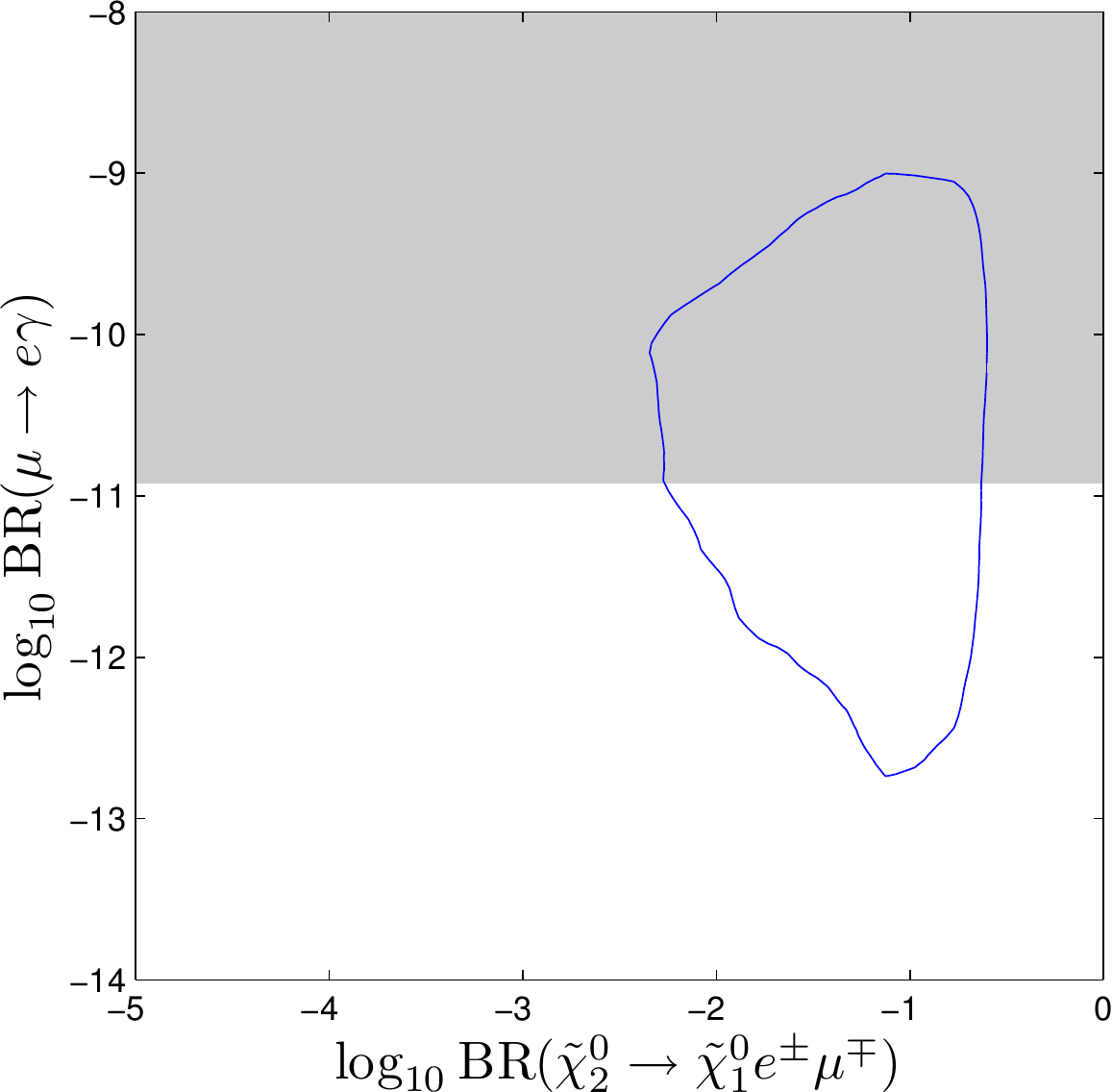}
\includegraphics[width=5cm]{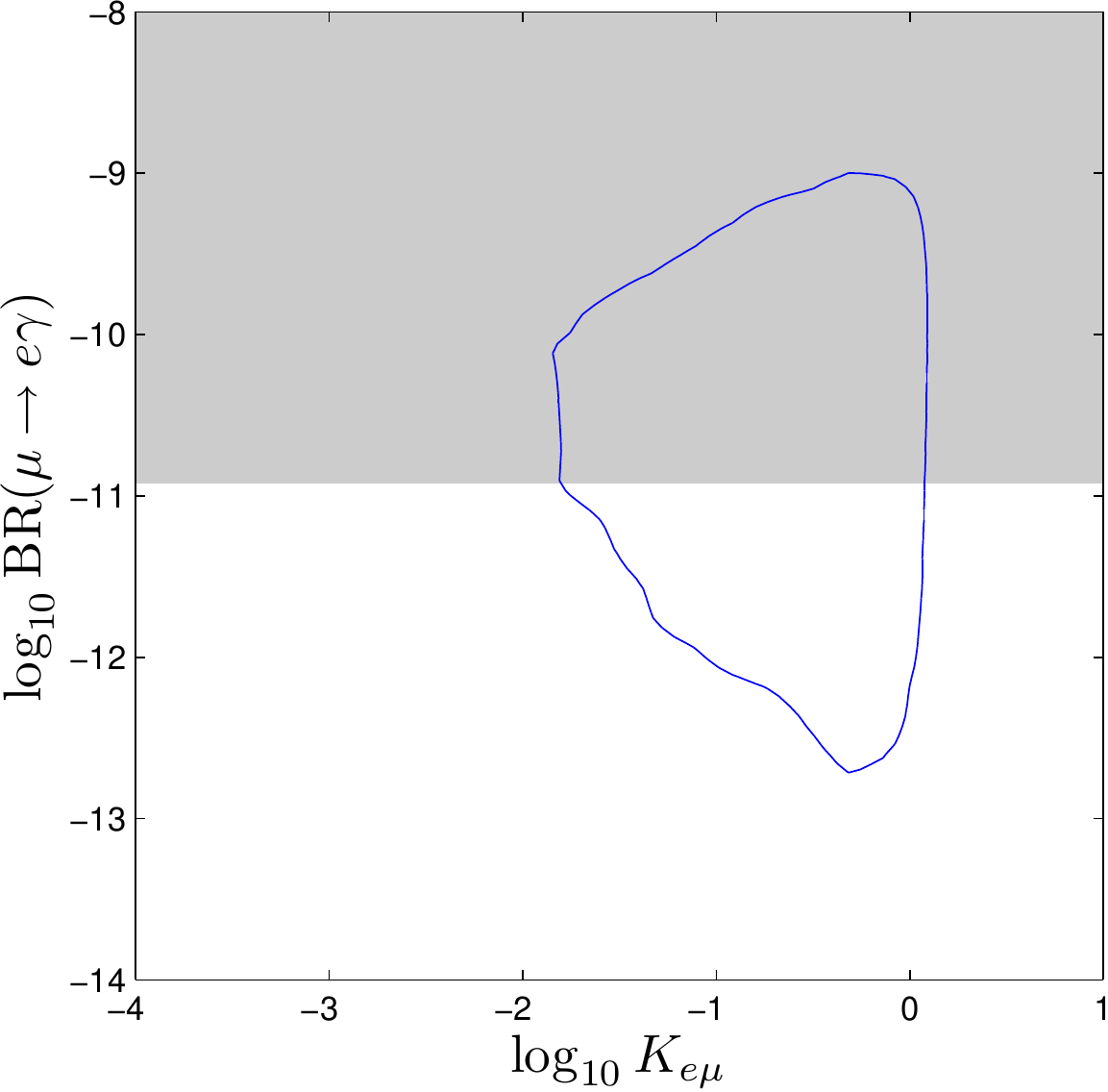}
\caption{ As Fig.~\ref{fig:chi20_decays_BS50}, but for $F^Z/M_*=1500~\textrm{GeV}$.
}\label{fig:chi20_decays_BS1500}
\end{figure}

For the remaining discussion, we fix $\tan\beta=5$, $F^T/2R=1500~\textrm{GeV}$ (i.e.~$m_{\tilde{g}}\sim3~\textrm{TeV}$). 
It turns out that, even for small $F^Z/M_*$, the LFV effects in the SUSY decays can be large, particularly in the $e-\mu$ sector, while still satisfying the current $\textrm{BR}(\mu\rightarrow e \gamma)$ bound. The details depend on the signs in both $\lam{E}_{ij}$ and $\lambda_{ij}^{m^2_L}$. We will focus on the $\tilde{\chi}_2^0\rightarrow \tilde{\chi}_1^0 l_i l_j$ decays. 
Following \cite{Andreev:2006sd}, we define the observable $K_{ij}$  as 
\be
K_{ij}=\frac{\textrm{BR} (\tilde{\chi}_2^0 \rightarrow l_i^\pm l_{j\neq i}^\mp \tilde{\chi}_1^0 ) }{\textrm{BR} (\tilde{\chi}_2^0 \rightarrow l_i^\pm l_i^\mp \tilde{\chi}_1^0)+ \textrm{BR}( \tilde{\chi}_2^0 \rightarrow l_j^\pm l_j^\mp \tilde{\chi}_1^0)}\,.
\label{eq:Kij}
\ee
to quantify the rate of LFV in the $\tilde{\chi}_2^0 \rightarrow l_i^\pm l_{j}^\mp \tilde{\chi}_1^0$ decays.
Figures \ref{fig:chi20_decays_BS50} and \ref{fig:chi20_decays_BS1500} show the PDFs of $\tilde{\chi}_2^0\rightarrow \tilde{\chi}_1^0 l_i l_j$ decay branching ratios, with $F^Z/M_*=50~\textrm{GeV}$ and $F^Z/M_*=1500~\textrm{GeV}$, respectively. For $F^Z/M_*=50~\textrm{GeV}$, it is in particular LFV in the $e-\mu$ sector that can be large enough to give observable effects at the LHC, while at $F^Z/M_*=1500~\textrm{GeV}$, LFV in all three sectors can be sizeable (though LFV still tends to be largest in the $e-\mu$ sector). The distributions shown in Fig.~\ref{fig:chi20_decays_BS50} and \ref{fig:chi20_decays_BS1500} represent the typical behaviour as far as BRs of SUSY particles are concerned, but the sign combination is chosen such that it slightly favours small $\textrm{BR}(\mu\rightarrow e \gamma)$. Note that, as opposed to Figure \ref{fig:BRs}, there is no strong correlation between $\textrm{BR}(\tilde{\chi}_2^0 \rightarrow e^\pm \mu^\mp \tilde{\chi}_1^0)$ and $\textrm{BR}(\mu\rightarrow e \gamma)$.

Let us now discuss possible LHC signatures. Collider signatures will again depend on the various possible mass orderings of the $\tilde{\chi}_1^0$ with respect to the $\tilde{e}_R$ and $\tilde{\mu}_R$. There are the following possibilities:
\begin{itemize}
\item
If $m_{\tilde{\chi}_1^0}<m_{\tilde{e}_R,\tilde{\mu}_R}$, then the $\tilde{\chi}_1^0$ is stable and a dark matter candidate (barring the $\widetilde G$ and $\tilde a$ LSP options). In the detector, the decay of $\tilde{\chi}_2^0$ produces OS dileptons plus $E_T^{\rm miss}$.

\item
If $m_{\tilde{e}_R}<m_{\tilde{\chi}_1^0}<m_{\tilde{\mu}_R}$, or if $m_{\tilde{\chi}_1^0}-m_{\tilde{\mu}_R}$ is sufficiently small, the $\tilde{\chi}_1^0$ decays mainly to $e\tilde{e}_R$. The $\tilde{\chi}_2^0$ therefore decays as $\tilde{\chi}_2^0 \rightarrow l_i^\pm l_{j}^\mp e\tilde{e}_R$.
 The signature will be OS dileptons plus an electron, and the track of $\tilde{e}_R$, without  $E_T^{\rm miss}$. 

\item
If $m_{\tilde{e}_R,\tilde{\mu}_R}<m_{\tilde{\chi}_1^0}$ and $|m_{\tilde{\mu}_R}-m_{\tilde{e}_R}|$ is not too small, 
the $\tilde{\chi}_1^0$ has a sizeable branching fraction to $e\tilde{e}_R$ and $\mu\tilde{\mu}_R$, and the next-to-lightest slepton decays to the lightest slepton within the detector. We will assume in the following that the lightest slepton is the $\tilde{e}_R$. 
The $\tilde{\mu}_R$ decays through the three-body decay $\tilde{\mu}_R\rightarrow e\mu\tilde{e}_R$.
The complete $\tilde{\chi}_2^0$ decay chains are thus $\tilde{\chi}_2^0\rightarrow l_i^\pm l_j^\mp \tilde{\chi}_1^0 \rightarrow l_i^\pm l_j^\mp  e \tilde{e}_R $ or 
$\tilde{\chi}_2^0\rightarrow l_i^\pm l_j^\mp \tilde{\chi}_1^0 \rightarrow l_i^\pm l_j^\mp [\mu^\pm\mu^\mp] e \tilde{e}_R $.
The signature of the $\tilde{\chi}_2^0$ decay will be OS dileptons plus $e$, and the track of $\tilde{e}_R$, without  $E_T^{\rm miss}$ and possibly with additional OS dimuons.

If the  $\tilde{\mu}_R$ is guaranteed to decay within the detector, there is no ambiguity in the observation of LFV. One can detect LFV by simply counting leptons: These events will have an odd number of leptons (3 or 5). Flavour conserving events will have an odd number of $e$ and an even number of $\mu$ and $\tau$, while LFV processes will have either an even number of $e$ and
an odd number of $\mu$ or $\tau$, or an odd number of $e$ and an odd number of $\mu$ and $\tau$.
The observable $K_{ij}$, defined in Eq.~\eqref{eq:Kij} to quantify LFV, can be directly measured by lepton counting :
\be
\begin{split}
K_{ij}&=\frac{\textrm{BR} (\tilde{\chi}_2^0 \rightarrow l_i^\pm l_{j\neq i}^\mp [\mu^\pm\mu^\mp] e \tilde{e}_R )}
{\textrm{BR}( \tilde{\chi}_2^0 \rightarrow l_i^\pm l_i^\mp [\mu^\pm\mu^\mp] e \tilde{e}_R) +
\textrm{BR} (\tilde{\chi}_2^0 \rightarrow l_j^\pm l_j^\mp [\mu^\pm\mu^\mp] e \tilde{e}_R ) }\\
&=\frac{N(l_i^\pm l_{j\neq i}^\mp [\mu^\pm\mu^\mp] e) }{N(l_i^\pm l_i^\mp [\mu^\pm\mu^\mp] e) + N(l_j^\pm l_j^\mp [\mu^\pm\mu^\mp] e)}~.                
\end{split}
\ee

\item
Finally, if $|m_{\tilde{\mu}_R}-m_{\tilde{e}_R}|$ is sufficiently small, both sleptons are stable within the detector.
The $\tilde{\chi}_2^0$ decays are therefore either $\tilde{\chi}_2^0\rightarrow l_i^\pm l_j^\mp \tilde{\chi}_1^0 \rightarrow l_i^\pm l_j^\mp  e \tilde{e}_R $
or $\tilde{\chi}_2^0\rightarrow l_i^\pm l_j^\mp \tilde{\chi}_1^0 \rightarrow l_i^\pm l_j^\mp  \mu \tilde{\mu}_R $.
The signature of the $\tilde{\chi}_2^0$ decay will be OS dileptons plus $e$ or $\mu$, and the track of $\tilde{e}_R$ or $\tilde{\mu}_R$, without $E_T^{\rm miss}$.
LFV in the $e-\tau$ or $\mu-\tau$ sectors may  be observed by detecting a single $\tau$ in these decays. 
The observables $K_{e\tau}$ and $K_{\mu\tau}$ can be inferred without ambiguity using  
\be
\begin{split}
K_{e\tau}&=\frac{\textrm{BR}( \tilde{\chi}_2^0 \rightarrow \tau^\pm e^\mp  e \tilde{e}_R  )}
{\textrm{BR}( \tilde{\chi}_2^0 \rightarrow \tau^\pm \tau^\mp  e \tilde{e}_R  )+
\textrm{BR}( \tilde{\chi}_2^0 \rightarrow e^\pm e^\mp  e \tilde{e}_R  )
}\\
&=\frac{N(\tau^\pm e^\mp e) }{N(\tau^\pm \tau^\mp e) + N(e^\pm e^\mp e)}~,                
\end{split}
\ee
\be
\begin{split}
K_{\mu\tau}&=\frac{\textrm{BR}( \tilde{\chi}_2^0 \rightarrow \tau^\pm \mu^\mp  \mu \tilde{\mu}_R  )}
{\textrm{BR}( \tilde{\chi}_2^0 \rightarrow \tau^\pm \tau^\mp  \mu \tilde{\mu}_R  )+
\textrm{BR}( \tilde{\chi}_2^0 \rightarrow \mu^\pm \mu^\mp  \mu \tilde{\mu}_R  )
}\\
&=\frac{N(\tau^\pm \mu^\mp \mu) }{N(\tau^\pm \tau^\mp \mu) + N(\mu^\pm \mu^\mp \mu)}~.
\end{split}
\ee
On the other hand, in the $e-\mu$ sector, one cannot detect LFV by simple lepton counting since $ee\mu$ or $e\mu\mu$ combinations can be produced both in flavour conserving and in flavour violating channels. However, the flavour-conserving channels give same-flavour dileptons with opposite signs (SFOS), while in the flavour-violating channels all sign combinations of the SF dileptons appear with equal probability.  To disentangle the two contributions, one should look out for same-sign dileptons, which can only appear through the flavour violating channels.
Neglecting the flavour violating effects of the right-handed slepton sector, which are $\mathcal{O}(10^{-2})$ (more precisely $\mathcal{O}(\epsilon^2)$) because of the hierarchical mixing, $K_{e\mu}$ is given by
\be
\begin{split}
K_{e\mu}&=\frac{2\times \textrm{BR}( \tilde{\chi}_2^0 \rightarrow e^\pm e^\pm \mu  \tilde{e}_R  ) }
{
\textrm{BR}( \tilde{\chi}_2^0 \rightarrow e^\pm e^\mp e  \tilde{e}_R  )+
\textrm{BR}( \tilde{\chi}_2^0 \rightarrow \mu^\pm \mu^\mp e  \tilde{e}_R  )-
 \textrm{BR}( \tilde{\chi}_2^0 \rightarrow \mu^\pm \mu^\pm e  \tilde{e}_R  )
}\\
&=\frac{2\times N(e^\pm e^\pm \mu) }{N(e^\pm e^\mp e)+N(\mu^\pm \mu^\mp e)-N(\mu^\pm \mu^\pm e)}
\end{split}
\ee
or
\be
\begin{split}
K_{e\mu}&=\frac{2\times \textrm{BR}( \tilde{\chi}_2^0 \rightarrow \mu^\pm \mu^\pm e  \tilde{\mu}_R  ) }
{
\textrm{BR}( \tilde{\chi}_2^0 \rightarrow \mu^\pm \mu^\mp \mu  \tilde{\mu}_R  )+
\textrm{BR}( \tilde{\chi}_2^0 \rightarrow e^\pm e^\mp \mu  \tilde{\mu}_R  )-
\textrm{BR}( \tilde{\chi}_2^0 \rightarrow e^\pm e^\pm \mu  \tilde{\mu}_R  )
}\\
&=\frac{2\times N(\mu^\pm \mu^\pm e) }{N(\mu^\pm \mu^\mp \mu)+N(e^\pm e^\mp \mu)-N(e^\pm e^\pm \mu)}
~.
\end{split}
\ee
The additional term in the denominator supresses the contributions to $\textrm{BR}( \tilde{\chi}_2^0 \rightarrow e^\pm e^\mp \mu  \tilde{\mu}_R  )$ [ $\textrm{BR}( \tilde{\chi}_2^0 \rightarrow \mu^\pm \mu^\mp e  \tilde{e}_R  )$]  coming from flavour violation. 
This ensures that the denominator contains only flavour conserving contributions.

\end{itemize}

\clearpage

\section{Conclusions}\label{conclusions}

We have studied flavour violation in supersymmetric models with a GUT-scale warped extra dimension. With matter fields located in the bulk, and the Higgs fields as well as the SUSY breaking hidden sector localized on the infrared brane, exponential wave function profiles can at the same time generate hierarchical fermion masses and mixings and somewhat suppress flavour changing neutral currents. 

However, we find that the constraints on FCNCs (in particular those on lepton flavour violation) are stringent enough to still rule out most generic models. For the concrete example of the holographic GUT model of NPT, several additional assumptions on the hidden sector are necessary in order to obtain a realistic spectrum and evade the experimental bounds. More specifically, there should be contributions to the SUSY breaking soft terms both from the radion superfield and from the brane-localized hidden sector fields; and the brane-induced trilinear soft terms should be small or zero (which could be enforced by symmetry). With these assumptions, substantial regions of the parameter space can give rise to realistic sparticle mass spectra while avoiding unacceptably large lepton flavour violation.

In these surviving regions of parameter space, the LHC phenomenology depends on whether the soft terms are predominantly induced by the radion, or whether the contributions from the radion and from the brane-localized hidden sector fields are comparable. We have given an account of the expected mass spectra and  LHC signatures in both cases. 

Generically, the bounds on lepton flavour violation, in particular BR$(e\to\mu\gamma)$, force the spectrum to be heavy, with squark and gluino masses well above 1~TeV. One the one hand this might explain why no signal of SUSY has yet been observed at the LHC operating at $\sqrt{s}=7$~TeV. 
On the other hand it means that, should the setup studied here be realized in Nature, it will require the high-luminosity run at 14~TeV to explore it. Moreover, a detailed experimental study of this scenario will most likely require precision measurements at even higher energy and/or luminosity (LHC upgrade) \cite{LHCupgrade}, or at a multi-TeV $e^+e^-$ collider~\cite{clic}.

\section*{Acknowledgements} 
We thank Werner Porod and Florian Staub for help with questions on SPheno, 
and Benjamin Fuks for providing cross sections with Madgraph. 
F.B.~thanks LPSC Grenoble for hospitality and support during various stages of this project. This research was supported in part by the French ANR project {\tt ToolsDMColl}, BLAN07-2-194882.

\begin{table}
\centering
\begin{tabular}{|c|c|c|c|c|}
\hline 
Point & A & B & C & D \tabularnewline
\hline
\hline 
$m_{\tilde{\chi}_{1}^{0}}$ & $652$ & $655$ & $655$ & $670$\tabularnewline
\hline 
$m_{\tilde{\chi}_{2}^{0}}$ & $1224$ & $1235$ & $1235$ & $1258$\tabularnewline
\hline 
$m_{\tilde{\chi}_{3}^{0}}$ & $2870$ & $3116$ & $3124$ & $5096$\tabularnewline
\hline 
$m_{\tilde{\chi}_{4}^{0}}$ & $2872$ & $3117$ & $3125$ & $5097$\tabularnewline
\hline 
$m_{\tilde{\chi}_{1}^{\pm}}$ & $1224$ & $1235$ & $1235$ & $1259$\tabularnewline
\hline 
$m_{\tilde{\chi}_{2}^{\pm}}$ & $2873$ & $3118$ & $3126$ & $5097$\tabularnewline
\hline
\hline 
$m_{\tilde{l}_{1}}$ & $555\,(\sim\tilde{e}_{R})$ & $555\,(\sim\tilde{e}_{R})$ & $556\,(\sim\tilde{e}_{R})$ & $772\,(\sim\tilde{e}_{R})$\tabularnewline
\hline 
$m_{\tilde{l}_{2}}$ & $679\,(\sim\tilde{\mu}_{R})$ & $672\,(\sim\tilde{\mu}_{R})$ & $619\,(\sim\tilde{\mu}_{R})$ & $904\,(\sim\tilde{l}_{L\,1})$\tabularnewline
\hline 
$m_{\tilde{l}_{3}}$ & $993\,(\sim\tilde{\tau_1})$ & $1267\,(\sim\tilde{\tau_1})$ & $1096\,(\sim\tilde{\tau_1})$ & $914\,(\sim\tilde{\mu}_{R})$\tabularnewline
\hline 
$m_{\tilde{l}_{4}}$ & $1057\,(\sim\tilde{\mu}_{L})$ & $1580\,(\sim\tilde{\tau_2})$ & $1543\,(\sim\tilde{\tau_2})$ & $1012\,(\sim\tilde{l}_{L\,2})$\tabularnewline
\hline 
$m_{\tilde{l}_{5}}$ & $1057\,(\sim\tilde{e}_{L})$ & $1582\,(\sim\tilde{\mu}_{L})$ & $1573\,(\sim\tilde{\mu}_{L})$ & $1037\,(\sim\tilde{l}_{L\,3})$\tabularnewline
\hline 
$m_{\tilde{l}_{6}}$ & $1069\,(\sim\tilde{\tau_2})$ & $1615\,(\sim\tilde{e}_{L})$ & $1581\,(\sim\tilde{e}_{L})$ & $2402\,(\sim\tilde{\tau}_{R})$\tabularnewline
\hline 
$m_{\tilde{\nu}_{1}}$ & $990$ & $1577$ & $1534$ & $901$\tabularnewline
\hline 
$m_{\tilde{\nu}_{2}}$ & $1053$ & $1579$ & $1569$ & $1009$\tabularnewline
\hline 
$m_{\tilde{\nu}_{3}}$ & $1054$ & $1604$ & $1579$ & $1034$\tabularnewline
\hline
\hline 
$m_{\tilde{d}_{1}}$ & $2721\,(\sim\tilde{b}_R)$ & $2867\,(\sim\tilde{d}_L)$ & $2867\,(\sim\tilde{d}_L)$ & $2785\,(\sim\tilde{b}_R)$\tabularnewline
\hline 
$m_{\tilde{d}_{2}}$ & $2752\,(\sim\tilde{s}_R)$ & $2888\,(\sim\tilde{s}_L)$ & $2888\,(\sim\tilde{s}_L)$ & $2796\,(\sim\tilde{s}_R)$\tabularnewline
\hline 
$m_{\tilde{d}_{3}}$ & $2753\,(\sim\tilde{d}_R)$ & $2911\,(\sim\tilde{b})$  & $2912\,(\sim\tilde{b}_1)$ & $2797\,(\sim\tilde{d}_R)$ \tabularnewline
\hline 
$m_{\tilde{d}_{4}}$ & $2873\,(\sim\tilde{d}_L)$ & $2985\,(\sim\tilde{s}/\tilde{b})$ & $2985\,(\sim\tilde{s}/\tilde{b}_1)$ & $2915\,(\sim\tilde{d}_L)$\tabularnewline
\hline 
$m_{\tilde{d}_{5}}$ & $2897\,(\sim\tilde{s}_L)$ & $2990\,(\sim\tilde{s}/\tilde{b}_2)$ & $2991\,(\sim\tilde{s}/\tilde{b}_2)$ & $2939\,(\sim\tilde{s}_L)$ \tabularnewline
\hline 
$m_{\tilde{d}_{6}}$ & $2943\,(\sim\tilde{b}_L)$ & $2992\,(\sim\tilde{d}_R)$ & $2992\,(\sim\tilde{d}_R)$ & $4358\,(\sim\tilde{b}_L)$ \tabularnewline
\hline 
$m_{\tilde{u}_{1}}$ & $2694\,(\sim\tilde{t}_1)$ & $2737\,(\sim\tilde{u}_{R\,1})$ & $2737\,(\sim\tilde{u}_{R\,1})$ & $2739\,(\sim\tilde{u}_R)$\tabularnewline
\hline 
$m_{\tilde{u}_{2}}$ & $2744\,(\sim\tilde{u}_{R\,1})$ & $2744\,(\sim\tilde{u}_{R\,2})$ & $2744\,(\sim\tilde{u}_{R\,2})$ & $2762\,(\sim\tilde{c}_{R})$\tabularnewline
\hline 
$m_{\tilde{u}_{3}}$ & $2776\,(\sim\tilde{u}_{R\,2})$ & $2818\,(\sim\tilde{t}_{1})$ & $2819\,(\sim\tilde{t}_{1})$ & $2914\,(\sim\tilde{d}_L)$\tabularnewline
\hline 
$m_{\tilde{u}_{4}}$ & $2872\,(\sim\tilde{u}_{L\,1})$ & $2867\,(\sim\tilde{u}_{L\,1})$ & $2866\,(\sim\tilde{u}_{L\,1})$ & $2938\,(\sim\tilde{c}_L)$\tabularnewline
\hline 
$m_{\tilde{u}_{5}}$ & $2897\,(\sim\tilde{u}_{L\,2})$ & $2888\,(\sim\tilde{u}_{L\,2})$ & $2888\,(\sim\tilde{u}_{L\,2})$ & $4104\,(\sim\tilde{t}_1)$\tabularnewline
\hline 
$m_{\tilde{u}_{6}}$ & $2967\,(\sim\tilde{t}_2)$ & $2998\,(\sim\tilde{t}_{2})$ & $2999\,(\sim\tilde{t}_2)$ & $4373\,(\sim\tilde{t}_2)$\tabularnewline
\hline 
$m_{\tilde{g}}$ & $3201$ & $3210$ & $3210$ & $3303$\tabularnewline
\hline
\hline 
$m_{h^{0}}$ & $118.0$ & $122.0$ & $122.0$ & $118.9$\tabularnewline
\hline 
$m_{H^{0}}$ & $1366$ & $864$ & $710$ & $1863$\tabularnewline
\hline 
$m_{A^{0}}$ & $1368$ & $865$ & $716$ & $1866$\tabularnewline
\hline 
$m_{H^{\pm}}$ & $1366$ & $869$ & $710$ & $1837$\tabularnewline
\hline
\end{tabular}\caption{Sample spectra for $F^T/2R=1.5~\textrm{TeV}$. Points A--C are representative for the radion-dominated scenario with low and high $\tan\beta$:   point~A has $\tan\beta=5$, while points B and C have $\tan\beta=30$ with respectively $m_{\tilde{\chi}_1^0}<m_{\tilde{\mu}_R}$ and $m_{\tilde{\chi}_1^0}>m_{\tilde{\mu}_R}$. Point~D is an example of a mixed brane-radion scenario with a neutralino LSP.}
\label{Tab:points_RM_1500}
\end{table}
\begin{table}
\centering
\begin{tabular}{|c|c|c|c|c|}
\hline 
Point & A' & B' & C' & D' \tabularnewline
\hline
\hline 
$m_{\tilde{\chi}_{1}^{0}}$ & $427$ & $429$ & $429$ & $439$\tabularnewline
\hline 
$m_{\tilde{\chi}_{2}^{0}}$ & $808$ & $816$ & $816$ & $832$\tabularnewline
\hline 
$m_{\tilde{\chi}_{3}^{0}}$ & $1994$ & $2155$ & $2156$ & $3430$\tabularnewline
\hline 
$m_{\tilde{\chi}_{4}^{0}}$ & $1997$ & $2157$ & $2157$ & $3431$\tabularnewline
\hline 
$m_{\tilde{\chi}_{1}^{\pm}}$ & $809$ & $816$ & $816$ & $832$\tabularnewline
\hline 
$m_{\tilde{\chi}_{2}^{\pm}}$ & $1997$ & $2157$ & $2158$ & $3432$\tabularnewline
\hline
\hline 
$m_{\tilde{l}_{1}}$ & $372\,(\sim\tilde{e}_{R})$ & $373\,(\sim\tilde{e}_{R})$ & $373\,(\sim\tilde{e}_{R})$ & $518\,(\sim\tilde{e}_{R})$\tabularnewline
\hline 
$m_{\tilde{l}_{2}}$ & $453\,(\sim\tilde{\mu}_{R})$ & $441\,(\sim\tilde{\mu}_{R})$ & $418\,(\sim\tilde{\mu}_{R})$ & $597\,(\sim\tilde{\mu}_{R})$\tabularnewline
\hline 
$m_{\tilde{l}_{3}}$ & $629\,(\sim\tilde{\tau}_1)$ & $862\,(\sim\tilde{\tau}_1)$ & $856\,(\sim\tilde{\tau}_1)$ & $611\,(\sim\tilde{l}_{L\,1})$\tabularnewline
\hline 
$m_{\tilde{l}_{4}}$ & $646\,(\sim\tilde{\tau}_2)$ & $1051\,(\sim\tilde{\mu}_{L})$ & $1046\,(\sim\tilde{\mu}_{L})$ & $683\,(\sim\tilde{l}_{L\,2})$\tabularnewline
\hline 
$m_{\tilde{l}_{5}}$ & $710\,(\sim\tilde{\mu}_{L})$ & $1054\,(\sim\tilde{e}_{L})$ & $1054\,(\sim\tilde{e}_{L})$ & $701\,(\sim\tilde{l}_{L\,3})$\tabularnewline
\hline 
$m_{\tilde{l}_{6}}$ & $711\,(\sim\tilde{e}_{L})$ & $1093\,(\sim\tilde{\tau}_2)$ & $1091\,(\sim\tilde{\tau}_2)$ & $1604\,(\sim\tilde{\tau}_{R})$\tabularnewline
\hline 
$m_{\tilde{\nu}_{1}}$ & $625$ & $1047$ & $1042$ & $606$\tabularnewline
\hline 
$m_{\tilde{\nu}_{2}}$ & $706$ & $1050$ & $1050$ & $679$\tabularnewline
\hline 
$m_{\tilde{\nu}_{3}}$ & $706$ & $1076$ & $1073$ & $697$\tabularnewline
\hline
\hline 
$m_{\tilde{d}_{1}}$ & $1879\,(\sim\tilde{b}_{R})$ & $1976\,(\sim\tilde{d}_{L})$ & $1976\,(\sim\tilde{d}_L)$ & $1922\,(\sim\tilde{b}_R)$\tabularnewline
\hline 
$m_{\tilde{d}_{2}}$ & $1899\,(\sim\tilde{s}_{R})$ & $1988\,(\sim\tilde{s}/\tilde{b}_{1})$ & $1988\,(\sim\tilde{s}/\tilde{b}_1)$ & $1930\,(\sim\tilde{s}_R)$\tabularnewline
\hline 
$m_{\tilde{d}_{3}}$ & $1900\,(\sim\tilde{d}_{R})$ & $1993\,(\sim\tilde{s}/\tilde{b}_{2})$  & $1993\,(\sim\tilde{s}/\tilde{b}_2)$ & $1930\,(\sim\tilde{d}_R)$ \tabularnewline
\hline 
$m_{\tilde{d}_{4}}$ & $1980\,(\sim\tilde{d}_{L})$ & $2048\,(\sim\tilde{s}_R)$ & $2048\,(\sim\tilde{s}_R)$ & $2009\,(\sim\tilde{d}_L)$\tabularnewline
\hline 
$m_{\tilde{d}_{5}}$ & $1995\,(\sim\tilde{s}_{L})$ & $2052\,(\sim\tilde{d}_R)$ & $2052\,(\sim\tilde{d}_R)$ & $2024\,(\sim\tilde{s}_L)$ \tabularnewline
\hline 
$m_{\tilde{d}_{6}}$ & $2024\,(\sim\tilde{b}_{L})$ & $2063\,(\sim\tilde{b})$ & $2064\,(\sim\tilde{b})$ & $2939\,(\sim\tilde{b}_L)$ \tabularnewline
\hline 
$m_{\tilde{u}_{1}}$ & $1846\,(\sim\tilde{t}_{1})$ & $1888\,(\sim\tilde{u}_{R\,1})$ & $1888\,(\sim\tilde{u}_{R\,1})$ & $1892\,(\sim\tilde{u}_R)$\tabularnewline
\hline 
$m_{\tilde{u}_{2}}$ & $1894\,(\sim\tilde{u}_{R})$ & $1891\,(\sim\tilde{u}_{R\,2})$ & $1891\,(\sim\tilde{u}_{R\,2})$ & $1907\,(\sim\tilde{c}_R)$\tabularnewline
\hline 
$m_{\tilde{u}_{3}}$ & $1914\,(\sim\tilde{c}_{R})$ & $1933\,(\sim\tilde{c}/\tilde{t})$ & $1933\,(\sim\tilde{c}/\tilde{t})$ & $2007\,(\sim\tilde{u}_L)$\tabularnewline
\hline 
$m_{\tilde{u}_{4}}$ & $1979\,(\sim\tilde{u}_{L})$ & $1975\,(\sim\tilde{u}_{L})$ & $1975\,(\sim\tilde{u}_{L})$ & $2022\,(\sim\tilde{c}_L)$\tabularnewline
\hline 
$m_{\tilde{u}_{5}}$ & $1994\,(\sim\tilde{c}_{L})$ & $1989\,(\sim\tilde{c}_{L})$ & $1989\,(\sim\tilde{c}_{L})$ & $2760\,(\sim\tilde{t}_1)$\tabularnewline
\hline 
$m_{\tilde{u}_{6}}$ & $2060\,(\sim\tilde{t}_{2})$ & $2079\,(\sim\tilde{t})$ & $2079\,(\sim\tilde{t})$ & $2962\,(\sim\tilde{t}_2)$\tabularnewline
\hline 
$m_{\tilde{g}}$ & $2198$ & $2202$ & $2202$ & $2265$\tabularnewline
\hline
\hline 
$m_{h^{0}}$ & $115.5$ & $119.7$ & $119.7$ & $116.5$\tabularnewline
\hline 
$m_{H^{0}}$ & $890$ & $599$ & $585$ & $1259$\tabularnewline
\hline 
$m_{A^{0}}$ & $893$ & $599$ & $586$ & $1260$\tabularnewline
\hline 
$m_{H^{\pm}}$ & $892$ & $604$ & $591$ & $1244$\tabularnewline
\hline
\hline
$\sigma(pp\to \tilde q\tilde q)$                              & $9.53$ & $8.97$ & $8.98$ & $8.65$ \tabularnewline
$\sigma(pp\to \tilde g\tilde g, \tilde g\tilde q)$       & $2.39$ & $2.29$ & $2.29$ & $1.91$ \tabularnewline
$\sigma(pp\to \tilde\chi\tilde\chi)$                         & $4.56$ & $4.19$ & $4.19$ & $3.81$ \tabularnewline
$\sigma(pp\to \tilde q\tilde\chi, \tilde g\tilde\chi)$  & $1.64$ & $1.63$ & $1.64$ & $1.44$\tabularnewline
$\sigma(pp\to \tilde l_i\tilde l_j, \tilde l_i\tilde\nu_j, \tilde\nu_i\tilde\nu_j)$ &  $4.01$ & $1.05$ & $1.33$ & $3.23$ \tabularnewline
\hline

\end{tabular}\caption{Sample spectra for $F^T/2R=1~\textrm{TeV}$. Points A'--C' are representative for the radion-dominated scenario with low and high $\tan\beta$:   point~A' has $\tan\beta=5$, while points B' and C' have $\tan\beta=30$ with respectively $m_{\tilde{\chi}_1^0}<m_{\tilde{\mu}_R}$ and $m_{\tilde{\chi}_1^0}>m_{\tilde{\mu}_R}$. Point~D' is an example of a mixed brane-radion scenario with a neutralino LSP. 
All masses are in GeV units. 
Production cross sections (in fb) at the LHC with $\sqrt{s}=14$~TeV, computed with {\tt MadGraph}~\cite{madgraph}, are also given.}
\label{Tab:points_RM_1000}
\end{table}

\clearpage

\appendix

\noindent{\Large\bf Appendix}

\section{Radius stabilization and SUSY breaking}\label{modsb}
In this Appendix we sketch a warped 5D model (following \cite{Luty:1999cz,Luty:2000ec}) in which the radion is stabilized, and in which both the radion superfield and some additional IR brane fields have non-vanishing $F$-term expectation values. It therefore provides a dynamical origin for the SUSY breaking field background on which we based our analysis. 

Consider first a warped extra dimension with two separate sectors: A super-Yang-Mills theory in the bulk and a super-Yang-Mills theory on the IR brane. In the infrared the degrees of freedom are a radion superfield $T$ and two non-abelian gauge superfields which will undergo gaugino condensation. The strong-coupling scales of the SYM theories are taken parametrically smaller than the KK scale. Later we will add an ``uplifting'' sector, consisting of a dynamical SUSY-breaking sector on the IR brane.

In units of the 4D reduced Planck mass $M_4=2.4\cdot 10^{18}$ GeV, the effective Lagrangian after gaugino condensation can be written as
\be\label{radionL}
{\cal L}=\int d^4\theta\,\ol\phi\phi\left(-3\,e^{-K/3}\right)+\int d^2\theta\,\phi^3\left(ae^{-bT}+c\right)\hc\,
\ee
Here the radion K\"ahler potential is 
\be\label{radionK}
K=-3\,\log\left[\frac{M_5^3}{k}\left(e^{k\pi(T+\ol T)}-1\right)\right]\,,
\ee
and $a$, $b$, $c$ are constants. While $a$ and $b$ come from bulk gaugino condensation, and are of order unity (or somewhat large since the theory is weakly coupled at the compactification scale), $c$ comes from the IR brane gaugino condensate and is exponentially small. Note that our conventions differ from those of \cite{Luty:2000ec} by the sign of $k$ and by a factor $\pi$ in the definition of the radion field.

Eqns.~\eqref{radionL} and \eqref{radionK} are written in a frame where the warp factor is unity in the IR and exponentially large on the UV brane. For consistency with our conventions in the main text, we perform a Weyl rescaling, which is a redefinition of the chiral compensator:
\be\label{weylresc}
\varphi=e^{\pi k T}\phi\,.
\ee
This gives
\be\label{radionL2}
{\cal L}=\int d^4\theta\,\ol\varphi\varphi\left(-3\,e^{-K/3}\right)+\int d^2\theta\,\varphi^3\left(ae^{-bT}+c\right)e^{-3k\pi T}\hc\,
\ee
with 
\be
K=-3\,\log\left[\frac{M_5^3}{k}\left(1-e^{-k\pi(T+\ol T)}\right)\right]\,.
\ee
It is convenient to define the warp factor superfield $\omega$ by the holomorphic field redefinition 
\be
\omega=\varphi e^{-k\pi T}
\ee 
(note that this does not just amount to undoing the Weyl rescaling of Eq.~\eqref{weylresc}; the chiral compensator is always normalized such that $\vev{\varphi}=1+F^\varphi\theta^2$, while here we are choosing a different way of parameterizing the radion). The Lagrangian becomes
\be\label{radionL3}
{\cal L}=-\frac{3M_5^3}{k}\int d^4\theta\,\left(\ol\varphi\varphi-\ol\omega\omega\right)+\int d^2\theta\,\left(a\,\omega^{3+\nu}\varphi^{-\nu}+c\,\omega^3\right)\hc\,,
\ee
where $\nu=b/k\pi$. This yields the $F$-terms
\be\begin{split}
{\ol F}^{\bar\omega}&=-\frac{k}{3 M_5^3}\left((3+\nu)a\,\omega^{2+\nu}+3c\,\omega^2\right)\,,\\
{\ol F}^{\bar\varphi}&=-\frac{k}{3 M_5^3}\,a\nu\,\omega^{3+\nu}\,,
\end{split}
\ee
and the scalar potential
\be
V=\frac{3 M_5^3}{k}\left(\left|F^\omega\right|^2-\left|F^\varphi\right|^2\right)\,.
\ee
At large warp factors, i.e.~small $|\omega|$, the $|F^\varphi|^2$ term in $V$ is subdominant, and the potential is minimized at a finite value of $\omega$,
\be
|\omega|=\left|\frac{3c}{(3+\nu)a}\right|^{1/\nu}\,.
\ee
This ansatz is self-consistent because $c$ is exponentially small. There is also a decompactification solution at $\omega\into 0$, which is however of no interest for us.

For the gravitino mass we find
\be
m_{3/2}^2=e^K|W|^2\approx\frac{k^3}{M_5^9}\left(\frac{\nu}{3+\nu}\right)^2|c|^2\,|\omega|^6\sim\left|F^\varphi\right|^2\,. 
\ee
Returning to the old variables, the radion $F$-term is 
\be
\frac{F^T}{2R}=\frac{1}{2\pi kR}\left(F^\varphi-\frac{F^\omega}{\omega}\right)\,.
\ee
$F^\omega/\omega$ vanishes to leading order, but the subleading terms turn out to be finite and are parametrically of the order ${\cal O}(\omega^{\nu+3})\sim{\cal O}(F^\varphi)$.

So far the vacuum energy density is negative, and the vacuum is a non-supersymmetric AdS minimum. This can be remedied by adding an additional sector which breaks supersymmetry dynamically on its own, in the rigid limit, thus providing a positive contribution to the cosmological constant. The details of such an ``$F$-term uplift'' have been worked out mainly in the context of effective field theories from type IIB flux compactifications (see e.g.~\cite{GomezReino:2006dk}). Our main interest here is the relative importance of the contributions to soft terms from the uplifting sector compared to the radion contributions.
In our normalization, including SUSY breaking IR brane fields $Z_I$ in this background as
\be
\Delta{\cal L}_{\rm brane}=\int d^4\theta\, e^{-2\pi kR}\sum_I \left|Z_I\right|^2+\int d^2\theta\,e^{-3\pi kR}W\left(Z_I\right)
\ee
will give an additional contribution to the scalar potential,
\be
\Delta V=e^{-2\pi kR}\,\sum_I \left|F^{Z_I}\right|^2\,.
\ee
To fine-tune the cosmological constant to zero, we thus need
\be
e^{-\pi kR} F^{Z_I}\sim F^\varphi\,.
\ee
Therefore, in this particular model, the brane-localized contributions to the soft terms dominate over the contributions from the gravitational sector, i.e.~the radion and compensator contributions. Since the warp factor is only moderately large in the scenarios we are considering in the main text, this model could still serve as an example for mixed brane-radion mediation. The gravitino is naturally the LSP, as is common in models of warped supersymmetry.

\section{Matching parameters to fermion masses and mixings}\label{appendix_epsilon_expansions}
\subsection{Preliminaries}
The superpotential of the MSSM contains the quark Yukawa terms
\be
W=\yuk{U}_{ij}\,H_u Q^i U^j+\yuk{D}_{ij}\,H_d Q^i D^j\,.
\ee
Both Yukawa matrices $\yuk{U}$ and $\yuk{D}$ may be diagonalized by bi-unitary transformations,
\be\label{SVD}
\yuk{U}_{\rm diag}={\cal U}_{UL}\; \yuk{U}\;{\cal U}_{UR}^\dag,\quad \yuk{D}_{\rm diag}={\cal U}_{DL}\; \yuk{D}\;{\cal U}_{DR}^\dag\,.
\ee
Here ${\cal U}_{UL}$ is a unitary matrix which diagonalizes the Hermitian matrix $\yuk{U}{\yuk{U}}^\dag$,
\be
{\cal U}_{UL}\left(\yuk{U}{\yuk{U}}^\dag\right){\cal U}_{UL}^\dag=\yuk{U}_{\rm diag}{\yuk{U}_{\rm diag}}^\dag\,,
\ee
and ${\cal U}_{DL}$ is a unitary matrix which diagonalizes the Hermitian matrix $\yuk{D}{\yuk{D}}^\dag$,
\be
{\cal U}_{DL}\left(\yuk{D}{\yuk{D}}^\dag\right){\cal U}_{DL}^\dag=\yuk{D}_{\rm diag}{\yuk{D}_{\rm diag}}^\dag\,,
\ee
The CKM matrix ${\cal V}_{\rm CKM}$ of quark mixings is given by
\be
{\cal V}_{\rm CKM}={\cal U}_{UL}{\cal U}_{DL}^\dag\,,
\ee
and the physical quark masses are given by the matrix entries of $\yuk{U}_{\rm diag}$ and $\yuk{D}_{\rm diag}$ multiplied by the appropriate Higgs expectation value. Three mixing angles, one phase, and six mass eigenvalues constitute the physical observables in the quark sector.

If there is only one small suppression parameter $\epsilon$, as in the simple example in Section \ref{hfroml}, the Yukawa matrices are
\be
\yuk{U}=\left(\begin{array}{ccc}
\lambda^\Ui_{11}\epsilon^4 & \lambda^\Ui_{12}\epsilon^3 & \lambda^\Ui_{13}\epsilon^2\\
\lambda^\Ui_{21}\epsilon^3 & \lambda^\Ui_{22}\epsilon^2 & \lambda^\Ui_{23}\epsilon\\
\lambda^\Ui_{31}\epsilon^2 & \lambda^\Ui_{32}\epsilon & \lambda^\Ui_{33}\\
\end{array}\right),\qquad 
\yuk{D}=\left(\begin{array}{ccc}
\lambda^\Di_{11}\epsilon^3 & \lambda^\Di_{12}\epsilon^3 & \lambda^\Di_{13}\epsilon^3\\
\lambda^\Di_{21}\epsilon^2 & \lambda^\Di_{22}\epsilon^2 & \lambda^\Di_{23}\epsilon^2\\
\lambda^\Di_{31}\epsilon & \lambda^\Di_{32}\epsilon & \lambda^\Di_{33}\epsilon\\
\end{array}\right)\,.
\ee
Here $\epsilon\approx 0.1$, and the $\lambda^{u,d}_{ij}$ of order unity. Note that to leading order in $\epsilon$, the structure of  $y^{u}y^{u\dag}$ and  $y^{d}y^{d\dag}$ is similar, up to an overall $\epsilon^2$ factor:
\be
y^{u}y^{u\dag}=\left(\begin{array}{ccc}
r_\Ui \epsilon^4 & s_\Ui \epsilon^3 & t_\Ui \epsilon^2\\
s_\Ui^*\epsilon^3 & u_\Ui \epsilon^2 &v_\Ui \epsilon\\
t_\Ui^*\epsilon^2 & v_\Ui^*\epsilon & w_\Ui \\
\end{array}\right),\qquad 
y^{d}y^{d\dag}=\epsilon^2\left(\begin{array}{ccc}
r_\Di\epsilon^4 & s_\Di\epsilon^3 & t_\Di\epsilon^2\\
s_\Di^*\epsilon^3 & u_\Di\epsilon^2 &v_\Di\epsilon\\
t_\Di^*\epsilon^2 & v_\Di^*\epsilon & w_\Di\\
\end{array}\right)\,.
\ee
Here we have defined
\be\begin{split}
r_\Ui &=|\lambda_{13}^\Ui|^2+|\lambda_{12}^\Ui|^2\epsilon^2+|\lambda_{11}^\Ui|^2\epsilon^4,\quad s_\Ui =\lambda_{13}^\Ui\lambda_{23}^{\Ui*}+\lambda_{12}^\Ui\lambda_{22}^{\Ui*}\epsilon^2
+\lambda_{11}^\Ui\lambda_{21}^{\Ui*}\epsilon^4,\\
t_\Ui &=\lambda_{13}^\Ui\lambda_{33}^{\Ui*}+\lambda^\Ui_{12}\lambda_{32}^{\Ui*}\epsilon^2
+\lambda^\Ui_{11}\lambda_{31}^{\Ui*}\epsilon^4,\quad u_\Ui =|\lambda_{23}^\Ui|^2+|\lambda_{22}^\Ui|^2\epsilon^2+|\lambda_{21}^\Ui|^2\epsilon^4,\\
v_\Ui &=\lambda_{23}^\Ui\lambda_{33}^{\Ui*}+\lambda_{22}^\Ui\lambda_{32}^{\Ui*}\epsilon^2
+\lambda_{21}^\Ui\lambda_{31}^{\Ui*}\epsilon^4,\quad
w_\Ui =|\lambda_{33}^\Ui|^2+|\lambda_{32}^\Ui|^2\epsilon^2+|\lambda_{31}^\Ui|^2\epsilon^4,
\end{split}
\ee
and
\be\begin{split}
 r_\Di&=|\lambda_{11}^\Di|^2+|\lambda_{12}^\Di|^2+|\lambda_{13}^\Di|^2,\quad s_\Di=\lambda_{11}^\Di\lambda_{21}^{\Di*}+\lambda_{12}^\Di\lambda_{22}^{\Di*}+\lambda_{13}^\Di\lambda_{23}^{\Di*},\\
t_\Di&=\lambda^\Di_{11}\lambda_{31}^{\Di*}+\lambda^\Di_{12}\lambda_{32}^{\Di*}+\lambda_{13}^\Di\lambda_{33}^{\Di*},\quad
u_\Di=|\lambda_{21}^\Di|^2+|\lambda_{22}^\Di|^2+|\lambda_{23}^\Di|^2,\\
v_\Di&=\lambda_{21}^\Di\lambda_{31}^{\Di*}+\lambda_{22}^\Di\lambda_{32}^{\Di*}+\lambda_{23}^\Di\lambda_{33}^{\Di*},\quad
w_\Di=|\lambda_{31}^\Di|^2+|\lambda_{32}^\Di|^2+|\lambda_{33}^\Di|^2\,.
\end{split}
\ee

\subsection{Fermion masses}\label{sect:fermion_masses}

A hermitian matrix of the form
\be\label{epsmatrix}
M=\left(\begin{array}{ccc}
r \epsilon^4& s\epsilon^3 & t\epsilon^2\\
s^*\epsilon^3 & u\epsilon^2 &v\epsilon\\
t^*\epsilon^2 & v^*\epsilon & w\\
\end{array}\right)
\ee
has the eigenvalues
\be\begin{split}
\mu_1&=w+\frac{|v|^2}{w}\epsilon^2+\frac{1}{w}\left(\frac{|v|^2}{w^2}(uw-|v|^2)+|t|^2\right)\epsilon^4+{\cal O}(\epsilon^6),\\
\mu_2&=\frac{uw-|v|^2}{w}\epsilon^2+\frac{1}{w}\left(\frac{|sw-tv^*|^2}{uw-|v|^2}-\frac{|v|^2}{w^2}(uw-|v|^2)\right)\epsilon^4+{\cal O}(\epsilon^6),\\
\mu_3&=\frac{1}{u}\left((ru-|s|^2)-\frac{|sv-tu|^2}{uw-|v|^2}\right)\epsilon^4+{\cal O}(\epsilon^6)\,.
\end{split}
\ee
Note that in the case of $M=y^{u}y^{u\dag}$, all of the ``minors'' $(ru-|s|^2)$, $(sw-tv^*)$, $(sv-tu)$ and $(uw-|v|^2)$ are ${\cal O}(\epsilon^2)$ (unless there is some fine-tuned cancellation between the $\lambda_{ij}^\Ui$). Their ratios are therefore ${\cal O}(1)$, and expressions such as $(sw-tv^*)^2/(uw-|v|^2)$ or $(sv-tu)^2/(uw-|v|^2)$ are ${\cal O}(\epsilon^2)$. In particular, the smallest eigenvalue $\mu_3=y_u^2$ is only generated at higher order, namely at ${\cal O}(\epsilon^8)$.

For the case of real $\lambda$, we thus find the Yukawa couplings
\be\begin{split}\label{yuks}
y_t&=|\lambda_{33}^\Ui|+\frac{(\lambda_{23}^\Ui)^2+(\lambda_{32}^\Ui)^2}{2\,|\lambda_{33}^\Ui|}\epsilon^2+{\cal O}(\epsilon^4),\\
y_c&=\frac{|\lambda_{33}^\Ui\lambda_{22}^\Ui-\lambda_{32}^\Ui\lambda_{23}^\Ui|}{|\lambda_{33}^\Ui|}\epsilon^2+{\cal O}(\epsilon^4),\\
y_u&={\cal O}(\epsilon^4),\\
y_b&=\sqrt{w_\Di}\,\epsilon+{\cal O}(\epsilon^3),\\
y_s&=\sqrt{u_\Di-\frac{v_\Di^2}{w_\Di}}\,\epsilon^2+{\cal O}(\epsilon^4),\\
y_d&={\cal O}(\epsilon^3)\,.
\end{split}
\ee
In many grand-unified models, including our benchmark model, the matrix $\lambda^\Ui_{ij}$ is symmetric because it originates from a ${\bf 10}_i{\bf 10}_j{\bf 5}_H$ coupling (in $\SU 5$ notation), so $\lambda^\Ui_{23}=\lambda^\Ui_{32}$.

For the more general case that there are three distinct $c$-parameters in the up-type sector, one should parameterize the wave function suppression in a more general way, by allowing for several distinct suppression factors. Defining
\be
\epsilon_i\equiv e^{-\pi kR(|c_{{\cal T}i}|-1/2)},
\ee
the up-type Yukawa matrices have the structure
\be
\yuk{U}\sim(\epsilon_3)^2\left(\begin{array}{ccc}
\left(\frac{\epsilon_1}{\epsilon_3}\right)^2 & \frac{\epsilon_1\,\epsilon_2}{(\epsilon_3)^2} & \frac{\epsilon_1}{\epsilon_3} \\
\frac{\epsilon_1\,\epsilon_2}{(\epsilon_3)^2} & \left(\frac{\epsilon_2}{\epsilon_3}\right)^2 & \frac{\epsilon_2}{\epsilon_3}\\
\frac{\epsilon_1}{\epsilon_3} & \frac{\epsilon_2}{\epsilon_3} & 1
\end{array}\right)
\ee
where we did not explicitly write any $\lambda$ factors. Assuming that $\epsilon_3\gg\epsilon_2\gg\epsilon_1$, the Yukawa couplings of the first two generations are, to leading order, 
\be\begin{split}
y_t&=|\lambda_{33}^\Ui|(\epsilon_3)^2+\frac{(\lambda_{23}^\Ui)^2+(\lambda_{32}^\Ui)^2}{2\,|\lambda_{33}^\Ui|}(\epsilon_2)^2+\ldots,\\
y_c&=\frac{|\lambda_{33}^\Ui\lambda_{22}^\Ui-\lambda_{32}^\Ui\lambda_{23}^\Ui|}{|\lambda_{33}^\Ui|}(\epsilon_2)^2+\ldots
\end{split}
\ee
In the down-type sector, with the assumption $c_{{\cal F}1}\approx c_{{\cal F}2}\approx c_{{\cal F}3}$, all that needs to be done is to set 
\be
\epsilon=e^{-\pi kR(|c_{{\cal F}i}|-1/2)}
\ee
in the last three of Eqns.~\eqref{yuks}.

\subsection{CKM matrix}

Again for the case of a single small parameter $\epsilon$, the matrix $M$ of Eq.~\eqref{epsmatrix} is diagonalized by
\be\label{UL}
{\cal U}_L=\left(\begin{array}{ccc}
1 -\frac{1}{2}|\zeta|^2\,\epsilon^2& -\zeta\,\epsilon & \frac{sv-tu}{uw-|v|^2}\epsilon^2\\
\zeta^*\epsilon & 1-\frac{1}{2}(|\gamma|^2+|\zeta|^2)\,\epsilon^2 & -\gamma\,\epsilon \\
\frac{t^*}{w}\,\epsilon^2 & \gamma^*\,\epsilon & 1-\frac{1}{2}|\gamma|^2\,\epsilon^2
\end{array}\right)+{\cal O}(\epsilon^3)\,,
\ee
where
\be
\gamma=\frac{v}{w},\qquad\zeta=\frac{sw-tv^*}{uw-v^2}.
\ee
Applying this to calculate ${\cal U}_{UL}$ and ${\cal U}_{DL}$, we find for the CKM matrix
\be\begin{split}
&{\cal V}_{\rm CKM}={\cal U}_{UL}{\cal U}_{DL}^\dag=\,\\
&\left(\begin{array}{ccc}1-\frac{1}{2}\left(|\zeta_\Di|^2+|\zeta_\Ui|^2-2\zeta_\Ui\zeta_\Di^*\right)\epsilon^2 & (\zeta_\Di-\zeta_\Ui)\epsilon & \left(\frac{t_\Di}{w_\Di}-\gamma_\Di\zeta_\Ui+\frac{s_\Ui v_\Ui-t_\Ui  u_\Ui }{u_\Ui w_\Ui -|v_\Ui |^2}\right)\epsilon^2 \\
-(\zeta_\Di^*-\zeta_\Ui^*)\epsilon & \genfrac{}{}{0pt}{}{1-\frac{1}{2}\bigl(|\gamma_\Di|^2+|\gamma_\Ui |^2-2\gamma_\Ui^*\gamma_\Di\quad}{\qquad+|\zeta_\Di|^2+|\zeta_\Ui |^2-2\zeta_\Ui^*\zeta_\Di\bigr)\epsilon^2}& (\gamma_\Di-\gamma_\Ui )\epsilon \\
\left(\frac{t_\Ui^*}{w_\Ui }-\gamma_\Ui^*\zeta_\Di^*+\frac{s_\Di^*v_\Di^*-t_\Di^*u_\Di}{u_\Di w_\Di-|v_\Di|^2}\right)\epsilon^2 & -(\gamma_\Di^*-\gamma_\Ui^*)\epsilon & 1-\frac{1}{2}\left(|\gamma_\Di|^2+|\gamma_\Ui |^2-2\gamma_\Ui \gamma_\Di^*\right)\epsilon^2
\end{array}\right)\\
&\qquad+{\cal O}(\epsilon^3)\,.
\end{split}
\ee
Note that the leading terms in $\gamma_\Ui $ and $\zeta_\Ui $ have a rather simple form,
\be
\gamma_\Ui =\frac{\lambda_{23}^\Ui}{\lambda_{33}^\Ui}+{\cal O}(\epsilon^2),\quad \zeta_\Ui =
\frac{\lambda_{12}^\Ui\lambda_{33}^\Ui-\lambda_{13}^\Ui\lambda_{32}^\Ui}{\lambda_{22}^\Ui\lambda_{33}^\Ui-\lambda_{23}^\Ui\lambda_{32}^\Ui}+{\cal O}(\epsilon^2)\,,
\ee
while $\gamma_\Di$ and $\zeta_\Di$ are fairly complicated when expressed in terms of the $\lambda_{ij}^\Di$.

In terms of our parameters, for all $\lambda$ real, the CKM mixing angles $\theta_{12}$ and $\theta_{23}$ are then approximately given by
\be
\sin\theta_{12}=(\zeta_\Di-\zeta_\Ui )\epsilon,\qquad\sin\theta_{23}= (\gamma_\Di-\gamma_\Ui )\epsilon\,. 
\ee

\end{document}